\documentclass[12pt, a4paper]{article}
\pdfoutput=1
\usepackage[utf8]{inputenc}
\usepackage{fancybox}
\usepackage[T1]{fontenc}
\usepackage{lmodern, textcomp}
\usepackage{bm}
\usepackage[american]{babel}
\usepackage{csquotes}

\usepackage{eurosym}
\usepackage[nottoc]{tocbibind}
\usepackage{authblk}
\usepackage{fancyhdr}
\usepackage{xspace}

\usepackage{graphicx}
\usepackage{subcaption}
\usepackage{float}
\usepackage{makecell}
\usepackage{multirow}
\usepackage{multicol} 
\usepackage{marginnote}
\usepackage[multiple]{footmisc}

\usepackage[usenames, dvipsnames, svgnames, table]{xcolor}

\usepackage[backend=bibtex8, citestyle=numeric-comp, maxbibnames=99,
			sorting=none, sortcase=false, sortcites=true, giveninits=true]{biblatex}

\usepackage{amsmath, amsfonts, amssymb}
\usepackage{mathtools}
\usepackage{mathrsfs}
\usepackage{cancel}
\usepackage{braket}
\usepackage{tensor}
\usepackage{slashed}
\usepackage{siunitx}
\usepackage{listings}

\usepackage{stmaryrd}
\usepackage{caption}
\usepackage{relsize,exscale}
\usepackage{array}
\usepackage[toc,page]{appendix}

\usepackage{enumerate}

\DeclareSymbolFont{largesymbols}{OMX}{cmex}{m}{n}

\setcellgapes{1pt}
\makegapedcells
\newcolumntype{R}[1]{>{\raggedleft\arraybackslash }b{#1}}
\newcolumntype{L}[1]{>{\raggedright\arraybackslash }b{#1}}
\newcolumntype{C}[1]{>{\centering\arraybackslash }b{#1}}

\newcommand{\Tr}{\mathrm{Tr}}

\newtheorem{theorem}{Theorem}
\newtheorem{definition}{Definition}

\newtheorem{proposition}{Proposition}
\newtheorem{remark}{Remark}

\newcommand{\SU}{\mathrm{SU}}
\newcommand{\SO}{\mathrm{SO}}

\newcommand{\beq}{\begin{equation}}
\newcommand{\eeq}{\end{equation}}
\newcommand{\bea}{\begin{eqnarray}}
\newcommand{\eea}{\end{eqnarray}}
\definecolor{mygray}{gray}{0.3}

\newcommand{\bes}{\begin{eqnarray}}
\newcommand{\ees}{\end{eqnarray}}

\newcommand\restr[2]{{
  \left.\kern-\nulldelimiterspace 
  #1 
  \vphantom{\big|} 
  \right|_{#2} 
  }}

\newcommand{\U}{\mathrm{U}}
\usepackage{multicol}
\usepackage{amssymb}

\usepackage[heightrounded, top=3.5cm, bottom=3.5cm, left=3cm, right=3cm, headheight=14pt]{geometry}

\usepackage[pdftex, breaklinks=true, linktocpage,
	colorlinks=true, urlcolor=blue, linkcolor=blue, citecolor=red
]{hyperref}

\newcommand{\email}[1]{\href{mailto:#1}{\nolinkurl{#1}}}
\newcommand{\emailfoot}[1]{\thanks{\email{#1}}}

\usepackage{marginnote}
\newcounter{draftcommentcnt}
\NewDocumentCommand{\draftcomment}{s O{red} m}{%
	\def\margnote{\IfBooleanTF{#1}{\marginnote}{\marginpar}}%
	\stepcounter{draftcommentcnt}%
	\textcolor{#2}{#3}%
	\margnote{\textcolor{#2}{$\Leftarrow$ \arabic{draftcommentcnt}}}%
}

\fancypagestyle{preprint}{
	\fancyhf{}

	\fancyhead[R]{\textsf{\preprint}}
}

\numberwithin{equation}{section}

\hypersetup{
	pdftitle={Stochastic dynamics for Group Field Theories},
}

\title{Stochastic dynamics for group field theories}

\author[1]{Vincent Lahoche\emailfoot{vincent.lahoche@cea.fr}}
\author[1,2]{Dine Ousmane Samary\emailfoot{dine.ousmanesamary@cipma.uac.bj}}

\affil[1]{%
	Université Paris Saclay, \textsc{Cea}, \textsc{List}, Gif-sur-Yvette, F-91191, France
}

\affil[2]{%
	Faculté des Sciences et Techniques (ICMPA-UNESCO Chair)
	\protect\\
	Université d'Abomey-Calavi, 072 BP 50, Bénin
}

\begin{document}
\maketitle

\begin{abstract}
Phase transitions with spontaneous symmetry breaking are expected for a long time ago for group field theories as a basic feature of the geometogenesis scenario. The following paper aims to investigate the stochastic dynamics of such a field theory, supported by a fictitious “time” inducing a stochastic process described through a Langevin equation, from which the randomness of the tensor field  will be a consequence. Our aim is essentially to propose a non-perturbative renormalization group framework, focusing on an Abelian model, and as a first step on equilibrium dynamics. We then consider the quartic melonic theory in rank $5$, such that the equilibrium state is a just-renormalizable equilibrium theory, and elaborate a non-perturbative renormalization group formalism exploiting the effective vertex expansion techniques for the Martin-Siggia-Rose path integral. We provide some exact expressions in the melonic sector, and a significant effort is made on the pedagogy of the presentation making this work as autonomous as possible.

\end{abstract}
\vspace{1cm}

\hrule
\pdfbookmark[1]{\contentsname}{toc}

\pagebreak
\tableofcontents
\bigskip
\hrule
\pagebreak

\vspace{1cm}

\section{Introduction}

For more than one decade, group field theories
(GFTs) are considered as a promising way to
address the quantum gravity conundrum.
Mathematically, GFTs are fields theories defined
on $d$-copies of a group manifold $\mathrm{G}$, called \textit{group structure}, and distinguish themselves from standard quantum field theories (QFTs) by the specific non-locality of their interactions \cite{Freidel_2005,baratin2012ten,https://doi.org/10.48550/arxiv.1110.5606,https://doi.org/10.48550/arxiv.gr-qc/0607032,https://doi.org/10.48550/arxiv.1210.6257}. In the point of view of quantization, the particles (quanta) associated to group fields are interpreted as elementary excitation of the gravitational field, which, instead of being characterized by concepts like energy, polarization and so on, are characterized by topological and geometrical data. The quantized space-time is of dimension $d$, and the elementary excitation is interpreted as $(d-1)$-simplices with labeled faces. The interactions between these fields dictate the way the faces are “stuck” to each other according to these labels, to give effective $d$ -- simplices. Thus, the non-local structure of the interaction tells us how dual $(d-1)$-simplices are built and glued together. The structure group on the other hand has to reflect the local symmetry group of the dual spacetime. This interpretation can be motivated by the relation between GFTs and covariant approaches of loop quantum gravity (LQG) like \textit{spin-foams} \cite{Perez_2013}. Indeed, GFTs have been historically introduced in the context of the LQG \cite{Ashtekar_2021,rovelli_2004,Thiemann:2007pyv,https://doi.org/10.48550/arxiv.1001.3668} as a clever way to resume SF quantum amplitudes. Hence, on one hand, GFTs can be approached from the quantification of the classical general relativity (GR), which naturally leads to quantum states encoding discrete geometry as triangulation. It must be noticed that alternatively, GFT can be viewed as a second quantized version of LQG \cite{https://doi.org/10.48550/arxiv.1310.7786,oriti2015group}. Finally, the choice of the group structure is imposed by this connection with LQG, as the local group of space-time symmetries ($\SO(3,1)$ with the Lorentz signature, $\SO(4)$ for Euclidean quantum gravity, but other groups can be considered as toy models like $\SU(2)$ for Euclidean $3$D gravity, $\U(1)$ or $\mathbb{R}$).
\medskip

On the other hand, GFT can be approached directly through the prism of discrete random geometry, the continuum limit for quantum space-time being recovered as a phase transition in the model. For 2D, the most popular approach in this direction is random matrix models (RMM). In RMM, Feynman amplitudes provide weights for discrete triangulation, the way the elementary ‘‘triangles” are glued together being imposed by the interactions between matrix fields. The main feature of RMM is the existence of a topological $1/N$ expansion, controlled by the genus $g(\Delta)$ of the dual triangulation $\Delta$, and thus dominated by planar diagrams with $g=0$. Critical properties and continuum limits of RMM are essentially consequences of this basic property \cite{https://doi.org/10.48550/arxiv.1510.04430,Francesco_1995,Seiberg_2007}. Random tensors models (RTM) \cite{guruau2017random,rivasseau2016random,Gurau_2016} are an attempt to extend the success of RMM to dimension higher than $2$. The decisive step in this direction was the discovery by Gurau in 2009 of the existence of a power counting for colored random tensors, which admits an $1/N$ expansion analogous to RMM, controlled by a generalization (but unfortunately no topological) of the genus and called \textit{Gurau degree} $\varpi$. It has been shown that the existence of such a power count is related to an internal structure group, typically $\U(N)$ (or $\mathrm{O}(N)$, see \cite{benedetti20191,carrozza2018large,carrozza2022melonic,carrozza2016n}), leaving the interactions invariant \cite{bonzom2012random,Gurau_2016}. The leading order diagrams, having vanishing Gurau degree, are called melons, and critical properties of melonic sector, as well as double scaling limits, have been investigated for RTMs \cite{bonzom2011critical,bonzom2014double,dartois2013double}. With this respect, GFTs can be viewed as generalized RTM, with group-valued rather than discrete indices. This leads to a restrictive class of RTM, called Tensorial Group Field Theories (TGFTs), which are GFT whose interactions have the same non-local structure as RTMs, said \textit{tensorial interactions}. Note that some additional symmetries like \textit{closure} (Gauss) constraint or \textit{Plebanski} constraint have to be considered in the first point of view \cite{jercher2022emergent,baratin2011quantum,baratin2012group,lahoche2016renormalization}. The closure constraint for instance is a specific kind of gauge symmetry, which requires that physical fields solutions are invariant under the global right translation of the group elements \cite{carrozza2014renormalization,carrozza2014renormalization2}. Imposing it at the quantum level, a Feynman amplitude looks like a partition function for a gauge theory on a random lattice fixed by the cellular complex defined by the Feynman graph, with flat discrete connections. \medskip

The main challenging issue for GFTs remains how a smooth space-time structure corresponding to classical GR can be recovered by summing a very large number of quantum states having a very large number of quanta \cite{https://doi.org/10.48550/arxiv.1302.2849,https://doi.org/10.48550/arxiv.2110.08641,https://doi.org/10.48550/arxiv.1807.04875,https://doi.org/10.48550/arxiv.2112.02585}, and to this aim, the renormalization group (RG) is generally considered as the powerful tool to address this issue. RG is a general concept in physics to tackle the large-scale description of systems involving a very large number of (microscopic) interacting degrees of freedom \cite{ZinnJustinBook2,Zinn-Justin:1989rgp}. There are many incarnations of this idea in physics and  all of them aim to extract the large-scale regularities of a system, replacing its full description with an approximate but effective theory, keeping only relevant features of the original quantum (or statistical) microscopic states. In the Wilsonian point of view, RG is constructed from a partial integration procedure, integrating out ‘‘rapid" modes to construct an effective physics for ‘‘slow" modes, keeping fixed the large distance physics. Two different strategies have been considered for constructing RG flow for GFTs. The first one is based on lattice renormalization, viewing spin foams as a direct space regularization of quantum gravity amplitudes \cite{https://doi.org/10.48550/arxiv.1409.1450}. The other approach is based on local field theories and renormalization techniques. Indeed, the existence of a power counting for TGFTs provides a novel notion of a locality called \textit{traciality}, reflecting the way the divergences can be factorized out of some tensorial interaction. Renormalization ‘‘à la Wilson” requires identifying ‘‘slow” (infrared) and ‘‘rapid” (ultraviolet) modes. For TGFTs defined as enriched RTM with group valued indices, no such distinction exist between UV and IR modes. Indeed, usual GFT models or RTMs suggest that theories have to be ultra-local, with propagator equals to the identity matrix or suitable projectors, ensuring the global $\U(N)$ (or $\mathrm{O}(N)$) invariance of RTMs. The triviality of the propagator is surely appropriate for simplicial quantum gravity perspectives, but does not allow for the definition of a proper notion of scale. However, this poses a difficulty, because, without such a suitable notion of scale, no distinction exists between fluctuating degrees of freedom. There are no “infrared” (IR) or “ultraviolet” (UV) degrees of freedom, and any partial integration procedure ‘‘à la Wilson” imposes to arbitrarily fix what are IR and UV. This point of view has been considered in a series of papers both using perturbative and nonperturbative RG technics \cite{https://doi.org/10.48550/arxiv.1701.03029,Eichhorn_2020,Eichhorn_2019,Eichhorn_2019,Eichhorn_2014,Eichhorn_2013,Br_zin_1992}, but to date, there is no consensus about the reliability of the resulting RG flow -- see \cite{Lahoche:2020pjo,Lahoche:2019ocf}.
\medskip

A solution considered in the literature consists in modifying the propagator by adding a “Laplacian” type term (defined on the considered structure group) to the Gaussian kernel, whose non-trivial spectrum then provides a non-ambiguous notion of scale \cite{Carrozza_2014}. This Laplace type propagators may be viewed as a regulation, that affects only UV degrees of freedom but disappears in the IR, leading to an effective, dynamically generated ultralocal theory for the RG flow. Moreover, the presence of such a Laplacian can be motivated by the computation of radiative corrections to GFTs, which require such a Laplacian as a counter-term to be well-defined as the cut-off in large momenta is removed \cite{ben2011radiative}. There is vast literature and active research on this topic, exploring both the perturbative and nonperturbative aspects of the TGFTs \cite{Carrozza_2014, Carrozza_2015aaa,Carrozza_2015a,Carrozza_2016ccc,Geloun_2018,https://doi.org/10.48550/arxiv.1111.4997,Geloun_2016,Benedetti_2015,Benedetti_2016,Lahoche_2017bb,Carrozza_2017a,Carrozza_2017,carrozza2015discrete}. These investigations generally seek to reveal fixed point solutions and phase transitions. Such solutions have been found for some models, although very dependent on the approximation scheme used to solve the flow equations in the nonperturbative regime. At first, it seemed that the existence of such fixed point solutions and second-order phase transitions was a quasi-universal feature for TGFTs \cite{Geloun_2016,Benedetti_2015}. However, our recent works based on methods outperforming standard vertex expansion showed that it is not the case \cite{Lahoche_2019bb,Lahoche_2019a,Lahoche_2021c,Lahoche_2020b}. These methods consider both Ward identities and an effective vertex expansion (EVE) technique, which takes into account the full momenta dependence of the effective vertex and formally resumes branched sectors as the melonic one \cite{lahoche2021no,Lahoche_2020b,Lahoche_2021c,Lahoche_2020d,Lahoche_2019a,Lahoche_2019bb,Lahoche:2020pjo}. The existence of second-order phase transitions in the phase space of TGFTs is expected to be a basic requirement for geometrogenesis-type scenarios, where the semi-classical spacetime is assumed to emerge as an ‘‘inter-theoretical process” from a Bose-Einstein condensation. It is worth mentioning that the hypothesis of the existence of such condensates has allowed the rapid development of a whole literature exploring models of quantum cosmology \cite{oriti2018black,gielen2018cosmological,de2017dynamics,kegeles2018inequivalent,oriti2017universe,oriti2016horizon,oriti2015generalized,gielen2014quantum,gielen2014homogeneous,marchetti2021effective,jercher2022emergent,gielen2022effective,oriti2017bouncing,oriti2016emergent}.
\medskip

In this paper, we address the problem of quantization of GFTs through a first-order stochastic Langevin-type equation, such that equilibrium configurations match with standard path integral quantization for pure gravity models. The introduction of this  equation can be considered a purely mathematical exercise, but can also have real physical meaning. It can be seen as a way to dynamically (i.e., out of equilibrium) address the phase transitions revealed by the RG for equilibrium theory. Indeed, in general, although the phase transitions are discussed in equilibrium, assuming the existence of an observer outside the system and adjusting adiabatically the parameters of the theory, a realistic description of the transition requires a non-equilibrium approach to this phenomenon. However, such a non-equilibrium treatment is generally a difficult task \cite{livi2017nonequilibrium,https://doi.org/10.48550/arxiv.2110.08641}. In the quantum gravity context, an additional question that one is entitled to ask concerns the choice of the variable identified as a “time”. This question is closely related to the issue of time in classical and quantum gravity. Indeed, already at the classical level, general relativity does not allow in general to isolate a physical variable as a preferred time in the infinity of possible choices of “coordinated time”, and experimentally this is always the evolution of a physical variable, for a given problem, which defines a particular notion of the clock. Notice that this point of view is in agreement with the standard relational interpretation, where space and time are understood as relative special configurations of some fields, used as ‘‘clock” and ‘‘rulers” \cite{https://doi.org/10.48550/arxiv.2110.08641}. An intriguing relation between the choice of a physical time and the definition of equilibrium states has been proposed in the series of works \cite{rovelli_2004,rovelli1993statistical,rovelli1993statistical2,rovelli2011thermal,menicucci2011clocks,martinetti2003diamond}. For instance, it is shown that the statistical properties of the cosmic radiation background reveal a preferred time, which happens to be the ‘‘cosmic time" considered in the literature. A discussion for quantum theories is given in \cite{connes1994neumann}, where authors consider the one-parameter group of automorphism underlying by Von-Neumann algebra of quantum systems through Tomita-Takesaki theorem, as a single-out time flow. Thermal equilibrium states have been already considered for GFTs, especially in the context of cosmology, see \cite{https://doi.org/10.48550/arxiv.2010.15445} and references therein for an extended discussion about relational functional dynamics. However, except for these special configurations, no preferred time is expected at the fundamental level for a background-independent quantum theory of gravitation. Indeed, the structured space-time manifold is assumed to be entirely dissolved at the phase transition points where collective states of gravity quantum modes are not suitably described by Bose-Einstein condensates, and the concept of ‘‘direction” disappears as the concept of a smooth manifold. Phase transition in the GFTs is for this reason generally understood as a change of the theoretical paradigm -- i.e., as the identification of some regions of the phase space where the collective behavior of quantum gravity atoms can be approached with an effective, semi-classical theory, as a quantum gravity condensate for current cosmological solutions discussed previously. A way to recover a notion of temporality even approximates is through the notion of relational (space)-time, which is already found in classical GR. The contiguity relations between fields allow us to define space-time properties and, in particular, the coupling with gravitation defines the metric field. In that point of view, matter fields can be used to construct material frames locally, one of them playing the role of physical time. Let us recall how that works in the classical setting. Consider a theory involving $N$ fields $\phi_i(x_0,\vec{x})$, for $i=1, \cdots, N$ and $(x_0,\vec{x})$ are arbitrary coordinates for space-time evens. In concrete experiments, clocks and other reference frames are defined as specific configurations of four of these fields, which are assumed to behave as classically as to define such a reference frame and we denote them as $\phi_0, \cdots, \phi_3$. To be a good reference frame, we assume that locally space-time coordinates $x^{\mu}$ can be expressed uniquely in terms of the four numbers $\phi_\mu$. In particular, $x^{0}=F_0(\phi_0, \phi_1, \phi_2, \phi_3)$, and, we can express the equations of motions for the remaining fields in terms of the physical coordinates $(\phi_0, \vec{\phi}\,)$ \cite{rovelli_2004,connes1994neumann}. This relational viewpoint can be expected to survive when the gravitational field is described in a quantum manner, at least in a certain regime. An auxiliary matter field could play the role of a clock, as long as one can neglect the quantum character of this field. One can expect that such a regime would allow describing the (relational) dynamics of space-time towards or from the emergence of classical space-time, but the transition point (geometrogenesis), where the quantum nature of all fields cannot be neglected, breaks the dynamical description. For more details on the concept of emergence of time in quantum gravity, the reader may consult for instance \cite{https://doi.org/10.48550/arxiv.1807.04875,https://doi.org/10.48550/arxiv.2110.08641}. Recent application of relational time for TGFTs in the context of quantum cosmology can be found in \cite{marchetti2021effective,wilson2019relational,li2017group,https://doi.org/10.48550/arxiv.2112.02585,marchetti2021phase}.
\medskip

In the mathematical formalism presented in this paper, the GFT is quantized by a stochastic equation. This dynamics can be rewritten as a functional integral by the method detailed in the section \ref{secPath}, where time appears formally as a scalar variable, and the corresponding field theory is identified with a GFTs on the structure group $\mathrm{G}^{\times d} \times \mathbb{R}$, considered for instance in \cite{marchetti2021phase,wilson2019relational} as describing a scalar field coupled to gravity. In this paper, intended to be the first of a series, we provide the foundations of the formalism, focusing on equilibrium dynamics as a benchmark. We consider an Abelian TGFT without closure constraint, whose structure group will be $\U(1)$, and whose equilibrium states will correspond to a GFT without matter degree of freedom, just-renormalizable in rank $d=5$. This choice may appear somewhat artificial, since the GFTs usually considered as physically realistic candidates for quantum gravity are formulated on non-Abelian groups and subject to additional constraints, such as the closure condition or Plebanski’s constraint \cite{baratin2012group}. Our aim in this paper, however, is primarily to demonstrate how the renormalization group (RG) methods we have developed can be extended to a new setting, namely, a stochastic GFT, and to explore what this perspective offers in contrast with the traditional framework, using a model where these techniques are technically more manageable. This work should therefore be regarded as the first step in a broader program of investigations. In the present contribution, we focus mainly on the combinatorics of interactions in the construction of the RG.\footnote{It is worth noting that an Abelian GFT with a trivial propagator and a momentum cut-off $\Lambda$ is essentially a rank-5 random tensor model in disguise. In this sense, one may loosely refer to it as a “pure gravity model” in equilibrium, insofar as no matter degrees of freedom are included.}.
Our approach in this paper is essentially based on the nonperturbative RG (NPRG) formalism \cite{Morris_1994a,MORRIS_1994,Berges_2002,Delamotte_2012,Dupuis_2021}, and construct approximate solutions of the exact Wetterich equation using both effective vertex expansion (EVE) recently introduced in the GFTs context \cite{Lahoche_2019bb,Lahoche_2019a,Lahoche_2020d,Lahoche_2021c,Lahoche_2020b,Lahoche:2018oeo,Baloitcha:2020lha}, and Ward's identities to determine the derivative of the effective vertices with respect to the external momentum, involved in the computation of the anomalous dimension. The resulting equations are identical to the equilibrium case as soon as no coarse graining over time is done, what is expected because of the equilibrium assumption. The flow equations as coarse-grainig over frequencies is considered are also shortly investigated. Note that Besides we focus on TGFTs, we expect the same formalism should be used to investigate analogue regimes for theories with trivial propagators (i.e., without intrinsic scaling), as RMMs and RTMs, which will be the topic of forthcoming work, and well as the more difficult out of equilibrium regime. 
\pagebreak

\paragraph{Outline.} In the section \eqref{themodel} we define the model and provide the path integral approach allowing us to well define the functional renormalization group applicable with the so-called Wetterich equation. In section \eqref{sec3} we introduce the FRG formalism and the time reflection symmetry and causality which allows for coarse-grain by modifying the original Langevin equation. This also helps to add in the Langevin equation a driving force which depends non locally on the standard time on the classical trajectory and preserves causality. We also provide the scaling dimension of the model. In section \eqref{sec4}, we use standard local potential approximation to construct solutions of the exact RG equation. We consider two approximations, the first is the crude truncation and the second comes from the effective vertex expansion in the leading order melonic approximation. In section \eqref{WTI} we study the symmetry of our model given by the Ward identities and provide the rigorous analysis of compatibility with the flow equation and the optimal choices of the regulators. Section \eqref{sec6} is deserving for discussions on numerical investigations. We conclude our work in section \eqref{sec7}.

\section{Stochastic group field theories}\label{themodel}
In this section, we define the models and conventions used in the rest of the paper. We also derive a path integral representation and a few formal properties, which we will exploit in the next section devoted to the renormalization group. The reader may consult \cite{livi2017nonequilibrium,Zinn-Justin:1989rgp,ZinnJustinBook2} for more details about formal computations of the Langevin equation.

\subsection{The model}
A group field $\varphi$ is a field defined on $d$-copies of a group manifold $\bm{\mathrm{G}}$:
\begin{equation}
\varphi: (g_1,\cdots, g_d)\in (\bm{\mathrm{G}})^{\times d} \to \varphi(g_1,\cdots,g_d)\in \mathbb{K} \,.
\end{equation}
Usually $ \mathbb{K} = \mathbb{C}, \mathbb{R}$. In this paper, we focus on complex group fields, $\mathbb{K} \equiv \mathbb{C}$. To shorten the notations, we will denote by $\bm{g}:= (g_1,\cdots, g_d)$ the elements of $(\bm{\mathrm{G}})^{\times d}$ and by $\varphi(\bm{g})$ the value taken by the field at the point $\bm{g}$. We generally assume $\varphi$ to be a square-integrable function, and the standard $L^2((\bm{\mathrm{G}})^{\times d})$ inner product:
\begin{equation}
(\varphi, \varphi^\prime):= \int d\bm{g} \,\bar{\varphi}(\bm g) \varphi^\prime(\bm g)
\end{equation}
is assumed to be bounded: $\Vert \varphi \Vert := (\varphi, \varphi) < \infty$. In these notations, $\bar{\varphi}$ designates the standard complex conjugation of $\varphi$ and:
\begin{equation}
d\bm g:= dg_1dg_2\cdots dg_d\,,
\end{equation}
where $dg_\ell$ is the Haar measure over $\textbf{G}$. We will suppose that this field is moreover a dynamic variable, depending on a parameter $t\in \mathbb{R}$ called the “time”. The evolution of the field is postulate to satisfy  the dissipative Langevin equation, which is given by the following:
\begin{equation}
\boxed{
\dot{{\varphi}}(\bm g,t)=- \Omega \frac{\partial }{\partial \bar{\varphi}(\bm g,t)} \mathcal{H}[\varphi,\bar{\varphi}]+\eta(\bm g,t)\,,}\label{langevin}
\end{equation}
where $\eta(\bm g,t)$ is a random group field, playing the role of white noise, $\Omega > 0$ is a time scale, the notation "dot" means $d/dt$ and $\mathcal{H}$, the Hamiltonian, defines the deterministic parts of the equation.

\begin{remark}
Before turning to the definition of the model, let us make a general remark concerning the perspective adopted in this work. The notion of time introduced here is not the standard one usually employed in the GFT literature, where relational time is implemented through clock fields, typically leading to second derivatives with respect to this variable \cite{marchetti2021phase}. In contrast, the present framework may suggest a kind of “non-relativistic” evolution. We stress, however, that such terminology is not really meaningful in this context: general covariance is not built into our construction and would in any case require the introduction of additional degrees of freedom, as in more complete and physically realistic GFT models. One could speculate that the Langevin equation might emerge as a “slow-roll” approximation, akin to stochastic inflation, for a more fundamental equation respecting some form of covariance. Yet such an interpretation would be premature in the absence of further investigation. For the purposes of this paper, we will therefore refrain from speaking about covariance or about a “non-relativistic” regime. Instead, we simply regard equation \ref{langevin} as defining a toy model that enables us to study fluctuations around an equilibrium state with respect to which a preferred notion of time is defined.\end{remark}
\medskip

\begin{figure}
\begin{center}
\includegraphics[scale=1]{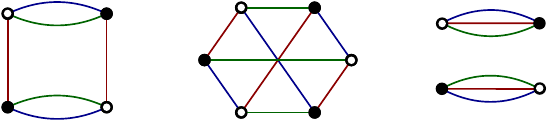}
\end{center}
\caption{Example of tensorial invariants for $d=3$}\label{figBubbles}
\end{figure}

\begin{figure}
\begin{center}
\includegraphics[scale=1]{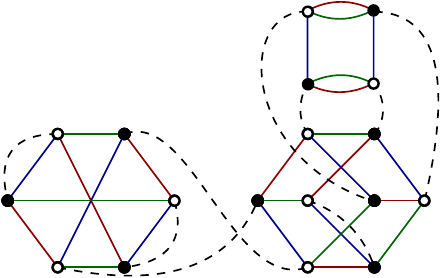}
\end{center}
\caption{A typical Feynman graph for $d=3$, with 3 vertices and 9 propagator edges.}\label{figFeynman}
\end{figure}

Without additional constraint on the random field $\varphi$, we assume that the probability measure for $\eta(\bm g,t)$ is:
\begin{equation}
d\rho(\eta):= \frac{1}{z_0}\, \exp \left(-\frac{1}{\Omega}\int dt d\bm g\, \bar{\eta}(\bm g,t)\eta(\bm g,t) \right)\, d[\eta]\,,
\end{equation}
where $d[\eta]:=\prod_{\bm g,t} d\eta(\bm g,t)d\bar{\eta}(\bm g,t)$ is the formal functional measure defining path integral, and the normalization $z_0$ being such that:
\begin{equation}
\langle \eta(\bm g, t) \bar{\eta}(\bm g^\prime, t^\prime) \rangle_\eta =\Omega\, \delta(\bm g^\prime (\bm g)^{-1}) \delta(t-t^\prime)\,,\label{noisecorrelation}
\end{equation}
the notation $\langle X \rangle_\eta$ meaning average over $\eta$ with probability density $d\rho(\eta)/d[\eta]$, and:
\begin{equation}
\delta(\bm g^\prime (\bm g)^{-1}) := \prod_{\ell=1}^d \delta(g_\ell^\prime g_\ell^{-1})\,,
\end{equation}
where $\delta(g_\ell g_\ell^{-1})$ denotes the standard Dirac delta over $\bm{\mathrm{G}}$,
\begin{equation}
\int dg \, \delta(g^\prime (g)^{-1}) f(g)=f(g^\prime)\,,
\end{equation}
for some function $f$. The Hamiltonian $\mathcal{H}$ will be designed such that equilibrium configurations (see Section \ref{secPath}) reproduce the generalized Gibbs states used in standard definitions of GFTs. With the previous definition, long time equilibrium states (i.e., the probability density for a field configuration) must behave like $P[\varphi,\bar{\varphi}]\sim e^{-2\mathcal{H}[\varphi,\bar{\varphi}]}$, accordingly to the usual definition provided that $\mathcal{H}$ is nothing but the microscopic action for group field. Because we focus on the TGFT formalism, we expect $\mathcal{H}$ is the sum of two contributions:
\begin{enumerate}
\item A kinetic part $\mathcal{H}_{\text{kin}}$, involving a non-trivial kernel depending on the Laplace-Beltrami operator $\Delta_{\bm g}$ over the manifold $(\bm{\mathrm{G}})^{\times d}$:
\begin{equation}
\mathcal{H}_{\text{kin}}[\varphi, \bar{\varphi}]:=\int d\bm g\, \bar{\varphi}(\bm g,t) (-\Delta_{\bm g}+m^2)\varphi(\bm g,t)\,,
\end{equation}
for some coupling constant $m^2$ defining a \textit{mass scale}.

\item An interaction $\mathcal{H}_{\text{int}}$, which expands in power of fields. The terms involved in that expansion, as the interaction Hamiltonian itself are furthermore assumed to be invariants under unitary transformations $U:L^2(\bm{\mathrm{G}})\to L^2(\bm{\mathrm{G}})$ defined as:
\begin{equation}
\varphi(\bm g) \to \varphi^\prime(\bm g):=\int d\bm{g}^\prime\, \left[\prod_{\ell=1}^d U_\ell(g_\ell,g^\prime_\ell) \right] \varphi(\bm{g}^\prime)\,.\label{unitarytrans}
\end{equation}
This defines a particular non-locality for interactions, called “tensoriality”, and terms involved in the expansion of $\mathcal{H}$ are \textit{tensorial invariants}.
\end{enumerate}
These invariants admit an elegant representation in terms of $d$-colored bipartite regular graphs. The receipt is the following:
\begin{enumerate}
\item To each field $\varphi$ and $\bar{\varphi}$ we assign a black and white dot respectively, with $d$ half-colored edges hooked to them, materializing the $d$ group variables $g_1,\cdots, g_d$:
\begin{equation*}
\vcenter{\hbox{\includegraphics[scale=0.8]{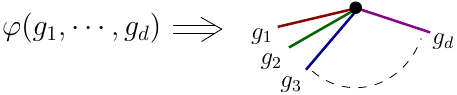}}}\qquad \vcenter{\hbox{\includegraphics[scale=0.8]{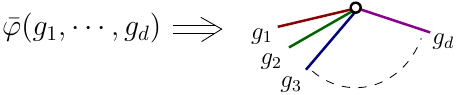}}}\,.
\end{equation*}
\item Colored edges are then hooked together, accordingly with their respective colors, between black and white dots only.
\end{enumerate}
On Figure \ref{figBubbles} we show some examples for $d=3$. To provide an explicit example, the first diagram reads explicitly as:
\begin{equation}
\vcenter{\hbox{\includegraphics[scale=0.75]{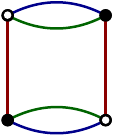}}}\equiv \int \prod_{i=1}^3 dg_i dg_i^\prime \varphi(g_1,g_2,g_3) \bar{\varphi}(g_1,g^\prime_2,g^\prime_3) \varphi(g^\prime_1,g^\prime_2,g^\prime_3)\bar{\varphi}(g_1^\prime,g_2,g_3)\,,
\end{equation}
assuming that the red edge corresponds to color $1$. As illustrated by the last example in Figure \ref{figBubbles}, graphs can be connected or not, and in this case, they are the product of connected graphs. We call \textit{a bubble} such a connected graph, made of a single piece. We moreover assume that $\mathcal{H}_{\text{int}}$ expands as:
\begin{equation}
\mathcal{H}_{\text{int}}[\varphi,\bar{\varphi}]=\sum_b \lambda_b \Tr_b[\varphi(t),\bar{\varphi}(t)]\,,\label{HamiltonianInt}
\end{equation}
the sum running over bubbles $b$ involving more than $2$ fields, and $\Tr_b[\varphi,\bar{\varphi}]$ denotes the corresponding tensorial invariant.
Long time equilibrium states (i.e., the probability density that a group field $\varphi$ has a given value in the ‘‘volume”\footnote{We use the notation "$d$" for the functional measure of time-dependent states, and the notation "$D$" for equilibrium, time-independent configurations.} $D[\varphi]:=\prod_{\bm g} d\varphi(\bm g) d\bar{\varphi}(\bm g)$, if it exists, must behave like (see Section \ref{secPath}):
\begin{equation}
\rho[\varphi,\bar{\varphi}]= \frac{1}{Z[\{\lambda_b\}]}\,e^{- 2\mathcal{H}[\varphi,\bar{\varphi}]}\,,\label{equilibrium}
\end{equation}
the partition function $Z[\{\lambda_b\}]$ which normalizes the state is given by the formal path integral over field configurations:
\begin{equation}
Z[\{\lambda_b\}]:= \int D[\varphi] \, e^{-2\mathcal{H}[\varphi,\bar{\varphi}]}\,.\label{normequilibrium}
\end{equation}
 Hence, the time variable is related to the definition of equilibrium states given by \eqref{equilibrium}, accordingly with the point of view of \cite{rovelli1993statistical,rovelli1993statistical2}, and we are aiming to study small perturbations with respect to it, described by the Langevin equation \eqref{langevin}. The perturbative expansion of the partition function organizes as a sum over quantum amplitudes that we denote as $A(G)$, labeled with vacuum Feynman graphs $G$. An example of such a Feynman graph is provided by Figure \ref{figFeynman}, the dotted edges materializing Wick contractions with free propagator $C(\bm g, \bm g^\prime)$,
\begin{equation}
C(\bm g, \bm g^\prime):=\, \frac{1}{2}\int_{1/\Lambda^2}^{+\infty} d\alpha e^{-\alpha m^2}\prod_{\ell=1}^d K_\alpha(g^\prime_\ell (g_\ell)^{-1})\,,
\end{equation}
for some UV cut-off $\Lambda$, $K_{\alpha}(g^\prime (g)^{-1})$ denotes the heat kernel, solution of the equation:
\begin{equation}
\frac{\partial}{\partial \alpha} K_{\alpha} (g)=\Delta_g K_{\alpha} (g)\,,
\end{equation}
with boundary conditions $K_{\alpha\to 0}(g^\prime (g)^{-1}) = \delta(g^\prime (g)^{-1})$.
Feynman graphs like the one pictured in Figure \ref{figFeynman} look like bipartite regular $(d+1)$-colored graphs, attributing the color ‘‘0” to the dotted edges. A very important notion for such a graph is the faces, and we recall the definition here:
\begin{definition}
A face $f$ is a bi-colored cycle (including color $0$), indexed by a couple $(\ell,\ell^\prime), \ell\neq \ell^\prime$. Such a cycle may be open (open face) or closed (closed face). The boundary of a face, $\partial f$ is the set of colored edges along the cycle.
\end{definition}\label{deffaces}
In the rest of this article, we will focus on the compact Abelian group $\bm{\mathrm{G}}= \U(1)$, and we normalize the Haar measure as:
\begin{equation}
\int d\bm g =1\,.
\end{equation}
The group is isomorphic to the unit circle, and each element of the group can be represented by $g\in \U(1) \equiv e^{i \theta} \in S_1$, where $\theta \in [0, 2\pi[$. Irreducible representations of the group are therefore $e^{i p \theta}$ for $p \in \mathbb{Z}$, and the standard Peter-Weyl theorem allows decomposing functions over the group manifold $\U(1)$ along this basis. For the heat kernel, for instance, we have:
\begin{equation}
K_{\alpha} (g) \big\vert_{g\equiv e^{i \theta}}= \sum_{p\in \mathbb{Z}} e^{-\alpha p^2} e^{ip \theta}\,,
\end{equation}
and the propagator in the Fourier representation reads:
\begin{equation}
C(\bm{p},\bm{p}^\prime)=\frac{1}{2}\frac{\delta_{\bm{p}\bm{p}^\prime}}{\bm{p}^2+m^2}\,,\label{equilibriumpropa}
\end{equation}
where $\bm{p}\in \mathbb{Z}^d$, $\bm{p}^2:=\sum_{\ell=1}^d p_{\ell}^2$ and $\delta_{\bm{p}\bm{p}\,^\prime}:=\prod_{\ell=1}^d \delta_{p_\ell p_\ell^\prime}$. This theory has the property to be power countable, and we have the following statement \cite{samary2014just,lahoche2015renormalization,carrozza2014renormalization2,carrozza2014renormalization}:
\begin{proposition}\label{propositionpower}
Let $A(G)$ the regularized Feynman amplitude associated with a Feynman diagram $G$, with $L(G)$ dotted edges and $F(G)$ closed faces of type $(0\ell)$, $\ell\in \llbracket 1,d \rrbracket$. Its dependence on the UV cut-off $\Lambda$ is given by:
\begin{equation}
\vert A(G) \vert \sim \Lambda^{\omega(G)}\,,
\end{equation}
where:
\begin{equation}
\omega(G)=-2L(G)+F(G)\,.
\end{equation}
\end{proposition}
Leading order graphs are those for which $\omega$ is optimal. The diagrams that make this counting optimal are called melons, and can be defined by a simple recursion, see \cite{carrozza2014renormalization2,Lahoche:2018oeo} and section \ref{nonbranching}. Melonic graphs are those for which the number of faces is maximal as $L(G)$ fixed. We can show that for these diagrams the number $V(G)$ of vertices is related to the numbers $F(G)$ and $L(G)$ by \cite{carrozza2014renormalization2,Lahoche:2018oeo}:
\begin{equation}
F(G)=(d-1)(L(G)-V(G)+1)\,.
\end{equation}
Hence, defining $\rho:=(d-1)(L(G)-V(G)+1)-F$, it can be established that the power counting can be read as:
\begin{equation}
\omega(G)=\sum_k ((d-3)k-(d-1))v_k(G)+(d-1)-\frac{N(G)}{2}(d-3)-\rho(G)\,.\label{powerCounting}
\end{equation}
In that equation, $v_k(G)$ denotes the number of bubbles with valence $2k$ (with $k$ white nodes), and $N(G)$ is the number of external edges.
\begin{definition}\label{definitionMelon}
For melonic diagrams $\rho(G)=0$, and it can be proved that $\rho(G)>0$ otherwise.
\end{definition}
The theory will be \textit{power-counting just renormalizable} if and only if $(d-3)k-(d-1)=0$. In particular, the sixtic model is just-renormalizable for $d=4$. In this paper, we will focus on the melonic quartic model, which is just-renormalizable for $d=5$, where power-counting reads as:
\begin{equation}
\boxed{
\omega_{\text{melon}}=4-N(G)\,.}
\end{equation}
In particular, only $2$ and $4$-points diagrams are power-counting divergent and require to be renormalized. The melonic diagrams have the property to be \textit{contractible}. This property invites the definition of a locality principle, and tensorial invariants which are connected and contractible are said to be local in that point of view. We will speak of \textit{traciality} to designate this specific notion of locality. This principle of locality allows us to define counter-terms, and we can show the following theorem \cite{lahoche2015renormalization,samary2014just}:
\begin{theorem}
The quartic melonic model in $d=5$ is just-renormalizable, and divergences can be removed with counter-terms for mass, quartic couplings and field strength normalization.
\end{theorem}\label{th1}
Explicitly, the renormalizable Hamiltonian reads as:
\begin{equation}
\boxed{
\mathcal{H}[\varphi,\bar{\varphi}]:=\int d\bm g\, \bar{\varphi}(\bm g) (-\Delta_{\bm g}+m^2)\varphi(\bm g)+\lambda \sum_{\ell=1}^5 \,\vcenter{\hbox{\includegraphics[scale=0.8]{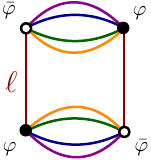}}}\,,}\label{model}
\end{equation}
where we attributed the same coupling constant for all the quartic interactions.
\medskip

Traciality allows us to think of locality in that context, as related to tensorial invariance, and we adopt the following definition in this paper:
\begin{definition}\label{locality}
Any tensorial invariant whose graph is a bubble is said to be local. In the same way, any function which expands as a sum of terms labeled with bubbles only will say to be local.
\end{definition}

\subsection{Dynamic action and path integral}\label{secPath}
We denote as $\bm q(t):=\{ \varphi(\bm g,t), \bar{\varphi}(\bm g,t)\}$ a given position for the random complex field in the functional space. Due to the randomness of the white noise $\eta(\bm g,t)$, trajectories can be suitably described through a probability distribution:
\begin{equation}
P(\bm q, t;\bm q^\prime, t^\prime)=\big\langle \delta(\bm q(t)-\bm q) \big\rangle_\eta\,, \label{eqProba}
\end{equation}
for $t>t^\prime$, assuming the initial condition $\bm{q}(t^\prime)=\bm{q}^\prime$. In the rest of the paper, we will use the shortest notation $P(\bm q, t)$ for $P(\bm q, t;\bm q^\prime, t^\prime)$, disregarding the initial state. The Langevin equation \eqref{langevin} being local in time, equation \eqref{eqProba} define a Markov-process whose evolution follows a Fokker-Planck equation \cite{ZinnJustinBook2}:
\begin{equation}
\frac{\partial }{\partial t} P(\bm q,t)=\Omega \hat{H} P(\bm q,t)\,,
\end{equation}
with:
\begin{equation}
\hat{H}:=\int d\bm g \left(\frac{\delta^2 }{\delta \varphi(\bm g)\partial \bar{\varphi}(\bm g)}+2\frac{\delta^2 \mathcal{H}}{\delta \varphi(\bm g)\partial \bar{\varphi}(\bm g)} +\frac{\delta \mathcal{H}}{\delta \varphi(\bm g)}\frac{\delta}{\delta \bar{\varphi}(\bm g)}+\frac{\delta \mathcal{H}}{\delta \bar{\varphi}(\bm g)}\frac{\delta}{\delta \varphi(\bm g)}\right)\,.
\end{equation}
This equation admits a long time equilibrium solution, if it exists, this solution is given by:
\begin{equation}
\rho(\bm q):=\lim\limits_{t \to +\infty} P(\bm q,t;\bm {q}^\prime, t^\prime)\,,
\end{equation}
which corresponds to stationary solutions of the Fokker-Planck equation, and it is easy to check that: $\rho(\bm q)\sim e^{-2\mathcal{H}}$, accordingly with equation \eqref{equilibrium}. This equilibrium solution exists provided that it is normalizable, i.e., that the integral \eqref{normequilibrium} exists. The transition probability $P(\bm q,t;\bm {q}^\prime, t^\prime)$ can be represented as a path integral. We introduce it here with some details -- see \cite{ZinnJustinBook2,livi2017nonequilibrium} for a complement. The basic ingredient is the following formal relation\footnote{Note that the Langevin equation that we consider is a first order differential equation, admitting a single causal solution, see \cite{ZinnJustinBook2}. This uniqueness can be challenged for some non-equilibrium configurations, whose classical action admits a large number of minima. }
\begin{equation}
1\equiv \int d\bm q \,\delta \left(\dot{\varphi}+\delta_{\bar{\varphi}} \mathcal{H}^\prime-\eta \right)\delta \left(\dot{\bar{\varphi}}+\delta_{\varphi}\mathcal{H}^\prime-\bar{\eta} \right)\, (\det \mathcal{M})^2\,,\label{identitynice}
\end{equation}
where $\mathcal{H}^\prime:=\Omega \mathcal{H}$, $\mathcal{M}$ is the operator-matrix with entries:
\begin{equation}
\mathcal{M}(\bm{g}^\prime,t^\prime; \bm g,t):= \delta(\bm g^\prime (\bm g)^{-1})\frac{d}{dt}\delta(t-t^\prime)+\frac{\delta^2 \mathcal{H}^\prime}{\delta \bar{\varphi}(\bm g,t)\delta \varphi(\bm g^\prime,t^\prime)}\,.\label{defM}
\end{equation}
We can then use this representation of the identity to determine the classical action associated with the Markov process, by rewriting the generating function:
\begin{equation}
Z[J,\bar{J}]:= \bigg\langle \exp \left( \int dt d\bm g J(\bm g,t) \bar{\varphi}(\bm g,t)+\int dt d\bm g \bar{J}(\bm g, t) \varphi(\bm g,t) \right) \bigg\rangle_\eta\,.
\end{equation}
Note that, because of the normalization for the averaging over $\eta$, we must have $Z[J=0,\bar{J}=0]=1$, fixing the normalization; moreover $\bm q(t)$ is assumed to be a solution of the motion equation for some initial conditions. It can be suitable to take the initial condition for $t^\prime=-\infty$, to ensure that the distribution is in equilibrium (if it exists). We introduce the shortest notation $\bm J(t)=\{\bar{J}(\bm g,t),J(\bm g,t)\}$ and we define the dot product:
\begin{equation}
\bm J \cdot \bm q := \int dt d\bm g \bar{\varphi}(\bm g,t) J(\bm g,t) +\int dt d\bm g \bar{J}(\bm g, t) \varphi(\bm g,t)\,.
\end{equation}
Introducing the identity \eqref{identitynice} in the previous equation, it becomes:
\begin{align}
Z[J,\bar{J}]= \int d\bm q d\rho(\bm{\eta})\, e^{\bm J \cdot \bm q} \delta \left(\dot{\varphi}+\delta_{\bar{\varphi}}\mathcal{H}^\prime-\eta \right)\delta \left(\dot{\bar{\varphi}}+\delta_{\varphi}\mathcal{H}^\prime-\bar{\eta} \right)\, (\det \mathcal{M})^2\,.
\end{align}
The delta functions can be easily integrated out, and to compute the determinant we can use the well-known formula $\det \mathcal{M}=\exp(\Tr\ln(\mathcal{M}))$. One can easily check that:
\begin{equation}
\det \mathcal{M} \sim \exp \left(\theta(0) \int dt d\bm g \,\frac{\partial^2 \mathcal{H}^\prime}{\partial \bar{\varphi}(\bm g,t)\partial \varphi(\bm g,t)} \right)\,.
\end{equation}
In this equation, the choice of this function as the inverse of $\partial/\partial t$ is required by causality (which is expected from the Langevin equation). However, a problem arises because $\theta(0)$ is undefined. There are two allowed solutions, depending on if we use Îto or Stratonovich prescription for computing time discretized version of path integrals \cite{ZinnJustinBook2,canet2011general,mannella2022ito}:
\begin{enumerate}
\item In the Îto sense, we evaluate the integrand at the left end point.
\item In the Stratonovich sense, we evaluate the integrand at the “middle” point.
\end{enumerate}
Each of these choices corresponds to a different convention for $\theta(0)$. Thus, $\theta(0)=0$ for Îto, and $\theta(0)=1/2$ for Stratonovich. In this paper, we will work within \^Ito convention, and we set $(\det \mathcal{M})^2=1$ in the calculations, leading to:
\begin{equation}
Z[J,\bar{J}]= \int d\bm q\, e^{-\frac{1}{\Omega}\int dt d\bm g (\dot{\varphi}+\delta_{\bar{\varphi}}\mathcal{H}^\prime)(\dot{\bar{\varphi}}+\delta_{\varphi}\mathcal{H}^\prime)+\bm J \cdot \bm q}\,.\label{step1}
\end{equation}
As a final step, we introduce a complex intermediate group field, $\bm \chi$, called \textit{response field}, such that $Z[J,\bar{J}]$ can be rewritten as, using basic properties of Gaussian integration:
\begin{equation}
Z[J,\bar{J},\jmath,\bar{\jmath}]= \int d\bm q d\bm{\chi}\, e^{-\Omega^2 S[\bm q,\bm \chi]+\bm J \cdot \bm q+ \bm \jmath \cdot \bm \chi}\,. \label{generatingfunctional0}
\end{equation}
where we introduced a source $\bm \jmath=(\jmath,\bar{\jmath}\,)$ for the response field, and where the complex classical action $S[\bm q,\bm \chi]$ is given by:
\begin{align}
\nonumber \Omega^2 S[\bm q,\bm \chi]:=\int dt d\bm g\,\bigg[\Omega\bar{\chi}(\bm g,t) \chi(\bm g,t)+&i\bar{\chi}(\bm g,t)\left(\dot{\varphi}+\Omega\delta_{\bar{\varphi}} \tilde{\mathcal{H}}\right)(\bm g, t)\\
&+i\left(\dot{\bar{\varphi}}+\Omega\delta_{\varphi}\tilde{\mathcal{H}}\right)(\bm g,t)\chi(\bm g,t) \bigg]\,.\label{classicaction0}
\end{align}
It will be useful in the following to work in the Fourier representation. We will note $T_{\bm p}(\omega)$ (resp. $\bar{T}_{\bm p}(\omega)$) the Fourier components of $\varphi(\bm g,t)$ (resp. $\bar{\varphi}(\bm g,t)$), where $\bm p \in \mathbb{Z}^d$, such that:
\begin{equation}
\varphi(\bm g,t)=\int_{-\infty}^{+\infty} \frac{d\omega}{\sqrt{2\pi}} e^{-i\omega t}\sum_{\bm p \in \mathbb{Z}^d} T_{\bm p}(\omega) \prod_{\ell=1}^d e^{i p_\ell \theta_\ell}\,,
\end{equation}
where $e^{i\theta_{\ell}} := g_\ell$. In that way, the full Hamiltonian $\tilde{\mathcal{H}}$ reads:
\begin{align}
&\tilde{\mathcal{H}}[T,\bar{T}]:= \sum_{\bm p \in \mathbb{Z}^5} \int_{-\infty}^{+\infty} d\omega \, \bar{T}_{\bm p}(\omega)(\bm p^2+m^2) T_{\bm p}(\omega)\\\nonumber
&+\frac{\lambda}{2\pi} \sum_{\ell=1}^5 \sum_{\{\bm p_i \}} \int \prod_{i=1}^4 d\omega_i \delta(\omega_1+\omega_3-\omega_2-\omega_4) \mathcal{W}^{(\ell)}_{\bm p_1,\bm p_2,\bm p_3,\bm p_4} T_{\bm p_1}(\omega_1) \bar{T}_{\bm p_2}(\omega_2) T_{\bm p_3}(\omega_3) \bar{T}_{\bm p_4}(\omega_4)\,,
\end{align}
where we introduced the symbols $\mathcal{W}^{(\ell)}_{\bm p_1,\bm p_2,\bm p_3,\bm p_4} $ defined as:
\begin{equation}
\mathcal{W}^{(\ell)}_{\bm p_1,\bm p_2,\bm p_3,\bm p_4}:=\delta_{p_{1\ell}p_{4\ell}}\delta_{p_{2\ell}p_{3\ell}} \prod_{j \neq i} \delta_{p_{1j}p_{2j}} \delta_{p_{3j}p_{4j}}\,.\label{quartickernel}
\end{equation}
Hence, $S[\bm q,\bm \chi]$ splits as:
\begin{equation}
S=: S_{\text{kin}}+S_{\text{int}}\,,
\end{equation}
where,
\begin{align}
\nonumber S_{\text{kin}}=\sum_{\bm p \in \mathbb{Z}^5} \int_{-\infty}^{+\infty} d\hat{\omega} \bigg( \bar{\chi}_{\bm p}(\hat{\omega}) \chi_{\bm p}(\hat{\omega})&+i\bar{\chi}_{\bm p}(\hat{\omega})\left(-i\hat{\omega}+\bm p^2+m^2\right) T_{\bm p}(\hat{\omega})\\
&+i\bar{T}_{\bm p}(\hat{\omega}) \left(i\hat{\omega}+\bm p^2+m^2\right){\chi}_{\bm p}(\hat{\omega}) \bigg)\,,
\end{align}
and:
\begin{equation}
S_{\text{int}}=\frac{i\lambda}{4\pi} \sum_{\ell=1}^d\left(\, \vcenter{\hbox{\includegraphics[scale=0.8]{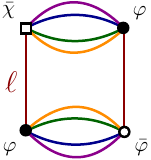}}}+\vcenter{\hbox{\includegraphics[scale=0.8]{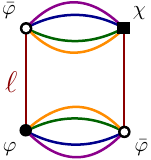}}}\,\right)\,,\label{Sint}
\end{equation}
where in the previous equation we introduced the graphical rule according to which the response fields $\chi$ and $\bar{\chi}$ will be materialized by black and white square nodes respectively, and where we introduced the dimensionless frequency $\omega \equiv \Omega \hat{\omega}$.
\medskip

The free propagator $C$ takes the form of a $2\times 2$ matrix, with components $C_{\chi\bar{\chi}}$, $C_{T\bar{\chi}}$, $C_{{\chi}\bar{T}}$ and $C_{T\bar{T}}$. It is easy to check that the response field does not propagate, i.e.,
\begin{equation}
C_{\chi\bar{\chi}}(\hat{\omega}, \bm p^2)=0\,. \label{freepropaCHI}
\end{equation}
Other components are given by:
\begin{equation}
C_{\bar{\chi}T}(\hat{\omega}, \bm p^2)=\frac{1}{\Omega^2}\frac{\hat{\omega}-i(\bm p^2+m^2)}{\hat{\omega}^2+(\bm p^2+m^2)^2}\,,\quad C_{\bar{T}{\chi}}(\hat{\omega}, \bm p^2)=-\frac{1}{\Omega^2}\frac{\hat{\omega}+i(\bm p^2+m^2)}{\hat{\omega}^2+(\bm p^2+m^2)^2}\,,\label{freepropa1}
\end{equation}
and:
\begin{equation}
C_{T\bar{T}}(\hat{\omega}, \bm p^2)=\frac{1}{\Omega^2}\frac{1}{\hat{\omega}^2+(\bm p^2+m^2)^2}\,.\label{freepropa2}
\end{equation}
The result \eqref{freepropaCHI} valid at order zero in the perturbative expansion survives to all orders, and is in fact an exact, non perturbative relation \cite{aron2010symmetries2}, meaning that component $\chi \bar{\chi}$ of the exact propagator $G$ (or equivalently the component $\bar{T} T$ of the mass matrix $\Gamma^{(2)}$) vanishes:
\begin{equation}
\boxed{
G_{\chi \bar{\chi}}(\hat{\omega},\bm p^2)=0\,.} \label{conditionG}
\end{equation}
The origin of this relation can be traced as follows. Let us consider $Z[J,\bar{J},\jmath,\bar{\jmath}]$ the generating functional \eqref{generatingfunctional0}. Let us add \textit{a linear} driving force $\frac{1}{\Omega}\int dt \sum_{\bm p} \bar{k}_{\bm p}(t) T_{\bm p}(t)+\text{c.c}$ to the Hamiltonian $\mathcal{H}$. This is equivalent to translating sources $\jmath$ and $\bar{\jmath}$ as:
\begin{equation}
\jmath_{\bm p} \to \jmath_{\bm p}-ik_{\bm p}\,,\quad \bar{\jmath}_{\bm p} \to \bar{\jmath}_{\bm p}+i\bar{k}_{\bm p}\,.
\end{equation}
Hence from the normalization conditions of the partition function we must have $Z[0,0,-ik,i\bar{k}]=1$, and therefore:
\begin{equation}
G_{\chi \bar{\chi}}= - \frac{\delta^2}{\delta k_{\bm p} \delta \bar{k}_{\bm p^\prime}} Z[0,0,-ik,i\bar{k}]\equiv - \frac{\delta^2 1}{\delta k_{\bm p} \delta \bar{k}_{\bm p^\prime}} =0\,.
\end{equation}
In the rest of this paper, we fix the original timescale such that $\Omega=1$, keeping the dependency over $\Omega$ explicit only for technical points.Such a choice simplifies all the expressions before.

\section{Functional renormalization group}\label{sec3}

In this section we introduce the formalism of the nonperturbative renormalization group as originally formulated by Wetterich and Morris \cite{Morris_1994a,MORRIS_1994,Berges_2002,Dupuis_2021}. This formalism is particularly well suited to deal with discrete models of quantum gravity, such as GFTs or random tensors. We will introduce this formalism for the dynamic GFT model introduced above. In this study, we  limit ourselves to the equilibrium dynamics.

\subsection{Regularization and flow equation}
The RG as conceived by Wilson and Kadanoff aims to interpolate between a microscopic model and a macroscopic, effective description. The effective description is constructed by integrating out quantum or thermodynamic fluctuation scale by scale, integrating out firstly the modes having a small
wavelength and ending with the ones having a large wavelength. This paradigm generally focuses on equilibrium physics. For non-equilibrium systems, temporal fluctuations can no longer be ignored. There are then two possible attitudes:
\begin{itemize}
\item Consider a coarse-graining only on the group variables (i.e., on the spectrum of the Laplacian $\Delta_{\bm g}$).

\item Or include the time to the notion of scale, and integrate partially on both the spectrum of the operators $-i \partial/\partial t$ and $\Delta_{\bm g}$.
\end{itemize}
We can still imagine partially integrating only on the $\omega$ frequencies by integrating on the whole spectrum of $\Delta_{\bm g}$. But we will not consider this possibility in the following. The possibility of a coarse-graining in frequency has been considered in \cite{duclut2017frequency} through the Wetterich framework, about non-equilibrium systems and in \cite{lahoche2021functional}, about a disordered Langevin type equation. Other approaches considering a frequency coarse-graining have been considered, notably for quantum mechanical problems \cite{synatschke2009flow,zappala2001improving}, inflation theory \cite{prokopec2018functional}, Brownian motion \cite{wilkins2020functional,wilkins2021functional2}, dissipative (open) quantum system \cite{jakobs2010nonequilibrium,schoeller2009perturbative,aoki2002nonperturbative} and references therein. The reader can also consult the recent review \cite{Dupuis_2021}. In this paper, we follow the same
strategy, and we focus on a coarse-graining both in frequency $\omega$ and momenta $\bm p$, that interpolates between two regimes:\\

\begin{enumerate}
\item The UV regime, where fluctuations are frozen and fields configurations are determined
by stationary points of the classical action $S$.\\

\item The IR regime, where fluctuations are all integrated out and field configurations
described through the effective action $\Gamma$, the Legendre transform of the Gibbs free energy.
\end{enumerate}
The standard procedure is to add a regulator to the classical action $S$, which has generally the form:
\begin{equation}
\Delta S_k= \sum_{\bm p\in \mathbb{Z}^d}\sum_{a,b} \,\int_{-\infty}^{+\infty} d\omega \,\bar{\Xi}_{a}(\bm p,\omega) R_{ab,k}(\bm p,\omega) {\Xi}_{b}(\bm p,\omega)\,,
\end{equation}
where ${\Xi}(\bm p,\omega)=(\chi_{\bm p}(\omega),T_{\bm p}(\omega))$. The \textit{regulator} $R_{ab,k}(\bm p,\omega)$ assumed to be a differentiable function of $k$, $\bm p$ and $\omega$. It behaves as a scale-dependent mass and is designed such that high energy modes concerning the scale $k$ (i.e., such that $\hat{\omega}/k^2, \bm p^2/k^2 \ll 1$) receive a small mass whereas low energy modes are essentially frozen, decoupling them from long range physics. In such a way, we are expecting to construct a smooth interpolation $\Gamma_k$ between microscopic physics described by classical action $\Omega^2 S$ for $k=\Lambda$ and macroscopic physics described by effective action $\Gamma$ -- the Legendre transform of the Gibbs free energy—for $k=0$. We introduce the mathematical definition of the effective average action $\Gamma_k$:
\begin{equation}
\Gamma_k[\bm M, \bm \sigma]+\Delta S_k[\bm M, \bm \sigma]=\bm M \cdot \bm J+ \bm \sigma \cdot \bm \jmath - W_k[\bm J, \bm \jmath\,]\,,\label{defeffectiveaction}
\end{equation}
where ${\bm M}:=(\bar M, M)$ and ${\bm \sigma}:=(\bar\sigma,\sigma)$ denote the classical fields i.e. 
\bea
M:=\frac{\delta W_k}{\delta  \bar J},\quad  \sigma:=\frac{\delta W_k}{\delta  \bar j}.
\eea
The microscopic scale $\Lambda$ is assumed to be large enough, and we will take the \textit{continuum limit} $\Lambda\to \infty$ in the computation of the $\beta$-function. For the equilibrium distributions, this limit makes sense because the model that we consider is just-renormalizable and asymptotically free \cite{samary2013beta,lahoche2015renormalization}. In the deep IR regime, for $k\sim 0$, one expects that regulator $R_{ab,k}$ almost vanish, ensuring that symmetries, in particular, should be ultimately restored, at least formally, for the exact RG equation \cite{Delamotte_2012}:
\begin{equation}
\frac{\partial}{\partial k} \Gamma_k= \Tr\, \frac{\partial \bm R_k}{\partial k} (\bm \Gamma_k^{(2)}+\bm R_k)^{-1}\,,\label{Wett}
\end{equation}
where capital bold letters designate $2\times 2$ matrix-valued functions and the trace runs over all the fields indices.  Note that the expression assumes implicitly that $\Omega=1$, and we define the effective propagator $\bm G_k$ as:
\begin{equation}
\bm G_k:= (\bm \Gamma_k^{(2)}+\bm R_k)^{-1}\,.\label{kdeppropa}
\end{equation}
The situation is however not so easy, because equation \eqref{Wett} cannot be solved exactly, even for very simple models, and approximations currently considered solving it, introduce a spurious dependency on the regulator for IR quantities \cite{pawlowski2007aspects,pawlowski2017physics}. In this paper, we will consider the \textit{minimal sensitivity prescription} (MSP) as a reliability criterion to quantify the dependency on the regulator, see \cite{canet2003optimization,duclut2017frequency,canet2011nonperturbative}. Methods usually considered for solving flow equations are called truncation and project them along a finite-dimensional subspace. The choice of this finite-dimensional subspace depends on physical constraints and symmetries expected to be unbroken along the flow, up to IR scales. This can be achieved by demanding that the regulator preserve the original symmetries of the classical action, i.e. that Ward-Takahashi (WT) identities remain unchanged along the flow \cite{zinn1975renormalization}. This condition however is usually too restrictive, and in many situations, symmetries are only restored in the deep IR, making the dependency on the regulator difficult to avoid. This is especially the case for gauge theories \cite{gies2012introduction}, another unconventional example being provided by RMMs and RTMs \cite{Lahoche:2020pjo,Lahoche:2019ocf}. In that paper, we only consider regulator compatibles with time-reversal symmetry preserved along the flow, but not only asymptotically. Time-reversal symmetry is expected because we assume to consider only \textit{equilibrium dynamics}, starting with a generalized Gibbs state $\rho(\bm q)$ and relaxing toward equilibrium \cite{ZinnJustinBook2}.

\subsection{Time reflection symmetry and causality}\label{sectioncausal}

    A way to construct coarse-graining is to modify the original Langevin equation \eqref{langevin}, adding to it a non-local driving force (see \cite{duclut2017frequency} for more detail):
\begin{equation}
\dot{T}_{\bm{p}}(t)=- \Omega \frac{\partial \mathcal{H}}{\partial \bar{T}_{\bm p}(t)}-f_{\bm{p}}(t,[\bm q(t)])+\eta_{\bm p}(t)\,,\label{langevin2}
\end{equation}
where the driving force $f$ is non-local in time and takes the form:
\begin{equation}
f_{\bm{p}}(t,[\bm q(t)]):= \int dt^\prime R_k^{(1)}(\bm p, t-t^\prime) {T}_{\bm{p}}(t^\prime)\,,
\end{equation}
where $R_k^{(1)}$ is assumed to be a real kernel. The effect of this force is to ‘‘freeze” IR contributions. In addition, we modify the noise correlation function, adding to it a non-local contribution introducing a \textit{short memory} in the system:
\begin{equation}
\langle \eta(\bm g, t) \bar{\eta}(\bm g^\prime, t^\prime) \rangle =\Omega\, \delta(\bm g^\prime (\bm g)^{-1}) \left[\delta(t-t^\prime)+\frac{1}{\Omega}R_k^{(2)}(t-t^\prime)\right]\,.\label{noisecorrelation2}
\end{equation}
Following the same steps as for the deduction of the generating functional \eqref{generatingfunctional0}, we find:
\begin{equation}
Z_k[J,\bar{J},\jmath,\bar{\jmath}\,]:= \int d\bm q d\bm{\chi}\, e^{-\Omega^2S[\bm q,\bm \chi]-\Delta S_k[\bm q,\bm \chi]+\bm J \cdot \bm q+ \bm \jmath \cdot \bm \chi}\,. \label{generatingfunctional1}
\end{equation}
where:
\begin{align}
\nonumber\Delta S_k[\bm q,\bm \chi]=&\sum_{\bm p \in \mathbb{Z}^d}\int d\omega \,\bigg( \bar{\chi}_{\bm p}(\omega) R_{k}^{(2)}(\bm p,\omega){\chi}_{\bm p}(\omega)\\
&+i R_k^{(1)}(\bm p,\omega) \bar{\chi}_{\bm p}(\omega) {T}_{\bm p}(\omega)+i R_k^{(1)}(\bm p,-\omega) \bar{T}_{\bm p}(\omega) {\chi}_{\bm p}(\omega)\bigg)\,,
\end{align}
which define the components of the bold matrix $\bm R_k$:
\begin{equation}
\bm R_k(\bm p,\omega):= \begin{pmatrix}
R_{k}^{(2)}(\bm p,\omega) & +i R_k^{(1)}(\bm p,\omega)\\
i R_k^{(1)}(\bm p,-\omega) &0
\end{pmatrix}\,, \label{equationregul}
\end{equation}
and where Fourier components of $R_k(\bm p,\omega)$ are defined as:
\begin{equation}
R_k(\bm p,t):=\frac{1}{2\pi} \int d\omega\, e^{-i\omega t} R_k(\bm p,\omega)\,.
\end{equation}
The partition function has the expected form. There are however two physical constraints to take into account. Causality and time-reversal symmetry, are closely related to the fluctuation-dissipation theorem (FDT) \cite{aron2010symmetries2,marconi2008fluctuation,kubo1966fluctuation}. Note that the Langevin equation \eqref{langevin} being of the first order, admits only one causal solution. We will describe the constraints on the regulator so that these physical conditions are preserved by the regularized theory, i.e. so that the effective models along the RG flow still describe an equilibrium dynamics compatible with causality. Note that this construction ensures that the component $\bar{\chi}\chi$ of propagator \eqref{kdeppropa} vanishes, $G_{k,\bar{\chi}\chi}=0$.
\medskip

\paragraph{Time-reversal symmetry and FDT.} The time-reflection symmetry is a direct consequence of this equilibrium dynamics, to which we will limit ourselves in this paper. It is realized by the following transformations on the fields $\Xi\to \Xi^\prime$ for the non-regularized theory ($\bm R_k=0$) as:
\begin{equation}
T_{\bm p}^\prime(t)=T_{\bm p}(-t)\,,\quad \chi^\prime_{\bm p}(t)=\chi_{\bm p}(-t)+\frac{2i}{\Omega} \dot{T}_{\bm p}(-t)\,, \label{transT1}
\end{equation}
and:
\begin{equation}
\bar{T}_{\bm p}^\prime(t)=\bar{T}_{\bm p}(-t)\,,\quad \bar{\chi}^\prime_{\bm p}(t)=\bar{\chi}_{\bm p}(-t)+\frac{2i}{\Omega} \dot{\bar{T}}_{\bm p}(-t)\,. \label{transT2}
\end{equation}
It is easy to see that these transformations leave the non-regularized classical action $S[\bm q, \bm \chi]$ invariant, within total derivatives. Moreover, the Jacobian of the transformation being equaled to $1$, the path integral defining the partition function is invariant as well for zero external sources. The transformations of the source terms into counterparts give a certain number of relations between observable, from which the classical FDT follows. Let us derive it from the expected invariance of the functional integral. The source term -- $\bm J \cdot \bm q+ \bm \jmath \cdot \bm \chi$ -- in \eqref{generatingfunctional0} transforms as:
\begin{equation}
\bm J \cdot \bm q+ \bm \jmath \cdot \bm \chi \to \tilde{\bm J} \cdot {\bm q}+ \tilde{\bm \jmath} \cdot {\bm \chi}-\frac{2i}{\Omega} \sum_{\bm p} \int dt \left(\bar{\tilde{\jmath}}_{\bm p}(t)\dot{T}_{\bm p}(t)+\tilde{\jmath}_{\bm p}(t) \dot{\bar{T}}_{\bm p}(t) \right)\label{transsources}
\end{equation}
up to total derivative contributions. We moreover introduced the notation $\tilde{X}(t):=X(-t)$. We introduce the following definitions:
\begin{equation}
R_{\bm{p}}(t,t^\prime):= \langle \bar{\chi}_{\bm{p}}(t) T_{\bm{p}}(t^\prime) \rangle \,, \qquad D_{\bm{p}}(t,t^\prime):=\langle \bar{T}_{\bm{p}}(t) T_{\bm{p}}(t^\prime) \rangle\,,
\end{equation}
and the transformation \eqref{transsources} leads to the FDT:
\begin{equation}
\boxed{
R_{\bm{p}}(t,t^\prime)-R_{\bm{p}}(-t,-t^\prime)=-\frac{2i}{\Omega} \frac{d}{dt} D_{\bm{p}}(t,t^\prime)\,.}
\end{equation}
Since $D_{\bm{p}}(t,t^\prime)$ is symmetric, and assuming translation invariance, i.e. $R_{\bm{p}}(t,t^\prime)\equiv R_{\bm{p}}(t-t^\prime)$ , this relation be rewritten as:
\begin{equation}
R_{\bm{p}}(t)=-\frac{2i}{\Omega} \theta(t)\frac{d}{dt} D_{\bm{p}}(t)\,.
\end{equation}
These relations can be converted to the Fourier representation as:
\begin{equation}
\boxed{
G_{\bar{\chi}T}(\hat{\omega}, \bm p^2)-G_{\bar{\chi}T}(-\hat{\omega}, \bm p^2)=2 \hat{\omega} G_{\bar{T}T}(\hat{\omega}, \bm p^2)\,.}\label{FPTF}
\end{equation}
It is easy to check that this relation is satisfied, at zero order, by free propagators \eqref{freepropa1} and \eqref{freepropa2}. Let us show how the regulator can be compatible with these physical constraints. We would like to construct a regulator $\Delta S_k$ which is compatible with the time reversal i.e. which is invariant under the transformations \eqref{transT1} and \eqref{transT2}. A calculation detailed in Appendix \ref{AppC} shows the we must have:
\begin{equation}
\boxed{
R_k^{(1)}(\bm p,t^\prime-t)-R_k^{(1)}(\bm p,t-t^\prime)-\frac{2}{\Omega}\dot{R}_{k}^{(2)}(\bm p,t^\prime-t)=0\,.} \label{relationregulator}
\end{equation}
In terms of Fourier components, this relation reads:
\begin{equation}
R^{(1)}_k(\bm p,\omega)-R^{(1)}_k(\bm p,-\omega)=-2i\hat{\omega} R_k^{(2)}(\bm p,\omega)\,.\label{eqR}
\end{equation}
For such a time-reversal symmetric regulator, FDT \eqref{FPTF} holds for all $k$.

\paragraph{Causality.} The driving force $f_{\bm{p}}(t,[\bm q(t)])$ added to the Langevin equation depending non-locally (in time) on the trajectory $\bm q(t)$, we must have to preserve causality:
\begin{equation}
\boxed{
R_k^{(1)}(\bm p, t-t^\prime) \propto \theta(t-t^\prime)\,.}
\end{equation}
\begin{figure}
\begin{center}
\includegraphics[scale=1]{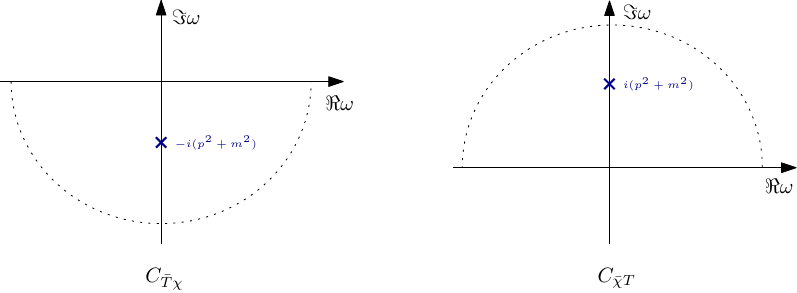}
\end{center}
\caption{Poles of the components $C_{\bar{T}{\chi}}$ and $C_{\bar{\chi}{T}}$ of the free propagator.}\label{figpoles}
\end{figure}
For the theory without a regulator, the free propagators satisfy non-trivial causality conditions, that can be investigated from the explicit expressions \eqref{freepropa1} and \eqref{freepropa2}. For instance, the component $C_{\bar{\chi}T}$ reads as:
\begin{equation}
C_{\bar{\chi}T}({\omega}, \bm p^2)=\frac{1}{{\omega}+i(\bm p^2+m^2)}\,,
\end{equation}
which has a single pole $\omega=-i(\bm p^2+m^2)$, in the lower half part of the complex part (see Figure \ref{figpoles}). Hence, the free $2$-point function:
\begin{equation}
\langle \bar{\chi}_{\bm p}(t) T_{\bm p}(t^\prime) \rangle = \int \frac{d\omega}{\sqrt{2\pi}}\frac{e^{i\omega (t-t^\prime)}}{{\omega}+i(\bm p^2+m^2)} \,,
\end{equation}
which vanish for $t-t^\prime > 0$ from residue theorem. Hence,
\begin{equation}
\langle \bar{\chi}_{\bm p}(t) T_{\bm p}(t^\prime) \rangle \propto \theta(t^\prime-t)\,.
\end{equation}
Note also that at zero moments, it is the mass that removes the ambiguity on the position of the poles\footnote{For a zero mass we should have to regularize with a parameter $\epsilon \to 0^+$ to guarantee causality.}. This causality will be an important condition to respect in the construction of the nonperturbative RG, and we will impose the effective two-point functions to satisfy them, asking that the poles of the functions $G_{k,\bar{T}{\chi}}$ and $G_{k,\bar{\chi}T}$ are respectively located in the half lower part and the half upper part of the complex plane, as in Figure \ref{figpoles}. This condition allows us to understand an important point. In the following sections, we will construct an approximation for the $\Gamma_k$ functional, through a truncation. Causality allows us to understand that this functional cannot contain independent contributions from the response fields $\chi$ and $\bar{\chi}$. In other words, it must necessarily have:
\begin{equation}
\Gamma_k\big\vert_{\chi=\bar{\chi}=0}=0\,,\quad \forall k\,,
\end{equation}
a property that we call \textit{heteroclicity}. We already know that this condition is realized initially for the action $S$, see \eqref{classicaction0}. To show that this contribution is zero, it is therefore sufficient to show that its flow is zero, in other words, that:
\begin{equation}
\frac{d}{dk}\Gamma_k\big\vert_{\chi=\bar{\chi}=0}=0\,,\quad \forall k\,.\label{conditiontrue}
\end{equation}
This is easy from \eqref{Wett}. The flow equation involves three contributions. The first one involves the product $$\frac{d}{dk}R^{(2)}_k(\bm p,t-t^\prime) \langle \bar{\chi}_{\bm p}(t) \chi_{\bm p}(t^\prime)\rangle\,,$$ vanishes due to the condition \eqref{conditionG}. The second contribution has the form $$\frac{d}{dk}R^{(1)}_k(\bm p,t-t^\prime) \langle \bar{\chi}_{\bm p}(t) T_{\bm p}(t^\prime)\rangle\,,$$ and vanishes because $R^{(1)}_k(t-t^\prime) \propto \theta(t-t^\prime)$ and $\langle \bar{\chi}(t) T(t^\prime)\rangle \propto \theta(t^\prime-t)$. The third contribution vanishes for the same reason.

\begin{remark}\label{remarkregul}
It is important to note that the condition \eqref{conditiontrue} is easy to check in the case of coarse-graining in time, as is the case in this paper. It is more subtle in the case where we practice coarse-graining only on moments and not on frequencies. In this case, one must return to the discrete version of the equations in the Îto prescription, see for instance \cite{canet2011general}. In this case, we show that the coincident time correlations must be replaced by regularized versions:
\begin{equation}
\langle \bar{\chi}_{\bm p}(t) T_{\bm p^\prime}(t^\prime) \rangle_{\epsilon} \equiv \langle \bar{\chi}_{\bm p}(t+\epsilon) T_{\bm p^\prime}(t^\prime) \rangle \,,\label{regulepsilon}
\end{equation}
which introduces a factor $e^{i\omega \epsilon}$ in the Fourier integrals. This factor ensures convergence of integrals in the upper or lower part of the complex integrals, and, the previous condition \eqref{conditiontrue} follows from the expected position of poles in the integrals, once again as a consequence of causality. Note moreover that the last condition is obvious in the supersymmetric formalism, quite natural in the Stratonovich sense. Supersymmetry, which is ensured by Ward-Takahashi identities for the quantum theory, implies the constant term flows vanish due to the cancellation of bosonic and fermionic loops \cite{lahoche2021functional}.
\end{remark}
\begin{remark}\label{remarkcausal}
The condition \eqref{conditiontrue} can be checked from perturbation theory as follows. Let us focus on the quartic melonic model. Figure \ref{figboundary} lists the expected boundaries for effective vertex functions, which can be generated in leading order from melonic diagrams.
\begin{figure}
\begin{center}
$\underset{a}{\includegraphics[scale=0.9]{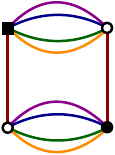}}\qquad \underset{b}{\includegraphics[scale=0.9]{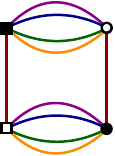}}\qquad
\underset{c}{\includegraphics[scale=0.9]{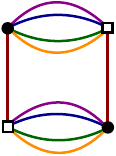}}\qquad \underset{d}{\includegraphics[scale=0.9]{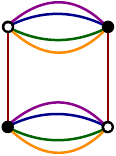}}$
\end{center}
\caption{List of boundaries which can be generated from initial conditions by Feynman diagrams. }\label{figboundary}
\end{figure}
Note that all the allowed configurations are not pictured in the Figure. For instance, there exist the same configuration as ($a$), obtained by reversing the black and white colors of the nodes. We will denote as ($\,\bar{a}$) this configuration. Note that some edges are their own type. Thus $d=\bar{d}$. Note moreover that only boundaries of type $a$ and $\bar{a}$ are involved in the classical action. At one loop, the boundary diagram $d$, which does not contain the response field, comes from the diagram pictured in Figure \ref{typicaldiag1loop}.
\begin{figure}
\begin{center}
\includegraphics[scale=1]{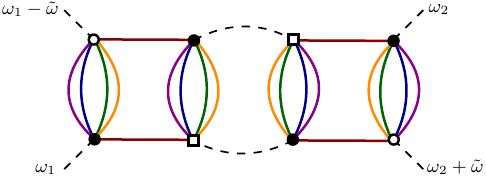}
\end{center}
\caption{The one-loop Feynman graph contributing to the 1PI effective vertex function corresponding to boundary $(d)$.}\label{typicaldiag1loop}
\end{figure}
For zero external momenta $\bm p$, the corresponding Feynman amplitude reads as:
\begin{equation}
\mathcal{A} \sim -\sum_{\bm q\in \mathbb{Z}^4} \int \frac{d\omega}{\sqrt{2\pi}}\,\frac{e^{i\omega \epsilon}}{{\omega}+i(\bm q^2+m^2)}\frac{e^{i(\omega-\tilde{\omega})\epsilon}}{(\omega-\tilde{\omega})+i(\bm q^2+m^2)} \,,\label{eqcausal}
\end{equation}
where, accordingly to the remark \eqref{remarkregul}, we introduced a factor $e^{i \omega \epsilon}$, $\epsilon\to 0^+$ (equation \eqref{regulepsilon} of remark \eqref{remarkregul}) and where $\tilde{\omega}$ denote the total external frequency. Introducing Feynman parameters \cite{peskin2018introduction}, the integrand reads:
\begin{align}
\nonumber \frac{e^{i\omega \epsilon}}{{\omega}+i(\bm q^2+m^2)}\frac{e^{i(\omega-\tilde{\omega})\epsilon}}{(\omega-\tilde{\omega})+i(\bm q^2+m^2)}& = \int_0^1 dx\, \frac{e^{i(2\omega-\tilde{\omega})\epsilon}}{[{\omega}+i(\bm q^2+m^2)-x\tilde{\omega}]^2}\\\nonumber
&=e^{i\tilde{\omega}\epsilon} \int_0^1 dx\, \frac{e^{2i\omega \epsilon}}{[{\omega}+i(\bm q^2+m^2)-x\tilde{\omega}]^2}\\\nonumber
&=e^{i\tilde{\omega}\epsilon} \frac{d}{d\tilde{\omega}} \int_0^1 \frac{dx}{x}\, \frac{e^{2i\omega \epsilon}}{{\omega}+i(\bm q^2+m^2)-x\tilde{\omega}}\,.
\end{align}
The unique pole is in the half lower part of the complex plane, and the integral over $\omega$ vanishes identically, in agreement with \eqref{conditiontrue}.
\end{remark}

\subsection{Renormalization and scaling dimension}

\subsubsection{Renormalized theory}\label{renth}
According to the theorem \ref{th1}, the equilibrium distribution of the quartic melonic model is just-renormalizable for $d=5$. Thus, it must be possible to make the perturbation theory for the equilibrium distribution $\rho(\bm q) \sim e^{-\mathcal{H}[\bm q]}$ finite at any order \footnote{Note that to agree with the convention in the literature and in appendix \ref{App1}, we canceled the global factor $2$ by a suitable redefinition of fields and couplings in this section.}, using a finite number of counter-terms. There are three of such a counter-terms $Z_m$, $Z_{\lambda}$ and $Z_{\infty}$, and renormalize respectively the mass $m^2\to Z_m m^2_r$, the coupling $\lambda \to Z_{\lambda}\lambda$ and the field $\varphi \to Z_{\infty}^{1/2} \varphi$ \cite{ZinnJustinBook2}. We will assume that these counter-terms are adjusted so that the continuous limit exists (the theory being asymptotically free, \cite{samary2013beta}). To simplify the notations we will simply call $\lambda$ and $m^2$ the coupling and mass parameters, including counter-terms, and we will note $\lambda_r$ and $m^2_r$ the renormalized (finite) versions of these parameters. Moreover, we will fix the finite part of $Z_{\infty}$ so that the effective propagator of the equilibrium theory behaves as\footnote{Avoiding IR fixed-points, see \cite{Lahoche:2018oeo} for more details.}
\begin{equation}
G_{\text{eq}}(\bm p^2)\sim  \frac{1}{\bm p^2+m^2_r}\label{2pointeqren}
\end{equation}
as $k\to 0$ and for $\bm p$ small enough. The regularized kinetic Lagrangian then reads as:
\begin{align}
\nonumber S_{\text{kin}}=\sum_{\bm p \in \mathbb{Z}^5} \int_{-\infty}^{+\infty} d\hat{\omega} \bigg( \bar{\chi}_{\bm p}(\hat{\omega}) \chi_{\bm p}(\hat{\omega})&+i\bar{\chi}_{\bm p}(\hat{\omega})\left(-i\hat{\omega}+Z_{\infty}\bm p^2+m^2\right) T_{\bm p}(\hat{\omega})\\
&+i \bar{T}_{\bm p}(\hat{\omega}) \left(i\hat{\omega}+Z_{\infty}\bm p^2+m^2\right){\chi}_{\bm p}(\hat{\omega}) \bigg)\,,\label{thoeryren}
\end{align}
disregarding the renormalization of the response field that we will consider later. The quartic interaction receives counter-terms as well, and in \eqref{Sint} we must replace $\lambda\to Z^2_{\infty}Z_\lambda \lambda_r$. For this model, the counter-terms $Z_{\infty}$ and $Z_\lambda$ can be formally computed, as the authors in \cite{Lahoche_2020b} showed. We recall their conclusions here for self-consistency:
\begin{proposition}
With the normalization condition \eqref{2pointeqren} and the renormalized coupling, $\lambda_r$ providing the correct $4$-point function at zero momenta, the counters terms $Z_{\infty}$ and $Z_\lambda$ are equal to all orders of the perturbation theory. Moreover,
\begin{equation}
Z_{\infty}^{-1}:=1-2\lambda_r A_{\infty}
\end{equation}
with $A_{\infty}$ given by
\begin{equation}
A_{\infty}:=\sum_{\bm{p} \in S_{\Lambda} \subset \mathbb{Z}^4}\, \left(\frac{1}{Z_{\infty}\bm{p}^2+Z_mm_r^2-\Sigma_{\infty}(\bm{p})}\right)^2\,,\label{Ainfty}
\end{equation}
where the sum is assumed to have some UV cut-off $\Lambda$ ($\lim_{\Lambda \to \infty} S_{\Lambda}= \mathbb{Z}^4$) and $\Sigma_{\infty}(\bm{p})$ has quartic and logarithmic divergences with respect to $\Lambda$.
\end{proposition}
The counter-terms in \eqref{Ainfty} cancels all the divergences in $\Sigma_{\infty}(\bm{p})$, except the global one of the sum, which corresponds to the last subtraction in the Zimmerman forest. Hence, $A_{\infty}$ behaves as $\ln(\Lambda)$.  More details can be found in appendix \ref{App1}.

\subsubsection{Scaling dimension}
In quantum theory in ordinary fields, the scaling dimension is closely related to renormalizability. An analogous notion can be defined for TGFTs {\color{blue}(see the references  \cite{Lahoche:2018oeo, Carrozza_2017a}, or \cite{carrozza2015discrete,})}, which accommodates the non-local nature of the interactions and the background independent definition of the theory. We have the following definition:
\begin{definition}
Let $b$ a bubble having $n(b)$ white vertices and $\mathbb{G}$ the set of $2$-points diagrams made of a single vertex of type $b$. The scaling dimension $\dim(b)$ is defined as:
\begin{equation}
\dim(b)=2-\underset{r\in \mathbb{G}}{\max} \,\,\omega(r)\,.\label{canonicaldim}
\end{equation}
\end{definition}
This definition in particular implies that $2$-points bubbles have dimension $2$, and in particular, the mass must have dimension $2$: $[m^2]=2$. In the same way, from \eqref{powerCounting}, the leading order $2$-point functions build with a single quartic melonic vertex are such that $\omega(r)=2$, and the canonical dimension vanish $[\lambda]=0$, in agreement with the just-renormalizability of the quartic model.

\section{Melonic Approximation}\label{sec4}

\subsection{Truncation and regulation}

Solving the exact RG equation \eqref{Wett} is a difficult task, even for simple problems, and requires approximations. Usually, these approximations take the form of truncation in the full theory space, which is the functional space of infinite dimension spanned by all allowed classical actions defined by the condition that the classical Hamiltonian $\mathcal{H}[\varphi,\bar{\varphi}]$ is a sum of connected invariants. The truncation will allow for the restriction of the phase space to a smaller domain where the equations will be easily solvable. The method that we propose, the effective vertex expansion (EVE), nevertheless allows capturing entire sectors, containing an infinite number of interactions, as well as the dependence of the effective vertices on the external momenta.
\medskip

We will choose the following ansatz for the effective average action $ \Gamma_k$:
\begin{align}
\nonumber \Gamma_k[M,\bar{M},\sigma,\bar{\sigma}]&=\Omega^2\sum_{\bm p \in \mathbb{Z}^5} \int_{-\infty}^{+\infty} d\hat{\omega} \bigg(Y(k) \bar{\sigma}_{\bm p}(\hat{\omega}) \sigma_{\bm p}(\hat{\omega})\\\nonumber
&+i\bar{\sigma}_{\bm p}(\hat{\omega})\left(-iY(k)\hat{\omega}+Z(k)\bm p^2+m^2(k)\right)M_{\bm p}(\hat{\omega})\\\nonumber
&+i \bar{M}_{\bm p}(\hat{\omega}) \left(iY(k)\hat{\omega}+Z(k)\bm p^2+m^2(k)\right){\sigma}_{\bm p}(\hat{\omega})\\
&+i \bigg(\bar{\sigma}_{\bm p}(\hat{\omega}) \frac{\delta \hat{\mathcal{H}}_{\text{int},k}}{\delta \bar{M}_{\bm p}(\hat{\omega})}+{\sigma}_{\bm p}(\hat{\omega}) \frac{\delta \hat{\mathcal{H}}_{\text{int},k}}{\delta {M}_{\bm p}(\hat{\omega})} \bigg)\bigg)\,,\label{AnsatzGamma}
\end{align}
where $Y(k)$ as $Z(k)$ and $m(k)$ look like  a kinetic  coupling and this parameter  ensures the time reversal symmetry with respect to the transformation \eqref{transT1} and \eqref{transT2}, and 
\begin{equation}
\hat{\mathcal{H}}_{\text{int},k}[M,\bar{M}]=\int d\hat{t}\sum_b k^{\dim(b)} Z^{n(b)}(k)\bar{\lambda}_b \Tr_b[\varphi(\hat{t}),\bar{\varphi}(\hat{t})]\,,\label{defexpansion}
\end{equation}
provided that the dimensionless time $\hat{t}$ is $\hat{t}:= \Omega t$. The sum runs over connected tensorial invariants, $\dim(b)$ is the scaling dimension of the bubble $b$ (see \eqref{canonicaldim}) and $n(b)$ the number of fields $\varphi(\bm g,t)$ involved in the interaction $b$. One can justify the truncation \eqref{AnsatzGamma} as follows. First, the time-reversal symmetry \eqref{transT1} implies that the quadratic term in $\bar{\sigma}_{\bm p}(\hat{\omega}) \sigma_{\bm p}(\hat{\omega})$ renormalizes as the linear terms $i\hat{\omega} \bar{\sigma}_{\bm p}(\hat{\omega}) M_{\bm p}(\hat{\omega})$ and $-i \hat{\omega} \bar{M}_{\bm p}(\hat{\omega}) \sigma_{\bm p}(\hat{\omega})$.
\medskip

\begin{remark}
The truncation \eqref{AnsatzGamma} is compatible with a symmetric phase approximation, i.e., with an expansion around zero vacuum field. In the symmetric phase, it is easy to check that $2$-point functions are diagonals in their momenta indices. Moreover, odd vertex function vanishes identically -- see \cite{Lahoche:2018oeo} for an extended discussion.
\end{remark}

 Let us move on to the choice of the regulator. For all our investigations we chose regulators $R_k^{(1)}$ and $R_k^{(2)}$ as a product of a pure frequency regulator with a momentum regulator. For the frequency regulator, we choose\footnote{Although we chose $\Omega=1$ above, we reintroduce $\Omega$ here to clarify the conventions.}:
\begin{equation}
R^{(1)}_k(\bm p,\omega)=\Omega k^2Z(k) \rho_k(\omega) r_k(\bm p^2) \,,
\end{equation}
and for the momentum regulator $r_k(\bm p^2)$ we choose the usual Litim regulator:
\begin{equation}
r_k(\bm p^2):=\alpha \left(1-\frac{\bm p^2}{k^2}\right)\theta(k^2-\bm p^2)\,.
\end{equation}
For the frequency regulator, we chose:
\begin{equation}
\rho_k(\omega):= \frac{k^2}{k^2-i\beta \omega/\Omega}\,.
\end{equation}
This choice has been considered in \cite{duclut2017frequency,lahoche2021functional}. It is causal (with a single pole in the lower part of the complex plane), and its Fourier transform behaves like $\sim e^{-k^2\Omega t/\beta} \theta (t)$. Numerical coefficients $\alpha, \beta$ should be numerically tuned from the \textit{ minimal sensitivity principle} (MSP), which assumes that an optimized flow induces a minimal dependence on the choice of the regulator \cite{duclut2017frequency,canet2011general,DePolsi:2022wyb}. Thus, by numerically computing the critical exponents and varying the parameters $\alpha$ and $\beta$, the MSP will fix their values at the points where the derivatives of the exponents concerning these parameters will vanish. In particular, for $\beta=0$, the coarse-graining is about momenta only, and we recover the standard RG without time regularization. In the rest of this paper, we will introduce dimensionless momenta $x$ and frequencies $y$, defined as:
\begin{equation}
\bm p^2 = k^2 x\,,\qquad \omega = \Omega Z(k) k^2 Y^{-1}(k) y\,.
\end{equation}
We furthermore define the renormalized $\beta$ as:
\begin{equation}
\beta= Z^{-1}(k) Y(k)\hat{\beta}\,,
\end{equation}
such that $\rho_k(\omega)$ transforms as:
\begin{equation}
\rho_k(\omega) \to \hat{\rho}(y)= \frac{1}{1-i\hat{\beta} y}\,,
\end{equation}
and:
\begin{equation}
R^{(1)}_k(\bm p,\omega) \to \hat{R}^{(1)}(x,y):= Z(k)\hat{\rho}(y) r(x)\,,
\end{equation}
where $r(x):=\alpha(1-x)\theta(1-x)$. The equation for $R_k^{(2)}$ can be derived from \eqref{eqR}, we have:
\begin{equation}
R_k^{(2)}(\bm p, \omega)=\frac{\Omega^2k^2}{2i\omega} Z(k) \left(\rho_k(-\omega)-\rho_k(\omega) \right) r_k(\bm p^2)\,,
\end{equation}
and we get:
\begin{equation}
R_k^{(2)}(\bm p, \omega)=: {\Omega}Y(k) \hat{\tau}(y) r_k(\bm p^2)\,,
\end{equation}
 where
\beq
\hat{\tau}(y)=-\frac{\hat{\beta}}{1+\hat{\beta}^2 y^2},
\eeq
which define a dimensionless function $\hat{R}^{(2)}(x,y)$ as:
\begin{equation}
\hat{R}^{(2)}(x,y):=Y(k) \hat{\tau}(y) r(x)\,.
\end{equation}
Derivatives with respect to $k$ can be easily computed, we get for $R^{(1)}_k(\bm p,\omega)$
\begin{align}
\nonumber k\frac{d}{dk} R^{(1)}_k(\bm p,\omega)=&(2+\eta)R^{(1)}_k(\bm p,\omega)-i Z(k)y \hat{\beta}\Omega k^2\frac{2- \eta_Y+\eta}{(1-i\hat{\beta} y)^2}r(x)\\
&+2\alpha \Omega k^2 Z(k) \frac{x}{1-i \hat{\beta}y} \theta (1-x) \,,
\end{align}
and for $R^{(2)}_k(\bm p,\omega)$:
\begin{align}
\nonumber k\frac{d}{dk} R^{(2)}_k(\bm p,\omega)=\,&\eta_YR^{(2)}_k(\bm p,\omega)+\Omega Y(k) \hat{\beta} \frac{(\eta_Y-\eta)(\hat{\beta}^2 y^2-1)-\hat{\beta}^2 y^2}{(1+\hat{\beta}^2 y^2)^2}r(x)\\
&-2\alpha\Omega Y(k) \frac{\hat{\beta}x}{1+\hat{\beta}^2 y^2}\theta(1-x)\,.
\end{align}
For future calculations we will define two dimensionless quantities:
\begin{align}
\mu_1(x,y):=&(2+\eta)\rho(y)r(x)-i y \hat{\beta}\,\frac{2-\eta_Y+\eta}{(1-i\hat{\beta} y)^2}r(x)+ 2\alpha \frac{x}{1-i \hat{\beta}y} \theta (1-x)
\end{align}
and:
\begin{align}
\mu_2(x,y):=&\eta_Y \hat{\tau}(y) r(x)+ \hat{\beta} \frac{(\eta_Y-\eta)(\hat{\beta}^2 y^2-1)-\hat{\beta}^2 y^2}{(1+\hat{\beta}^2 y^2)^2}r(x)-\frac{2\alpha\hat{\beta} x}{1+\hat{\beta}^2 y^2}\theta(1-x)\,,
\end{align}
where:
\begin{equation}
\boxed{
\eta:= \frac{1}{Z(k)} k \frac{d}{dk} Z(k)\,,\qquad \eta_Y:= \frac{1}{Y(k)} k \frac{d}{dk} Y(k)\,.}\label{defeta}
\end{equation}

\subsection{Melonic equations in the non-branching sector}\label{nonbranching}
In this section we will focus on a restricted sector of the theory, the non-branching melonic sector. We will finally derive the flow equations in this approximation. This sector is stable (at leading order) along the RG, and has shown its interest in the past \cite{Lahoche:2018oeo,pascalie2019large,samary2015correlation,samary2014closed,lahoche2015renormalization}. Note that in this section and in the following, $\Omega=1$ everywhere.

\subsubsection{Non-branching melons}
As we recalled in the first part, the most divergent diagrams are said \textit{melonics}. Strictly, melons are connected graphs and are then bubbles as well. For $d$-colored graphs, melons can be defined recursively as follows:
\begin{definition}\label{defmelons}
Any melonic bubble $b_\kappa$ of valence $\kappa$ may be deduced from the elementary melon $b_1$:
\begin{equation}
b_1:=\vcenter{\hbox{\includegraphics[scale=1]{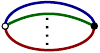} }}\,,
\end{equation}
replacing successively $\kappa-1$ colored edges (including maybe color ‘‘0") by $(d-1)$-dipole, the $(d-1)$-dipole insertion operator $\mathfrak{R}_{i}$ being defined as:
\begin{equation}
\vcenter{\hbox{\includegraphics[scale=1]{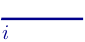} }}\underset{\mathfrak{R}_{i}}{\longrightarrow}\vcenter{\hbox{\includegraphics[scale=1]{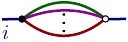} }}\,.
\end{equation}
In formula: $b_\kappa := \left(\prod_{\alpha=1}^{\kappa-1}\mathfrak{R}_{i_\alpha}\right) b_1$.
\end{definition}
For instance, the first bubble on Figure \ref{figBubbles} is a melon. For our nonperturbative investigations, we especially focus on a sub-sector of the melons, said \textit{non-branching}:
\begin{definition}\label{defnonbranch}
A non-branching melonic bubble of valence $\kappa$, $b_\kappa^{(\ell)}$ is labeled with a single index $\ell\in\llbracket 1,5\rrbracket$, and defined such that:
\begin{equation}
b_\kappa^{(\ell)}:= \left(\mathfrak{R}_{\ell}\right)^{\kappa-1}\,b_1\,.
\end{equation}
\end{definition}
Figure \ref{fig2} provides the generic structure of melonic non-branching bubbles in rank $3$. Note that the definition holds for diagrams involving square nodes.
\begin{center}
\begin{equation*}
\vcenter{\hbox{\includegraphics[scale=1]{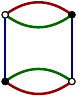} }} \,\underset{\mathfrak{R}_{i}}{\longrightarrow}\, \vcenter{\hbox{\includegraphics[scale=1]{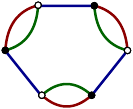} }}\,\cdots \underset{\mathfrak{R}_{i}}{\longrightarrow}\, \vcenter{\hbox{\includegraphics[scale=1]{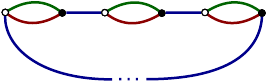} }}\underset{\mathfrak{R}_{i}}{\longrightarrow}\cdots
\end{equation*}
\captionof{figure}{Structure of the non-branching melons, from the smallest one $b_2$.} \label{fig2}
\end{center}
Another important concept is that of the boundary diagram. It concerns Feynman diagrams, such as the one shown in the figure \ref{figFeynman}. We have the following definition:
\begin{definition}\label{defboundary}
Let $G$ be a regular $(d+1)$-colored Feynman diagram with $2N$ external dotted edges. They are hooked to $2N$ black and white nodes, say externals, and the boundary diagram $\partial G$ of $G$ is the regular $d$-colored graph, discarding edges with color $0$ and such that:
\begin{enumerate}
\item Nodes of $\partial G$ are external nodes of $G$
\item Edges with color $\neq 0$ linking two external nodes are conserved.
\item Any open cycle made of colors $0$ and $i$ between two external nodes $n$ and $\bar{n}$ is replaced by a link of color $i$ in $\partial G$.
\end{enumerate}
\end{definition}
Figure \ref{figFeynman2} illustrates the mapping for a Feynman diagram in rank 3. Note that the boundary diagram is melonic, but branched in that example.

\begin{center}
\begin{equation*}
\vcenter{\hbox{\includegraphics[scale=0.8]{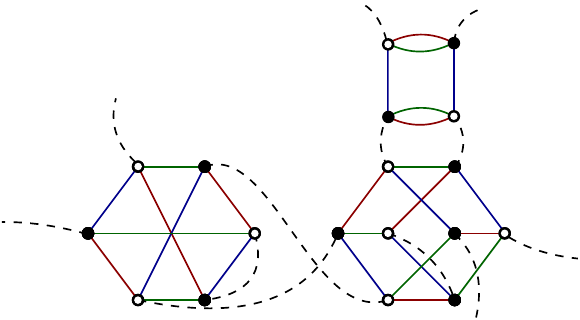}}}\quad \Huge{\rightarrow} \quad \vcenter{\hbox{\includegraphics[scale=1]{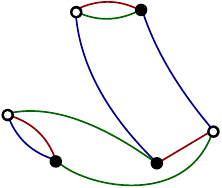}}}
\end{equation*}
\captionof{figure}{Illustration of the mapping $G\to \partial G$ for a 6-points Feynman diagram in rank 3.}\label{figFeynman2}
\end{center}

In the rest of this paper, we will work in the subspace of the theory space, generated by the non-branched melons. Thus, all $1$PI functions will be assumed to admit a Feynman series whose boundary diagrams are non-branched melons, and we will index each effective vertex by a bubble of this type. Because non-branching bubbles are labeled with a single color, the corresponding $2n$-points vertex functions decompose along $d$-components,
\begin{equation}
\Gamma_k^{(2n)}=\sum_{\ell=1}^d \Gamma_k^{(2n),(\ell)}\,,\label{decompositiond}
\end{equation}
each component $\Gamma_k^{(2n),(\ell)}$ being assumed to be labeled with non-branching melons of valence $2n$. We will now move on to the derivation of the flow equations in the non-branching sector. We only derive flow equations for $2$ and $4$ - points functions with zero external momenta and use melonic equations to close the hierarchy, expressing $6$-points functions in terms of $4$ and $2$ points ones.
\medskip

\subsubsection{Flow equations}\label{floweqsection}
\begin{figure}
\begin{center}
\includegraphics[scale=1.3]{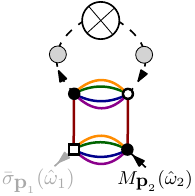}
\end{center}
\caption{Representation of the single-loop flow equation for ${\gamma}_{k,\bar{\sigma}M}^{(2)}$.}\label{figFlow1}
\end{figure}

Flow equations for different couplings can be obtained from the flow equation \eqref{Wett}, taking successive derivative with respect to classical fields $M, \bar{M}, \sigma, \bar{\sigma}$. We introduce the notations $\Xi=\{M,\sigma \}$, $\bar{\Xi}=\{\bar{M},\bar{\sigma}\}$ and:
\begin{equation}
\Gamma_{k,\bar{\Xi}^{a_1}\cdots \bar{\Xi}^{a_P}\cdots {\Xi}^{b_1} \Xi^{b_P}}^{(2P)}=\frac{\delta^{2P} \Gamma_k}{\delta \bar{\Xi}^{a_1}_{\bm p_1}(\hat{\omega}_1)\cdots \delta \bar{\Xi}^{a_P}_{\bm p_P}(\hat{\omega}_P)\cdots \delta {\Xi}^{b_P}_{\bm p_P^\prime}(\hat{\omega}_P^\prime)}\,,
\end{equation}
for $a_i,b_i=0,1$, $\Xi^0=M$, $\Xi^1=\sigma$. From truncation \eqref{AnsatzGamma}, we have:
\begin{equation}
\Gamma_{k, \bar{\sigma} \sigma}^{(2)}=Y(k) \delta_{\bm p_1\bm p_2} \delta(\hat{\omega}_1-\hat{\omega}_2)\,,
\end{equation}
and:
\begin{equation}
\Gamma_{k, \bar{\sigma} M \bar{M} M}^{(4),(\ell)}= \frac{i}{\pi} \pi^{(2)}_k(p_{1\ell}^2,p_{3\ell}^2)\left(\mathcal{W}^{(\ell)}_{\bm p_1,\bm p_2,\bm p_3,\bm p_4} +\bm p_2 \leftrightarrow \bm p_4\right)\delta(\hat{\omega}_1-\hat{\omega}_2+\hat{\omega}_3-\hat{\omega}_4)\,,\label{rencondGamma4}
\end{equation}
where the function $\pi^{(2)}_k(p_{1\ell}^2,p_{3\ell}^2)$ (depending on the square of external momenta) gives the momentum dependence of the vertex, with normalization condition:
\begin{equation}
\pi^{(2)}_k(0,0)=: \lambda(k)\,, \label{rencondcoupling}
\end{equation}
defining the effective quartic coupling at scale $k$. Note that we disregarded any dependence of $\pi^{(2)}_k$ on the frequency. Our truncation is then ultralocal for the time parameter. We moreover introduce the notation:
\begin{equation}
\Gamma_{k,\bar{\Xi}^{a_1}\Xi^{a_2}}^{(2)}=: \gamma_{k,\bar{\Xi}^{a_1}\Xi^{a_2}}^{(2)}(\bm{p}_1,\hat{\omega}_1) \delta_{\bm{p}_1\bm{p}_2}\delta(\hat{\omega}_1-\hat{\omega}_2)\,,\label{decomp2points}
\end{equation}
and we have:
\begin{equation}
\gamma_{k,\bar{\sigma}M}^{(2)}(\bm{p}=\bm{0},\hat{\omega}_1=0)=i\, m^2(k)\,,
\end{equation}
and:
\begin{equation}
\frac{d}{d\hat{\omega}_1}\gamma_{k,\bar{\sigma}M}^{(2)}(\bm{p}=\bm{0},\hat{\omega}_1=0):=Y(k)\,,\qquad \frac{d}{dp^2_i}\gamma_{k,\bar{\sigma}M}^{(2)}(\bm{p}=\bm{0},\hat{\omega}_1=0):=iZ(k)\,,\label{defZY}
\end{equation}
the last equation being valid for all $i=1,\cdots, d$, agrees with the isotropic assumption. The flow equation for ${\gamma}_{k,\bar{\sigma}M}^{(2)}$ can be deduced from \eqref{Wett}, taking derivatives with respect to $\bar{\sigma}$ and $M$. We obtain:
\begin{equation}
\dot{\gamma}_{k,\bar{\sigma}M}^{(2)}(\bm{p}_1,\hat{\omega}_1)\delta_{\bm{p}_1\bm{p}_2}\delta(\hat{\omega}_1-\hat{\omega}_2)=-\Tr\,\dot{\bm R}_{k} \bm G_{k} \Gamma_{k\bar{\sigma} M\bullet \bullet }^{(4)} \bm G_{k}\,,\label{eqflowMass}
\end{equation}
where we omitted momenta and frequencies to simplify the expression and the trace $\Tr$ runs both over momenta, frequencies and fields. The dots in the $4$-point functions $\Gamma_{k\bar{\sigma} M\bullet \bullet }^{(4)}$ designates allowed fields in the trace, and we introduced the notation:
\begin{equation}
\dot{X}:= k \frac{dX}{dk}\,.
\end{equation}
Equation \eqref{eqflowMass} can be pictured as in Figure \ref{figFlow1}, where dotted edges with grey circles materialize propagators, cross-circle materializes regulator contribution $\dot{\bm R}_{k} $, and we pictured the $4$-point function in order to make the index structure explicit. On the figure, arrows are oriented from barred to non-barred fields, and grey half edges materialize response fields.
Because $G_{k \bar{\chi} \chi}=0$ and $R_{k \bar{M} M}=0$ (see \eqref{equationregul}), and that from truncation \eqref{AnsatzGamma} the only non-vanishing field configuration for bullets in $\Gamma_{k\bar{\sigma} M\bullet \bullet }^{(4)}$ is $\Gamma_{k\bar{\sigma} M \bar{M} M}^{(4)}$, there are only two contributions allowed for internal fields. Hence, we get:
\begin{align}
\nonumber\dot{\gamma}_{k,\bar{\sigma}M}^{(2)}(\bm{p}_1,\hat{\omega}_1)\delta_{\bm{p}_1\bm{p}_2}\delta(\hat{\omega}_1-\hat{\omega}_2)&=-\sum_{i=1}^d\Bigg(\vcenter{\hbox{\includegraphics[scale=1]{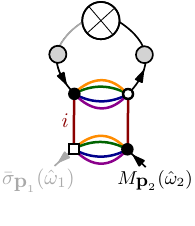}}}+\vcenter{\hbox{\includegraphics[scale=1]{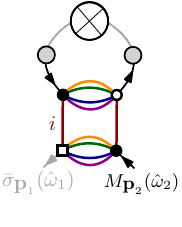}}}\\
& +\vcenter{\hbox{\includegraphics[scale=1]{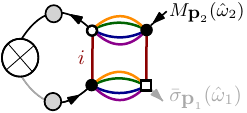}}}+\vcenter{\hbox{\includegraphics[scale=1]{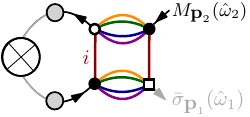}}}\Bigg)\,.\label{flowmassdiag}
\end{align}
The two last contributions create only one face, accordingly to definition \ref{deffaces}, and are therefore less relevant than the two first ones, which are melonics following definition \ref{defmelons}. Because we focus in this paper on the ultraviolet (UV) regime:
\begin{equation}
\Lambda \gg k \gg 1\,,
\end{equation}
for some UV cut-off $\Lambda$, the two last configurations in \eqref{flowmassdiag} can then be discarded at the leading order. Note that we do not include some numerical factors counting the number of corresponding configurations. For instance, the first configurations have to be multiplied by a factor $2$, counting the two allowed configurations for the internal (black or grey) solid edges:
\begin{equation}
G_{\bar{M}M}:=\vcenter{\hbox{\includegraphics[scale=1.2]{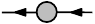}}}\,,\qquad G_{\bar{\sigma}M}:=\vcenter{\hbox{\includegraphics[scale=1.2]{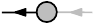}}}\,,
\end{equation}
the arrows being oriented from “barred” to “non-barred” fields. Diagrams of equation \eqref{flowmassdiag} can be easily translated in a formula, for instance (we count the diagram twice):
\begin{align}
\nonumber\vcenter{\hbox{\includegraphics[scale=0.8]{OneLoopMass2.pdf}}}=&\,- 2\delta_{\bm{p}_1\bm{p}_2}\delta(\hat{\omega}_1-\hat{\omega}_2) \frac{\pi^{(2)}_k(p_{11}^2,p_{11}^2)}{\pi }\\
& \times\sum_{\bm p}\delta_{p_1,p_{11}}\int {d\hat{\omega}}G_{k,\bar{\sigma} M}(\bm p^2,\hat{\omega})G_{k,\bar{M} M}(\bm p^2,\hat{\omega})\dot{R}^{(1)}_k(\bm p,\hat{\omega})\,.
\end{align}
In that equations, the components $G_{k,\bar{M} M}$ and $G_{k,\bar{\sigma} M}$ of the $2$-point function can be computed explicitly as:
\begin{equation}
G_{k,\bar{\sigma} M}(\bm p^2,\hat{\omega})=- \frac{1}{k^2}\frac{i}{Z(k)}\frac{1}{\hat{f}(x,-y)}
\end{equation}
and:
\begin{equation}
G_{k,\bar{M} M}(\bm p^2,\hat{\omega})=\frac{1}{k^4}\frac{Y(k)}{Z^2(k)}\frac{1+\hat{\tau}(y)r(x)}{\hat{f}(x,y)\hat{f}(x,-y)}\,,
\end{equation}
where $r(x):=\alpha (1-x) \theta(1-x)$, and for the truncation that we consider:
\begin{equation}
\hat{f}(x,y)=iy+x+\bar{m}^2+ \hat{\rho}(-y) r(x)\,, \label{truncationf}
\end{equation}
the dimensionless and renormalized mass and vertex function $\bar{m}^2$ and $\bar{\pi}_k^{(2)}$ being defined as (see \eqref{defexpansion}):
\begin{equation}
m^2=: Z(k) k^2 \bar{m}^2\,,\qquad {\pi}_k^{(2)}= Z^2(k) \bar{\pi}_k^{(2)}\,.
\end{equation}
Using these definitions, a straightforward calculation leads to the expression:
\begin{align}
\nonumber \vcenter{\hbox{\includegraphics[scale=0.8]{OneLoopMass2.pdf}}}&=i \bar{\pi}_k^{(2)}(p_{11}^2,p_{11}^2) \delta_{\bm{p}_1\bm{p}_2}\delta(\hat{\omega}_1-\hat{\omega}_2) Z(k) k^2 L_{21}(x_1)\,,
\end{align}
where $x_1:= p_1^2/k^2$ and assuming $k$ large enough sums can be replaced by integrals. Hence, we define:
\begin{equation}
L_{21}(x_1):= \frac{4}{\pi} \int_{\mathbb{R}^{d+1}} d\textbf{x}^\prime dy \delta(x_1^\prime-x_1) \mu_1(x^\prime,y) \frac{1+\hat{\tau}(y)r(x^\prime)}{\hat{f}(x^\prime,y)\hat{f}^2(x^\prime,-y)}\,,
\end{equation}
having introduced the notation $\textbf{x}\in \mathbb{R}^d$, with components $x_i$ and square length $x\equiv \sum_i x_i^2$. In the same way, we get the second diagram:
\begin{align}
\vcenter{\hbox{\includegraphics[scale=0.8]{OneLoopMass3.pdf}}}= i\bar{\pi}_k^{(2)}(p_{11}^2,p_{11}^2)\delta_{\bm{p}_1\bm{p}_2}\delta(\hat{\omega}_1-\hat{\omega}_2)Z(k)k^2 (-L_{22}(x_1))\,,
\end{align}
with:
\begin{equation}
L_{22}(x_1)=\frac{2}{\pi} \int_{\mathbb{R}^{d+1}} d\textbf{x}^\prime dy \delta(x_1^\prime-x_1) \frac{ \mu_2(x^\prime,y)}{\hat{f}(x^\prime,y)\hat{f}(x^\prime,-y)}\,.
\end{equation}
The flow equation for mass can be obtained by setting $\bm p_1=\bm 0$. From the normalization condition \eqref{rencondcoupling}, we get:
\begin{equation}
\boxed{
\beta_{m^2}=-(2+\eta){m}^2-d\bar{\lambda}\left(L_{21}(0)-L_{22}(0) \right)\,,}
\end{equation}
using the conventional notation in field theory $\beta_X:= \dot{X}$. In the symmetric phase moreover, where in particular $\Gamma_k^{(3)}=0$, the anomalous dimension $\eta_Y$ vanishes identically, as it can be easily checked from definitions \eqref{defeta} and \eqref{defZY}:
\begin{equation}
\boxed{
\eta_Y=0 \,. \quad \text{(In the symmetric phase)}}
\end{equation}
Let us detail the derivation of the flow equation for $\eta$. We will not be able to complete the derivation in this section, the end of the derivation being given in Section \ref{WTI}. From definition \eqref{defZY}, we have the self-consistency equation:
\begin{equation}
\boxed{
\eta=-\bar{\lambda}^{\prime}\left(L_{21}(0)-L_{22}(0) \right)-\bar{\lambda} \frac{d}{d p_1^2} \left(L_{21}(x_1)-L_{22}(x_1) \right)\bigg\vert_{x_1=0}\,,}\label{eqeta}
\end{equation}
where we defined,
\begin{equation}
\frac{d}{dp_1^2} \pi_k^{(2)}(p_1^2,p_1^2)\bigg\vert_{p_1=0}=:Z^2(k)k^{-2}\bar{\lambda}^{\prime}\,.\label{derivvertex}
\end{equation}
The equation for the $4$-point coupling $\lambda$ can be deduced from the renormalization condition \eqref{rencondGamma4}, setting external momenta and frequencies to zero. Deriving equation \eqref{Wett} one time for $\bar{\sigma}$, one time for $\bar{M}$ and two times concerning $M$ and setting external momenta and frequencies to zero, we get, using the same graphical representation as before:
\begin{align}
\nonumber&\frac{i\dot{\lambda}}{\pi} \delta(0)=-\vcenter{\hbox{\includegraphics[scale=0.8]{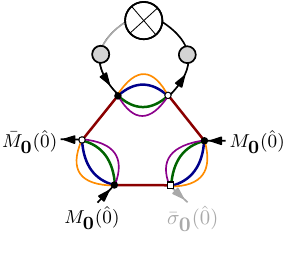}}}-\vcenter{\hbox{\includegraphics[scale=0.8]{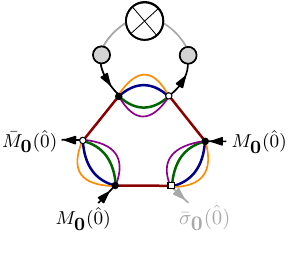}}}\\\nonumber
&+\vcenter{\hbox{\includegraphics[scale=0.8]{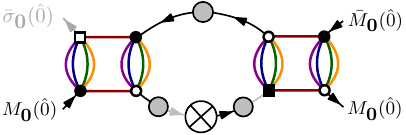}}}+\vcenter{\hbox{\includegraphics[scale=0.8]{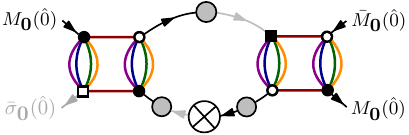}}}\\\nonumber
&+\vcenter{\hbox{\includegraphics[scale=0.8]{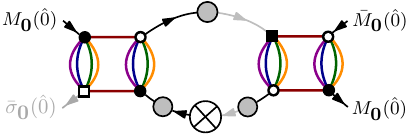}}}+\vcenter{\hbox{\includegraphics[scale=0.8]{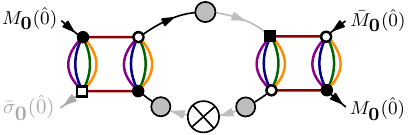}}}\\\nonumber
&+\vcenter{\hbox{\includegraphics[scale=0.8]{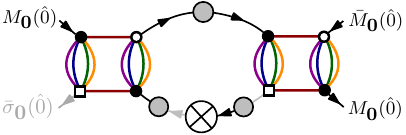}}}+\vcenter{\hbox{\includegraphics[scale=0.8]{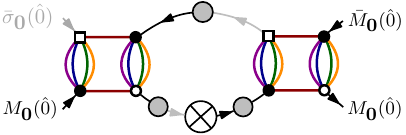}}}\\
&+\vcenter{\hbox{\includegraphics[scale=0.8]{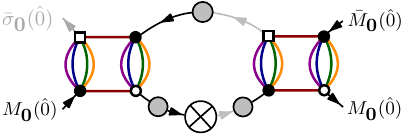}}}+\vcenter{\hbox{\includegraphics[scale=0.8]{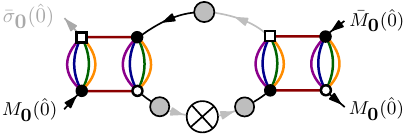}}}
\end{align}
This equation requires defining the $6$-point functions, as we defined the $4$-point ones (equation \eqref{rencondGamma4}). We need only the zero momenta function, which reads as follows:
\begin{equation}
\Gamma_{k, \bar{\sigma} M \bar{M}M \bar{M} M}^{(6),(\ell)}\bigg\vert_{\text{0}}=\frac{9i}{\pi^2}\kappa \delta(\hat{\omega}_1-\hat{\omega}_2+\hat{\omega}_3-\hat{\omega}_4+\hat{\omega}_5-\hat{\omega}_6)\,,\label{gamma6zero}
\end{equation}
    where $\kappa$ denotes the sixtic coupling constant.
For the two first diagrams, we get:
\begin{equation}
\vcenter{\hbox{\includegraphics[scale=0.8]{OneLoop4points1.pdf}}}+\vcenter{\hbox{\includegraphics[scale=0.8]{OneLoop4points2.pdf}}}=\frac{3i}{\pi} Z^2(k)\bar{\kappa}\delta(0) \left(L_{21}(0)-L_{22}(0) \right)\,,
\end{equation}
where from \eqref{canonicaldim}:
\begin{equation}
\kappa=: k^{-2} Z^3(k) \bar{\kappa}\,.\label{sansdimkappa}
\end{equation}
Indeed, the maximally divergent $2$-point diagram that we can build from a melonic $6$-point interaction has a divergent degree (see proposition \ref{propositionpower}):
\begin{equation}
\omega(r)=-2L+F=-2\times 2+2\times (d-1)=4\,,
\end{equation}
then $\dim(b)=-2$. The computation of diagrams involving $4$ points vertices requires being more careful. Let us compute the first one. Explicitly we have:
\begin{align}
\nonumber&\vcenter{\hbox{\includegraphics[scale=0.7]{OneLoopCoupling442.pdf}}}=-\delta(0) \frac{4i\lambda^2}{\pi^2} \\
&\qquad\times\sum_{\bm p}\delta_{p_1,p_{11}}\int {d\hat{\omega}}G_{k,\bar{\sigma} M}(\bm p^2,\hat{\omega})G_{k,\bar{\sigma} M}(\bm p^2,\hat{\omega})G_{k,\bar{M} M}(\bm p^2,\hat{\omega})\dot{R}^{(1)}_k(\bm p,\hat{\omega})\,,
\end{align}
which after some algebra can be rewritten as follows:
\begin{equation}
\vcenter{\hbox{\includegraphics[scale=0.7]{OneLoopCoupling442.pdf}}}=\delta(0)\frac{4i\bar{\lambda}^2}{\pi^2}Z^2(k)L_{31}\,,
\end{equation}
with:
\begin{equation}
L_{31}= \int_{\mathbb{R}^{4+1}} d\textbf{x} dy \frac{\mu_1(x,y)}{f^2(x,y)}\frac{1+\tau(y)r(x)}{f(x,y)f(x,-y)}\,.
\end{equation}
Each diagram can be computed in the same way. One can check for instance that the contribution of the first diagram equals one of the fifth diagrams, and after a tedious computation we get for $\beta_\lambda:= \dot{\bar{\lambda}}$:
\begin{align}
\boxed{
\beta_\lambda=-2\eta \bar{\lambda}-\frac{3}{2}\bar{\kappa}\left(L_{21}(0)-L_{22}(0) \right)+\frac{8\bar{\lambda}^2}{\pi} \left(L_{31}+\frac{1}{2}L_{32}-L_{33} \right)\,,}
\end{align}\label{flowphi4}
where:
\begin{align}
L_{32}&:= \int_{\mathbb{R}^{4+1}} d\textbf{x} dy \mu_1(x,y)\frac{1+\tau(y)r(x)}{f^2(x,y)f^2(x,-y)}\,,\\
L_{33}&:= \int_{\mathbb{R}^{4+1}} d\textbf{x} dy \frac{\mu_2(x,y)}{f^2(x,y)f(x,-y)}\,.
\end{align}
Note that to derive these equations we used the relation:
\begin{equation}
\boxed{
G_{k,\bar{\sigma} M}(\hat{\omega})=G_{k,\bar{M} \sigma}(-\hat{\omega})\,.}\label{symG}
\end{equation}
which are also trues for the bare propagators given by equation \eqref{freepropa1}.

\subsubsection{Structure equations}\label{structureeq}

The flow equation \eqref{flowphi4} for the quartic coupling $\lambda$ involves the sixtic coupling $\kappa$. Hence in principle, we are obliged to consider the flow equation for $\kappa$, which involves the octic couplings and so on. The infinite hierarchical structure does not stop, even if:
\begin{enumerate}
\item We stop it abruptly, imposing $\Gamma_k^{(2n)}=0$ up to some $n$ (crude truncation).
\item We are able to express $\Gamma_k^{(2n)}$, up to a given $n$, in terms of $\Gamma_k^{(2(n-1))}$, $\Gamma_k^{(2(n-2))}$ and so on.
\end{enumerate}
The first option has been widely considered for TGFTs \cite{Carrozza_2017,Carrozza_2017a,Benedetti_2015,Geloun_2016,Ben_Geloun_2015,Geloun_2018}, but some instability effects and incompatibilities with symmetry constraints have been noticed 
 in our previous works \cite{lahoche2021no, Lahoche_2020b,Lahoche_2019bb}, and the reliability of its predictions is still debated. The second option is more difficult to implement in general. It happens that we can close the hierarchy in this way by exploiting some constraints coming from the symmetries of the theory and which imply exact relations between effective vertices, such as the Ward identities (see for instance \cite{canet2011nonperturbative}). In this paper we follow the strategy developed in our previous work \cite{Lahoche:2018oeo}, where authors present a method exploiting the tree structure of leading order graphs, as melonic graphs, to get non-trivial relations between non-branching observable. Because melonic diagrams dominate the RG flow in the deep UV, this strategy is expected to out-perform the standard vertex expansion. Indeed, this method, called effective vertex expansion (EVE) allows closing the hierarchy and capture the full momenta dependence of effective vertices. We will now detail it here.
\medskip

\begin{figure}
\begin{center}
$\vcenter{\hbox{\includegraphics[scale=0.7]{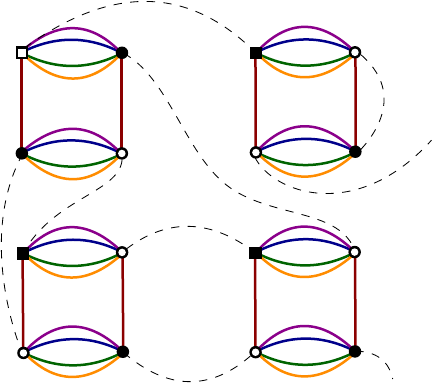}}}\qquad\Rightarrow\qquad \vcenter{\hbox{\includegraphics[scale=0.7]{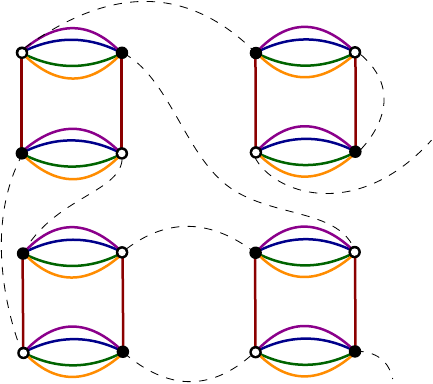}}}$
\end{center}
\caption{A typical Feynman graph $G$ (on left) and the corresponding normal graph $\bar{G}=\mathcal{F}(G)$ (on right).}\label{figFeynman3}
\end{figure}
If we consider the quartic model given by \eqref{model}, there are two kinds of quartic vertices, corresponding to vertices of type $a$ and type $\bar{a}$. Hence, a general Feynman graph for the model takes the form given by Figure \ref{figFeynman3} (on left), involving type $a$ and type $\bar{a}$ vertices. Note that we have no dotted edges liking square nodes because $G_{k \bar{\chi}\chi}=0$. We denote as $\mathcal{G}$ the set of Feynman graphs corresponding to this model. We moreover define $\mathcal{F}$ as the surjective map $\mathcal{F}: \mathcal{G} \to \bar{\mathcal{G}}$, which send any Feynman graph $G$ to a Feynman graph $\bar{G} \in \bar{\mathcal{G}}$ of the equilibrium theory \eqref{normequilibrium}. We denote as $\bar{\mathcal{G}}$ the set of Feynman graphs for the equilibrium theory. To be more precise, $\mathcal{F}$ acts on a given graph $G$ by replacing all square nodes with disk nodes, without changing their color. A write square becomes a white disk and a black square becomes a black disk, as illustrated on \ref{figFeynman3}. Moreover, propagators \eqref{freepropa1} and \eqref{freepropa2} for the dynamical theory are replaced by the propagator \eqref{equilibriumpropa} of the equilibrium theory, a rank $5$ Abelian GFT for $\U(1)$ structure group. Obviously, the inverse map $\mathcal{F}^{-1}$ is not one-to-one in general: $\mathcal{F}^{-1}(\bar{G})=(G_1, G_2\cdots, G_K)$, and we denote as \textit{K} the \textit{multiplicity of the graph $\bar{G}$}.
\begin{figure}
\begin{center}
\includegraphics[scale=1]{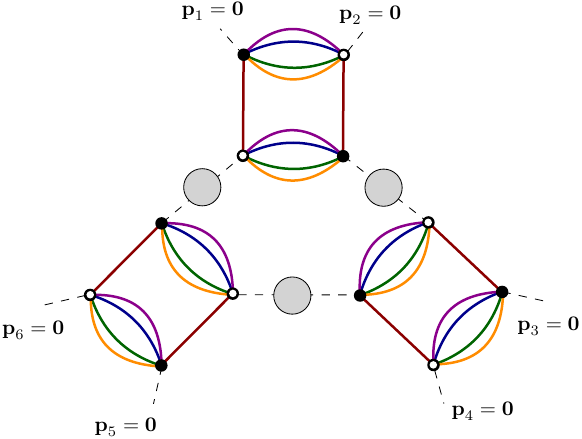}
\end{center}
\caption{The effective $6$-points graph $\bar{G}_0$. The external $4$-points vertex are indeed effective $4$-points functions, that we denote by their boundary graph.}\label{figG0}
\end{figure}
In our previous works, EVE has been considered for the field theory corresponding to the equilibrium state \eqref{equilibrium} in \cite{Lahoche_2020b,Lahoche:2018oeo,lahoche2021no}. The authors showed that melonic non-branched $6$-point function $\Gamma_k^{(6),(\ell)}$, corresponding to sixtic melonic boundaries with color $\ell$ (accordingly with the definition \eqref{decompositiond}) can be expressed in terms of the $4$-points and $2$-points functions, and for zero external momenta (which is what we need to close the hierarchy), this relation reads:
\begin{equation}
\Gamma_{k,\text{eq}}^{(6),(\ell)}\bigg\vert_{0}= (3!)^2 \times \mathcal{A}_{\bar{G}_0}\,,
\end{equation}
where $(3!)^2$ counts the number of different configurations for external momenta and the graph $\bar{G}_0$ is pictured in Figure \ref{figG0}. Note that $\bar{G}_0$ is not truly a Feynman graph but an \textit{effective graph}, where external $4$ point's vertex are effective $4$-points functions materialized by their boundary graphs and resuming an arbitrary number of graphs and where the interior $2$-point functions have been resumed as well (see equation \eqref{equilibriumpropa}):
\begin{equation}
\vcenter{\hbox{\includegraphics[scale=1.2]{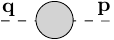}}}=\, \frac{\delta_{\textbf{p}\textbf{q}}}{\textbf{p}^2+ m^2-\Sigma_R(\textbf{p})}\,,
\end{equation}
where $\textbf{p}\in \mathbb{Z}^5$ and $\Sigma_R$ means that divergences of the self-energy have been canceled by counter-terms $Z_{\infty}$ and $Z_m$, accordingly with the renormalization condition \eqref{2pointeqren}. If we apply the inverse map $\mathcal{F}^{-1}$ to $\bar{G}_0$ we obtain a family of graphs $(G_1,\cdots, G_K)$, but having different boundaries (see definition \ref{defboundary}). Hence, if we restrict ourselves to the graphs having the same boundary $\partial G_0$ of $\Gamma_{k, {\sigma}\bar{M}M \bar{M} M\bar{M}}^{(6),(\ell)}\big\vert_{\text{0}}$, we focus on the set $S=\{ \mathcal{F}^{-1}[\bar{G}_0] \vert \partial G_i=\partial G_0 \,\forall G_i\in \mathcal{F}^{-1}[\bar{G}_0]\}$, where explicitly:
\begin{equation}
\partial G_0:=\vcenter{\hbox{\includegraphics[scale=1]{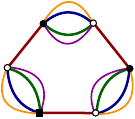}}}\,,
\end{equation}
and the zero momenta effective vertex function $\Gamma_{k, \bar{\sigma} M \bar{M}M \bar{M} M}^{(6),(\ell)}\big\vert_{\text{0}}$ decomposes as:
\begin{equation}
\Gamma_{k, {\sigma}\bar{M}M \bar{M} M\bar{M}}^{(6),(\ell)}\big\vert_{\text{0}}= 12\times \sum_{G\in S} \mathcal{A}_G\,,
\end{equation}
where we used the same notation $\mathcal{A}$ to denote the amplitude of the stochastic model, where $12=3!\times 2!$ counts the number of external momenta arrangements as before. It is easy to see that there are only three configurations for effective graphs, and graphically:
\begin{equation}
\Gamma_{k, {\sigma}\bar{M}M \bar{M} M\bar{M}}^{(6),(\ell)}\big\vert_{\text{0}}= 12\times \left(\vcenter{\hbox{\includegraphics[scale=0.31]{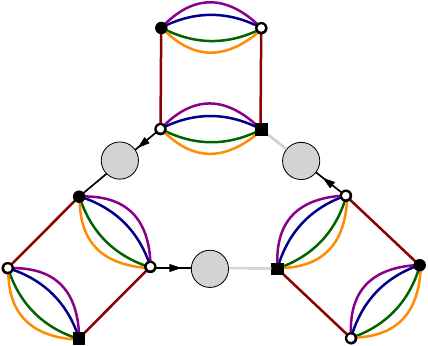}}}\quad +\quad \vcenter{\hbox{\includegraphics[scale=0.31]{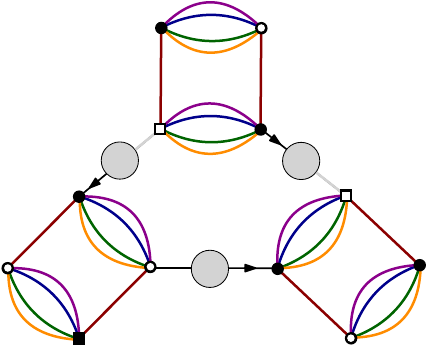}}} \quad +\quad \vcenter{\hbox{\includegraphics[scale=0.31]{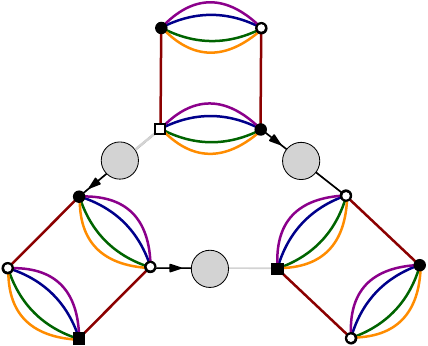}}} \right)\,,
\end{equation}
disregarding external momenta and frequencies for simplicity. Note moreover that, as before, external $4$-point vertices are effective $4$-points functions materialized by their boundaries. Explicitly, we get: 
\begin{equation}
\Gamma_{k, {\sigma}\bar{M}M \bar{M} M\bar{M}}^{(6),(\ell)}\big\vert_{\text{0}}= 12Z^3(k) k^{-2}i\left( \frac{\bar{\lambda}}{\pi}\right)^3\times \left( A I_1+BI_2+ C I_3 \right)\delta(0)\,,
\end{equation}
where $A$ and $B$ are numerical constants that we can determine by perturbation theory. It is easy to check that the loop integrals $I_1$, $I_2$ and $I_3$ are equals, and, using the integral approximation introduced before:
\begin{equation}
I_1:=  \int_{\mathbb{R}^4} d\textbf{x} \int dy\frac{1+\tau(y)r(x)}{f^3(x,y)f(x,-y)}\,.\label{loopsix}
\end{equation}
The perturbation theory leads to $A=B=C=16$, and we get from \eqref{gamma6zero} and \eqref{sansdimkappa}
\begin{equation}
\boxed{
\bar{\kappa}=\frac{64\bar{\lambda}^3}{2\pi} \int_{\mathbb{R}^4} d\textbf{x} \int dy\frac{1+\tau(y)r(x)}{f^3(x,y)f(x,-y)}\,.}\label{equationkappa}
\end{equation}
We are now in a position to derive the last piece of the puzzle, namely, the derivative of the effective vertex at zero external momenta \eqref{derivvertex}, required to compute the anomalous dimension.

\begin{remark}
It can be noticed that we used truncation to define the loop integral \eqref{loopsix}. The truncation is in principle expected to be valid only in a small region around $k$, namely on the support of $\dot{\textbf{R}}_k$. In \cite{Lahoche:2018oeo}, {\color{blue}we} showed that using truncation outside this support for divergent integral leads to dramatically wrong conclusions. In the same reference and \cite{Lahoche_2020b}, the {\color{blue} we} showed that it can be a good approximation for convergent integral, and in particular that this does not introduce Ward identity violations. We use this approximation scheme to compute our convergent integrals in this paper, as $I_1$ is.
\end{remark}

To conclude, not that in our truncation, the integral in \eqref{loopsix} can be computed analytically. For instance, without coarse-graining in frequency ($\hat{\beta}=0$) and setting $\alpha=1$, we have:
\begin{align}
\bar{\kappa}\vert_{\hat{\beta}=0}= \frac{4\pi ^2\bar{\lambda}^3 (\bar{m}^2 (\bar{m}^2+3)+3)}{(1+\bar{m}^2)^3} \,.
\end{align}
In the opposite limit, for $\alpha=0$ but $\hat{\beta}=1$, we get:
\begin{equation}
\bar{\kappa}\vert_{\alpha=0}=\frac{4\pi^2 \bar{\lambda}^3}{\bar{m}^2}\,.
\end{equation}
Both have an obvious infrared singularity for $\bar{m}^2=0$. Finally, for $\alpha=\hat{\beta}=1$, we get:
\begin{align}
\nonumber \bar{\kappa}\vert_{\alpha=1,\hat{\beta}=1}=\frac{4\pi ^2 \lambda ^3}{ (1+\bar{m}^2)^3} \Big(\bar{m}^2 (\bar{m}^2&+7)+2 (\bar{m}^2+1) (2 \bar{m}^2+3) \log (1+\bar{m}^2)\\
&-2 (1+\bar{m}^2) (2\bar{m}^2+3) \log (2+\bar{m}^2)+7\Big)\,.
\end{align}

\section{Anomalous dimension and WT identities}\label{WTI}

The aim of this section is to use WT identities to compute the last ingredients in the expression of anomalous dimension $\lambda^\prime$, defined in \eqref{derivvertex}. Since the interactions are invariants by construction under unitary transformations of the type \eqref{unitarytrans}, there must exist non-trivial Ward-Takahashi (WT) identities between effective vertex functions. These identities provide non-trivial relations that constrain the RG flow and allow us to compute the derivative of the effective vertices for their external momenta, which is exactly what we need to compute the anomalous dimension. These WT identities have been extensively discussed in the literature in the last years with this aim, see for instance \cite{Lahoche:2018oeo}, where authors investigate the equilibrium model \eqref{equilibrium}.

\subsection{WT identities for unitary symmetry}

Working in the Peter-Weyl basis, the unitary transformations \eqref{unitarytrans} act formally on fields components as:
\begin{equation}
T_{\bm p} \to T_{\bm p}^\prime = \sum_{q_i\in \mathbb{Z}} U_{p_i q_i} T_{\bm q}\big\vert_{q_j=p_j\,\forall j\neq i}\,,
\end{equation}
where:
\begin{equation}
\sum_{q\in \mathbb{Z}} U^\dagger_{pq} U_{q p^\prime}=\sum_{q\in \mathbb{Z}} U_{pq} U^\dagger_{q p^\prime}= \delta_{pp^\prime}\,.
\end{equation}
The interaction part of the Hamiltonian $\mathcal{H}$ is invariant under such a transformation. Indeed, invariance is only broken by the Laplacian term in the kinetic action. Let us consider now the classical action $S$ given by \eqref{classicaction0} and the generating functional $Z[J,\bar{J},\jmath, \bar{\jmath}\,]$, equation \eqref{generatingfunctional0}. The interacting part $S_{\text{int}}$ of the classical action is invariant if we transform both fields $\varphi$ and $\chi$:
\begin{equation}
T_{\bm p} \to T_{\bm p}^\prime = \sum_{\bm q\in \mathbb{Z}^d} \left[\prod_{i=1}^d U_{p_i q_i}^{(i)}\right] T_{\bm q}\,,\qquad \chi_{\bm p} \to \chi_{\bm p}^\prime = \sum_{\bm q\in \mathbb{Z}^d} \left[\prod_{i=1}^d U_{p_i q_i}^{(i)}\right] \chi_{\bm q}\,,
\end{equation}
where $\{ U^{(i)} \}$ are $d$--\textit{independent} unitary transformations. We consider infinitesimal transformations,
\begin{equation}
U=\mathrm{I}+\epsilon +\mathcal{O}(\epsilon^2)\,,
\end{equation}
where $\mathrm{I}$ is the identity matrix (with elements $\delta_{pq}$) and $\epsilon=-\epsilon^\dagger$ is along the natural representation of the Lie algebra of the unitary group. We furthermore define the operator $\hat{\epsilon}_i$, acting on the $i$--th component of fields as:
\begin{equation}
\hat{\epsilon}_i[T]_{\bm p} := \sum_{q_i} \epsilon_{p_i q_i} T_{\bm q} \big\vert_{q_j=p_j\,,j\neq i}\,.
\end{equation}
The global reparametrization invariance of the path integral defining the generating functional $Z[J,\bar{J},\jmath, \bar{\jmath}\,]$ means that:
\begin{equation}
\hat{\epsilon}_i[Z[J,\bar{J},\jmath, \bar{\jmath}\,]]=0\,,
\end{equation}
for all $i\in \llbracket 1,d\, \rrbracket$. We can expand this relation to first order in $\epsilon$:
\begin{equation}
0\equiv \int d\bm q d\bm \chi \left[\hat{\epsilon}_i[S[\bm q,\bm \chi]]+\hat{\epsilon}_i[\Delta S_k[\bm q,\bm \chi]]-\hat{\epsilon}_i[\bm J\cdot \bm q+\bm \jmath \cdot \bm \chi] \right]e^{-S[\bm q,\bm \chi]-\Delta S_k[\bm q,\bm \chi]+\bm J \cdot \bm q+ \bm \jmath \cdot \bm \chi}\,.\label{var1}
\end{equation}
We will compute each term of the variation separately, starting with the source terms:
\paragraph{Computation of $\hat{\epsilon}_i[\bm J\cdot \bm q+\bm \jmath \cdot \bm \chi]$.} The operator $\hat{\epsilon}_i$ acts linearly on each field, and after some arrangements we get:
\begin{align}
\nonumber\hat{\epsilon}_i[\bm J\cdot \bm q+\bm \jmath \cdot \bm \chi]=\int d{\omega}\sum_{\bm p, \bm p^\prime}& \prod_{j\neq i} \delta_{p_jp_j^\prime} [\bar{J}_{\bm p}({\omega}) T_{\bm p^\prime}({\omega})-\bar{T}_{\bm p}({\omega}) J_{\bm p^\prime}({\omega})\\
&+\bar{\jmath}_{\bm p}({\omega}) \chi_{\bm p^\prime}({\omega})-\bar{\chi}_{\bm p}({\omega}) \jmath_{\bm p^\prime}({\omega})]\epsilon_{p_ip_i^\prime}\,.
\end{align}

\paragraph{Computation of $\hat{\epsilon}_i[S[\bm q,\bm \chi]]$.} The variation splits in two contributions, for kinetic part and interactions:
\begin{equation}
\hat{\epsilon}_i[S[\bm q,\bm \chi]]=\hat{\epsilon}_i[S_{\text{kin}}[\bm q,\bm \chi]]+\hat{\epsilon}_i[S_{\text{int}}[\bm q,\bm \chi]]\,.
\end{equation}
The second contribution to interaction vanishes by construction. The kinetic action for the response field $\sum_{\bm p} \bar{\chi}_{\bm p} \chi_{\bm p}$ is invariant as well, and the corresponding variation vanish. This is also the case for contributions like $\hat{\omega} \sum_{\bm p} \bar{T}_{\bm p} \chi_{\bm p}$ and $m^2 \sum_{\bm p} \bar{T}_{\bm p} \chi_{\bm p}$. Finally, only the Laplacian contributes non-trivially to the variation, and we get:
\begin{equation}
\hat{\epsilon}_i[S[\bm q,\bm \chi]]=iZ_{\infty}\int d{\omega}\sum_{\bm p, \bm p^\prime} \prod_{j\neq i} \delta_{p_jp_j^\prime} \left[p_i^2-p_i^{\prime 2}\right](\bar{\chi}_{\bm p}({\omega}) T_{\bm p^\prime}({\omega})+\bar{T}_{\bm p}({\omega}) \chi_{\bm p^\prime}({\omega}) ) \epsilon_{p_ip_i^\prime}\,.
\end{equation}

\paragraph{Computation of $\hat{\epsilon}_i[\Delta S_k[\bm q,\bm \chi]]$.} The computation of the variation of the regulator follows the same strategy as for the kinetic action:
\begin{align}
\nonumber\hat{\epsilon}_i[\Delta S_k[\bm q,\bm \chi]]=&\int d{\omega}\sum_{\bm p, \bm p^\prime} \prod_{j\neq i} \delta_{p_jp_j^\prime} \bigg(i[R_k^{(1)}(\bm p,\omega)-R_k^{(1)}(\bm p^\prime,\omega)]\bar{\chi}_{\bm p}({\omega}) T_{\bm p^\prime}({\omega})\\\nonumber
&+i[R_k^{(1)}(\bm p,-\omega)-R_k^{(1)}(\bm p^\prime,-\omega)]\bar{T}_{\bm p}({\omega}) \chi_{\bm p^\prime}({\omega})\\
& +[R_k^{(2)}(\bm p,\omega)-R_k^{(2)}(\bm p^\prime,\omega)] \bar{\chi}_{\bm p}({\omega}) \chi_{\bm p^\prime}({\omega}) \bigg)\epsilon_{p_ip_i^\prime}\,.
\end{align}
\medskip

Taking into account all these contributions, the variation \eqref{var1} implies the relation:
\begin{align}
\nonumber0&=\int d{\omega}\sum_{\bm p, \bm p^\prime} \prod_{j\neq i} \delta_{p_jp_j^\prime} \bigg[\bigg(iZ_{\infty}\left[p_i^2-p_i^{\prime 2}\right]+i[R_k^{(1)}(\bm p,\omega)-R_k^{(1)}(\bm p^\prime,\omega)]\bigg) \\\nonumber
&\times\frac{\partial}{\partial \jmath_{\bm p}({\omega})} \frac{\partial}{\partial \bar{J}_{\bm p^\prime}({\omega})}
+\bigg(iZ_{\infty}\left[p_i^2-p_i^{\prime 2}\right]+i[R_k^{(1)}(\bm p,-\omega)-R_k^{(1)}(\bm p^\prime,-\omega)]\bigg)\\\nonumber
&\times \frac{\partial}{\partial \bar{\jmath}_{\bm p^\prime}({\omega})} \frac{\partial}{\partial {J}_{\bm p}({\omega})}+[R_k^{(2)}(\bm p,\omega)-R_k^{(2)}(\bm p^\prime,\omega)]\frac{\partial}{\partial \jmath_{\bm p}({\omega})}\frac{\partial}{\partial \bar{\jmath}_{\bm p^\prime}({\omega})}\\
&-\bigg(\bar{J}_{\bm p}({\omega}) \frac{\partial}{\partial \bar{J}_{\bm p^\prime}({\omega})}- J_{\bm p^\prime}({\omega})\frac{\partial}{\partial {J}_{\bm p}({\omega})}+\bar{\jmath}_{\bm p}({\omega})\frac{\partial}{\partial \bar{\jmath}_{\bm p^\prime}({\omega})}-\jmath_{\bm p^\prime}({\omega})\frac{\partial}{\partial {\jmath}_{\bm p}({\omega}) }\bigg)
\bigg] e^{W_k[J,\bar{J},\jmath,\bar{\jmath}\,]}\,,\label{Ward1}
\end{align}
where we introduced the free energy:
\begin{equation}
W_k[J,\bar{J},\jmath,\bar{\jmath}\,]:=\ln Z_k[J,\bar{J},\jmath,\bar{\jmath}\,]\,.
\end{equation}
We also introduce the following notation:
\begin{equation}
G^{(n+\bar{n};M+\bar{M})}_{p_1\cdots p_n, \bar{p}_1\cdots \bar{p}_{\bar{n}}; p_I\cdots p_M, \bar{p}_1\cdots \bar{p}_{\bar{M}}}:=\prod_{i=1}^n \frac{\partial}{\partial \bar{J}_{p_i}} \prod_{\bar{i}=1}^{\bar{n}} \frac{\partial}{\partial {J}_{\bar{p}_{\bar{i}}}} \prod_{I=1}^M \frac{\partial}{\partial \bar{\jmath}_{p_I}} \prod_{\bar{I}=1}^{\bar{M}} \frac{\partial}{\partial {\jmath}_{\bar{p}_{\bar{I}}}}W_k\,,
\end{equation}
where the notation $p_i$ means $p_i=(\bm p_i,\hat{\omega}_i)$. We furthermore introduce the following notations for classical fields:
\begin{equation}
M_{p}:=\frac{\partial W_k}{\partial \bar{J}_p} \,,\quad \bar{M}_{p}:=\frac{\partial W_k}{\partial {J}_p}\,, \quad \sigma_{p}:=\frac{\partial W_k}{\partial \bar{\jmath}_p}\,,\quad \bar{\sigma}_{p}:=\frac{\partial W_k}{\partial {\jmath}_p}\,.
\end{equation}
The equation \eqref{Ward1} then simplifies, and we deduce the following statement:
\begin{proposition}\label{propositionWard}
Observable of the equilibrium dynamical model satisfy the following Ward-Takahashi identity:
\begin{align}
\nonumber0&=\int d{\omega}\sum_{\bm p, \bm p^\prime} \prod_{j\neq i} \delta_{p_jp_j^\prime} \bigg[\left(iZ_{\infty}\left[p_i^2-p_i^{\prime 2}\right]+i[R_k^{(1)}(\bm p,\omega)-R_k^{(1)}(\bm p^\prime,\omega)]\right) \\\nonumber
&\times\bigg(G_{k,\bar{\sigma}M}^{(1;\bar{1})}(\bm p^\prime,{\omega};\bm p,{\omega})
+\bar{\sigma}_{\bm p}({\omega})M_{\bm p^\prime}({\omega})\bigg)\\\nonumber
&+\left(iZ_{\infty}\left[p_i^2-p_i^{\prime 2}\right]+i[R_k^{(1)}(\bm p,-\omega)-R_k^{(1)}(\bm p^\prime,-\omega)]\right)\\\nonumber
&\times\bigg(G_{k,\bar{M}{\sigma} }^{(\bar{1};1)}( \bm p,{\omega};\bm p^\prime,{\omega})+{\sigma}_{\bm p^\prime}({\omega})\bar{M}_{\bm p}({\omega})\bigg)\\\nonumber
&+[R_k^{(2)}(\bm p,\omega)-R_k^{(2)}(\bm p^\prime,\omega)] \left(G_{k,\bar{\sigma}\sigma}^{(0;{1}+\bar{1})}(\bm p^\prime,{\omega};\bm p,{\omega})+{\sigma}_{\bm p^\prime}({\omega})\bar{\sigma}_{\bm p}({\omega}) \right)\\
&-\bar{J}_{\bm p}({\omega}) M_{\bm p^\prime}({\omega})+ J_{\bm p^\prime}({\omega})\bar{M}_{\bm p}({\omega})-\bar{\jmath}_{\bm p}({\omega})\sigma_{\bm p^\prime}({\omega})+\jmath_{\bm p^\prime}({\omega})\bar{\sigma}_{\bm p}({\omega})
\bigg]\delta_{p_ip}\delta_{p_i^\prime p^\prime} \,.\label{Ward1}
\end{align}
\end{proposition}
We will exploit these identities in the melonic approximation, focusing on the non-branching sector of the theory, to compute $\lambda^\prime$. A technical complement is given in Appendix \ref{AppB}.

\subsection{Computation of $\lambda^\prime$ defined in \eqref{derivvertex}}\label{derivlambda}
We now move to the last Ward identity that we need to achieve our RG program. Applying the fourth derivative $\partial^4/\partial M_{\bm q}({\omega}_1)\partial \bar{\sigma}_{\bm{\bar{q}}}({\bar{\omega}}_1) \partial M_{\bm q^\prime}({{\omega}}_1^\prime) \partial \bar{M}_{\bm{\bar{q}^\prime}}({\bar{\omega}}_1^\prime)$. As the previous Ward identities provided a relation between the difference of $2$-points functions at different momenta and the $4$-point function, the Ward identities that we will derive in this section will provide non-trivial relations between $4$, $6$ points functions, and the difference between $4$-point kernels $\pi_k^{(2)}(p_1^2,p_2^2)$ with different momenta. As for previous relations, we have to note that Ward identities enjoy the same structure as flow equations considered in the previous section, and indeed play a symmetric role: Ward identities say how we escape to the purely local sector, i.e. how to move inside the theory space as momentum change, and RG equations say how the theory move as the cut-off change. Obviously, the Ward generators do not commute with the flow, because flow equations describe the flow of derivative coupling as well (as the anomalous dimension for instance).
\medskip

Using the same graphical representation as before, we obtain the equality:
\begin{align}
\nonumber&-\vcenter{\hbox{\includegraphics[scale=0.7]{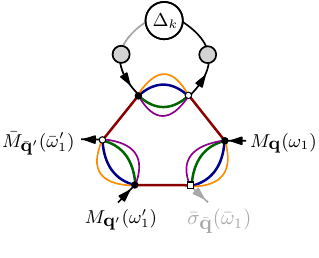}}}-\vcenter{\hbox{\includegraphics[scale=0.7]{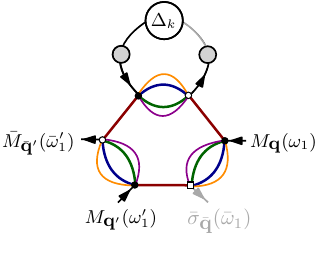}}}-\vcenter{\hbox{\includegraphics[scale=0.7]{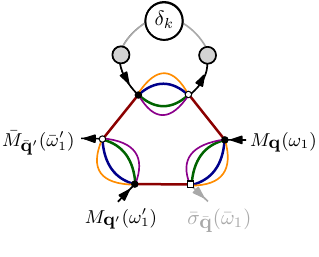}}}\\\nonumber
&+\vcenter{\hbox{\includegraphics[scale=0.7]{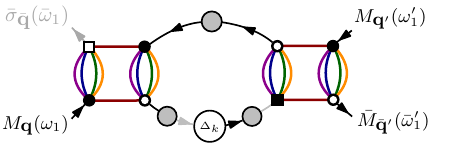}}}+\vcenter{\hbox{\includegraphics[scale=0.7]{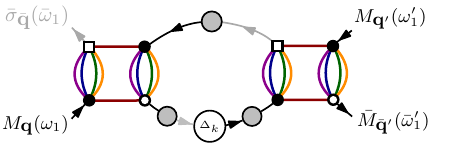}}}\\\nonumber
&+\vcenter{\hbox{\includegraphics[scale=0.7]{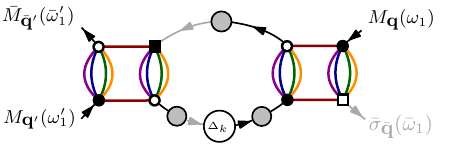}}}+\vcenter{\hbox{\includegraphics[scale=0.7]{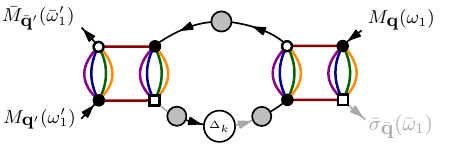}}}\\\nonumber
&+\vcenter{\hbox{\includegraphics[scale=0.7]{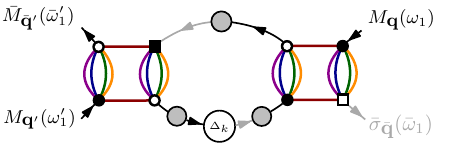}}}+\vcenter{\hbox{\includegraphics[scale=0.7]{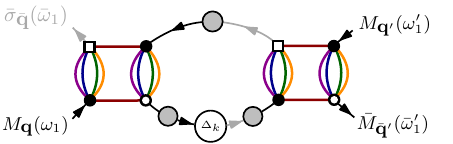}}}\\\nonumber
&+\vcenter{\hbox{\includegraphics[scale=0.7]{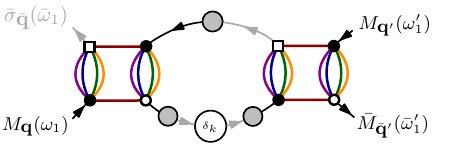}}}+\vcenter{\hbox{\includegraphics[scale=0.7]{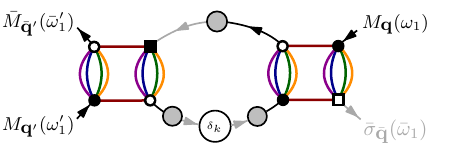}}}\\
&-\left(\vcenter{\hbox{\includegraphics[scale=0.7]{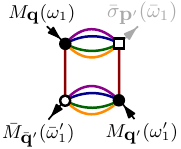}}}+\vcenter{\hbox{\includegraphics[scale=0.7]{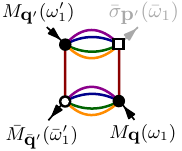}}}\right)\delta_{\bm{p}\bar{\bm{q}}}+\left(\vcenter{\hbox{\includegraphics[scale=0.7]{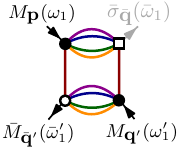}}}+\vcenter{\hbox{\includegraphics[scale=0.7]{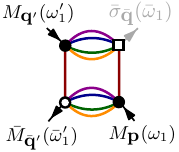}}}\right)\delta_{\bm{p}^\prime \bm{q}}=0
\end{align}
where permutations of external fields $M_{\bm q^\prime}({{\omega}}_1^\prime)$ and $M_{\bm q}({\omega}_1)$ are assumed when they are required, and where we introduced the kernels (following the convention of \eqref{equationregul}):
\begin{align}
\nonumber \Delta_k(\bm{p},\omega)&\equiv
\begin{pmatrix}
0&iZ_{\infty}\delta p^2+i\delta R_k^{(1)}(\bm p,\omega)\\
iZ_{\infty}\delta p^2+i\delta R_k^{(1)}(\bm p,-\omega)&0
\end{pmatrix} \delta_{p_ip}\delta_{p_i^\prime p^\prime}\\
&=:\Delta_k^\prime(\bm{p},\omega)\delta p^2 \,,
\end{align}
and:
\begin{equation}
\delta_k(\bm{p},\omega)\equiv \delta R_k^{(2)}(\bm p,\omega)\delta p^2 \delta_{p_ip}\delta_{p_i^\prime p^\prime}=:\delta_k^\prime(\bm{p},\omega)\delta p^2\,,
\end{equation}
the variation $\delta p:=p_i-p_i^\prime$ being at first order in the difference. Moreover, we assumed $\bm p^\prime \neq \bm q^\prime$ to cancel the last term involving two $4$-point diagrams, proportional to $\delta_{\bm p^\prime \bm q^\prime}$. This relation can be translated as a differential equation for $\pi_k^{(2)}$ as follows. We set $p_j=p_j^\prime =q_j=\bar{q}_j= 0 \, \forall j\neq i$, $p_i=p_i^\prime+1$, $\bm{q}^\prime=\bar{\bm{q}}^\prime=\bm 0$, $p_i= \bar{q}_i$ and $p_i^\prime=q_i$. With this configuration for the external momenta, the two last parentheses involve the difference:
\begin{equation}
\pi_k^{(2)}(0,p_i^2)-\pi_k^{(2)}(0,(p_i^\prime)^2)=\pi_k^{(2)}(0,p_i^2)-\pi_k^{(2)}(0,(p_i-1)^2) \,,
\end{equation}
which can be approached in the deep UV regime ($\Lambda \gg 1$) by a derivative (see the discussion before equation \eqref{equationderiv} in Appendix \ref{AppB}). Hence, setting $p_i=0$, we have:
\begin{equation}
\bigg[\pi_k^{(2)}(0,p_i^2)-\pi_k^{(2)}(0,(p_i-1)^2)\bigg]\bigg\vert_{p_i=0} \approx \frac{d}{dp_i^2}\pi_k^{(2)}(0,p_i^2)\bigg\vert_{p_i=0}\delta p^2\,,
\end{equation}
that can be rewritten as:
\begin{equation}
\frac{d}{dp_i^2}\pi_k^{(2)}(0,p_i^2)\bigg\vert_{p_i=0}=\frac{1}{2}\frac{d}{dp_i^2}\pi_k^{(2)}(p_i^2,p_i^2)\bigg\vert_{p_i=0} \,,
\end{equation}
assuming $\pi_k^{(2)}$ to be a symmetric function, and the Ward identity reads explicitly:
\begin{align}
- \frac{1}{2}\frac{i}{\pi}\frac{d}{dp_i^2}\pi_k^{(2)}(p_i^2,p_i^2)\bigg\vert_{p_i=0}=\frac{i\kappa}{2\pi^2}\mathcal{L}_{k,1} - 4i\left(\frac{\lambda}{\pi}\right)^2 \left(\mathcal{L}_{k,2}^{(1)}+\mathcal{L}_{k,2}^{(2)} \right) \,.
\end{align}
where we defined:
\begin{align}
\nonumber\mathcal{L}_{k,1}:=\int d\omega \sum_{\bm p \in \mathbb{Z}^{4}} \Bigg[&2\left(iZ_{\infty}+i \frac{d}{dp_1^2}R_k^{(1)}(\bm p,\omega)\right)G_{k,\bar{M} M}(\bm p,\omega) G_{k,\bar{\sigma} M}(\bm p,\omega)\\
&-\frac{d}{dp_1^2}R_k^{(2)}(\bm p,\omega) G_{k,\bar{\sigma} M}(\bm p,-\omega) G_{k,\bar{\sigma} M}(\bm p,\omega)\Bigg]\bigg\vert_{p_i=0}\,,
\end{align}
\begin{align}
\nonumber\mathcal{L}_{k,2}^{(1)}:=-i\int d\omega \sum_{\bm p \in \mathbb{Z}^{4}} &\left(iZ_{\infty}+i \frac{d}{dp_1^2}R_k^{(1)}(\bm p,\omega)\right)G_{k,\bar{M} M}(\bm p,\omega) G_{k,\bar{\sigma} M}(\bm p,\omega)\\
&\qquad \times \left( 2G_{k,\bar{\sigma} M}(\bm p,\omega)+G_{k,\bar{\sigma} M}(\bm p,-\omega)\right)\bigg\vert_{p_i=0}\,,
\end{align}
and
\begin{equation}
\mathcal{L}_{k,2}^{(2)}:=-i \int d\omega \sum_{\bm p \in \mathbb{Z}^{4}} \frac{d}{dp_1^2}R_k^{(2)}(\bm p,\omega) G_{k,\bar{\sigma} M}(\bm p,-\omega) G_{k,\bar{\sigma} M}^2(\bm p,\omega)\bigg\vert_{p_i=0} \,.
\end{equation}
Using dimensionless quantities and the definition \eqref{derivvertex},
\begin{align}
\boxed{
\bar{\lambda}^\prime= -\frac{\bar{\kappa}}{\pi} \mathcal{\bar{L}}_{k,1} +\frac{8\bar{\lambda}^2}{\pi} \left(\mathcal{\bar{L}}_{k,2}^{(1)}+\mathcal{\bar{L}}_{k,2}^{(2)} \right)}
\end{align}
where $\bar{\kappa}$ is given by equation \eqref{equationkappa} and, explicitly:
\begin{align}
\nonumber\mathcal{\bar{L}}_{k,1}:=2\int dy \int_{\mathbb{R}^4} d\bm x \Bigg[&\left(Z_{\infty}{Z}^{-1}(k)-\alpha \hat{\rho}(y) \theta(1-x)\right)\frac{1+\hat{\tau}(y)r(x)}{\hat{f}(x,y)\hat{f}^2(x,-y)}\\
&-\frac{1}{2}\alpha \hat{\tau}(y)\frac{ \theta(1-x) }{\hat{f}(x,y)\hat{f}(x,-y)}\Bigg]\,,
\end{align}
\begin{align}
\nonumber \mathcal{\bar{L}}_{k,2}^{(1)}:= -\int dy \int_{\mathbb{R}^4} d\bm x \Bigg[&\left(Z_{\infty}{Z}^{-1}(k)- \alpha \hat{\rho}(y) \theta(1-x)\right)\frac{1+\hat{\tau}(y)r(x)}{\hat{f}(x,y)\hat{f}^2(x,-y)}\\
&\times \left(\frac{2}{\hat{f}(x,-y)}+\frac{1}{\hat{f}(x,y)} \right)\Bigg]\,,
\end{align}
and:
\begin{equation}
\mathcal{\bar{L}}_{k,2}^{(2)}:=-\alpha\, \int dy \int_{\mathbb{R}^4} d\bm x\, \hat{\tau}(y)\frac{ \theta(1-x) }{\hat{f}(x,y)\hat{f}^2(x,-y)}\,.
\end{equation}
The previous expression can be further simplified. Indeed,  as discussed in the appendix \ref{AppB},  equation \eqref{Wardcontinuum} allows to replace $\mathcal{\bar{L}}_{k,1}$ by $\mathcal{\bar{L}}_{k,1} \approx -\pi/2\bar{\lambda}$. Moreover, repeating the argument given in our previous work \cite{Lahoche:2018oeo}, the sums involved in $ \mathcal{\bar{L}}_{k,2}^{(1)}$ and $ \mathcal{\bar{L}}_{k,2}^{(2)}$ being superficially convergent, the terms involving $Z_\infty$ has to be canceled in the continuum limit. Indeed, because we focus on the UV regime but so far to the IR scale, $Z_\infty/Z(k) \to 0$ as $1/\ln(\Lambda)$. Finally, using \eqref{equationkappa}:
\begin{align}
\boxed{
\bar{\lambda}^\prime= \frac{32\bar{\lambda}^2}{3\pi} \int_{\mathbb{R}^4} d\textbf{x} \int dy\frac{1+\tau(y)r(x)}{f^3(x,y)f(x,-y)}+\frac{8\bar{\lambda}^2}{\pi} \left(\mathcal{\bar{L}}_{k,2}^{(1)}\vert_{Z_\infty=0}+\mathcal{\bar{L}}_{k,2}^{(2)}\vert_{Z_\infty=0} \right)}\,.
\end{align}
We have then computed the last piece of the flow equation for the anomalous dimension, equation \eqref{eqeta}. The set of resulting equations closes the hierarchy of flow equations. 

\section{Discussion}\label{sec6}

It is instructive before to close this study to compare with equilibrium flow equations (see Appendix \ref{App1}) as a benchmark. As a first cheeking, it is easy to show that we recover the perturbative $\beta$-function as $\beta=0$. Indeed, 

\begin{equation}
\eta = 4 \pi^2 a \lambda+4 \pi^2 (1-a) \lambda + \mathcal{O}(\lambda^2)= 4 \pi^2 \lambda+ \mathcal{O}(\lambda^2)\,,
\end{equation}

\begin{equation}
L_{31}+\frac{1}{2}L_{32}-L_{33} = \frac{\pi^3}{2} + \mathcal{O}(m^2)\,.
\end{equation}
Then, 
\begin{equation}
\beta_\lambda = - 4 \pi^2 \lambda^2+\mathcal{O}(\lambda^3)\,.
\end{equation}
which is nothing but the equilibrium perturbative beta-function, ensuring asymptotic freedom of the theory. The fact that we recover the equilibrium $\beta$-function is not a surprise in itself, because of the equilibrium dynamics assumption, and the fact that the regulator preserve time translation invariance and time reversal symmetry. What is different is however is the interpretation of the flow equations. In particular, regions where the equilibrium potential is unbounded from below, for instance means that equilibrium state is non normalizable, and then the system remains out of equilibrium for a long time. Note that $\eta$ differs from the formula recalled in the appendix, because we used another (and more precise) approximation to compute the derivative of the vertex -- see \cite{Lahoche_2020b} for more details.\\ 

In our previous work (see \cite{Lahoche_2020b} for a summary), we have always considered the standard Litim's regulator i.e. for $\alpha=1$. We will take this opportunity to give an additional argument by evaluating the robustness of our major treatment, namely that there is no fixed point outside the Gaussian fixed point in the melonic sector. When $\beta=0$, for the value $\alpha=1$, we find a fixed point, for the value:

\begin{equation}
\bar{m}^2 \approx 0.67\,,\quad \bar{\lambda} \approx 0.37\,.
\end{equation}

\begin{figure}
\begin{center}
\includegraphics[scale=0.47]{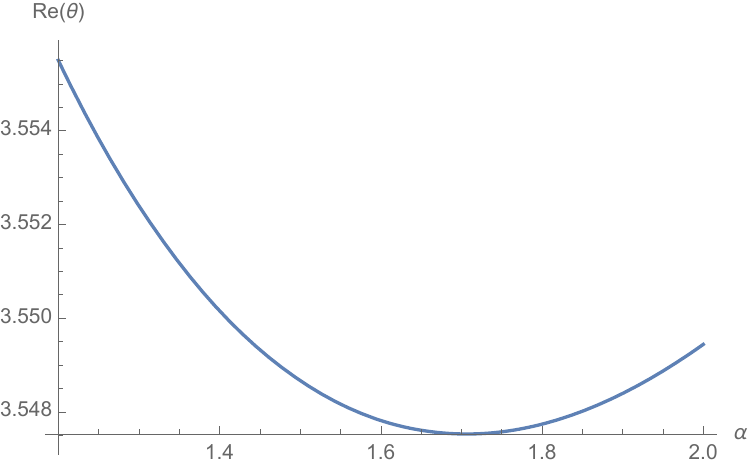}\qquad \includegraphics[scale=0.47]{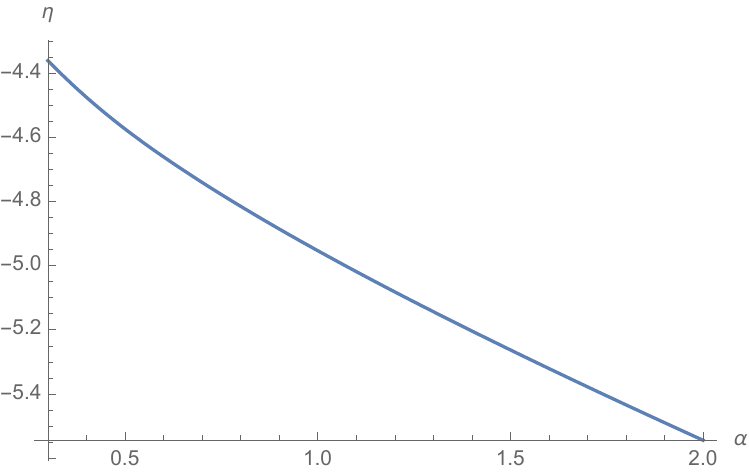}
\end{center}
\caption{Behavior of the relevant critical exponent and the anomalous dimension with respect to $\alpha$ ($\beta=0$).}\label{figsummarize}
\end{figure}

One might think so, but the corresponding value for the anomalous dimension $\eta \approx -5$ lies below the critical value $\eta =-2$, a good lower value below which the interpolation between UV and IR collapses (If $Z\sim k^{-2\delta}$, the product $k^2 Z$ can goes to zero when $k\to \infty$ if $\delta >1$). The fixed point is therefore not physically relevant, leaving only the Gaussian fixed point (in the melonic sector) smooth. This result was also approached in the infinite rank limit by exact methods in \cite{Lahoche_2021}. Let us now evaluate the dependence on the regulator on this result. Figure \ref{figsummarize} summarizes the behavior of the (real part of) critical exponent for the IR relevant direction of the fixed point, and the corresponding anomalous dimension, with respect to $\alpha$. The dependency on the regulator is minimal for $\alpha \approx 1.7$, but the anomalous dimension remains below the critical value $\eta >-2$. No relevant fixed point is expected for the equilibrium theory.

\section{Concluding remarks and outlooks}\label{sec7}

In this paper, we introduced the basics of a stochastic formalism for GFTs in equilibrium dynamics. In that regime, we were able to construct a nonperturbative RG formalism that takes into account time-reversal symmetry and causality of solutions of the stochastic equation \eqref{langevin}. We focused on a model which describes an Abelian stochastic complex field with rank $5$ and group structure $\U(1)$, whose equilibrium state is a just-renormalizable GFT for a pure gravity model. For this model, restricting ourselves to the melonic non-branching sector of the theory in the symmetric phase, we were able to construct an exact RG solution of the flow equation, closing the infinite hierarchy of the equation around just-renormalizable interactions. This strategy, mixing melonic equations and WT identities allows expressing vertex functions $\Gamma_k^{(2n)}$ for $n\geq 3$ in terms of $\Gamma_k^{(4)}$ and $\Gamma_k^{(2)}$, keeping by this way the full momenta dependence of the vertex, and to compute the derivative of the vertex for an external moment that plays an important role in the derivation of the anomalous dimension. Hence, the resulting equations describe the full non-branching sector and can be easily investigated numerically. Note that our construction focus on the deep UV regime and UV completion issue, and due to the compactness of the structure group $\U(1) \sim S_1$, symmetry restoration is expected in the deep IR due to the survival of zero modes. A way to solve this limitation should be to construct a kind of thermodynamic limit, sending the radius of the compact space $S_1$ to infinity, or to consider non compact group, see for instance \cite{lahoche2016renormalization,Benedetti_2015,benedetti2015critical,pithis2020phase,pithis2021no} and references therein. Although we focused on a toy model, disregarding some physical inputs in TGFTs, especially following the group structure which is not Abelian for realistic quantum gravity models, and in regard to some gauge symmetry like Gauss or Plebanski constraints, we expect that the general framework detailed in this paper could be suitable to investigate stochastic aspects of the best candidates for quantum gravity.  Finally, even if the ‘‘time" has been interpreted as a relational time, as the configurations of some matter fields, other matter fields could be added to the group fields, such that equilibrium states describe quantum gravity interacting with matter rather than a pure gravity regime. Another way of investigation concerns another current discrete approach to quantum gravity. For instance RTM. One can imagine a RTM described a dynamical tensorial variable $T_{i_1,\cdots, i_d}(t)$, by the same kind of equation like \eqref{langevin}. Such an equation will describe a stochastic tensor, and one may imagine many ways to approach its dynamics. \\

In this article, we have mainly focused on equilibrium dynamics. It is obvious that this approach is very limited, especially because the flow equations are similar to those for the equilibrium theory. The formalism, however, offers a new optimization lever through the frequency regulator \cite{duclut2017frequency}. However, the main interest of this work was essentially to discuss the methods and the general formalism. We will address in a sequel the more subtle case of non-equilibrium dynamics, whose physics remains the real added value of this approach.\\

\textbf{Acknowledgements:} The authors thank the anonymous referee for his comments which contributed to improve the presentation of this paper. 
\pagebreak

\appendix
\begin{center}
\begin{LARGE}
\textbf{Additional material}
\end{LARGE}
\end{center}
\section{Equilibrium state's RG}\label{App}

In this section, we review shortly the main results about RG for the equilibrium state \eqref{equilibrium}, which describes a pure gravity TGFT, for a complex group field with rank $d$. We focus on $d=5$, for an Abelian model with group structure $\U(1)$ and quartic melonic interactions \eqref{model}. This model has been largely investigated in the literature, see \cite{samary2014just,samary2013beta,Lahoche:2018oeo, Lahoche:2018ggd, Lahoche_2020b} and references therein. In this section we sketch the main lines of the reference \cite{Lahoche_2020b} that the reader may consult for more details. In the section \ref{App1} we provide a derivation of the relation between counter-terms for the wave function and coupling and a formal expression for them. Note that for this derivation we make use of the standard Schwinger-Dyson equation, considered in full detail in \cite{samary2015correlation,samary2014closed}, in contrast with the discussion given in \cite{Lahoche_2020b} based on the existence of a finite radius of convergence for the renormalized series (see also \cite{lahoche2015renormalization}). Note that in this section we define equilibrium state as $\rho \sim e^{-\mathcal{H}}$, without the factor 2 that can be canceled by a global redefinition of fields.

\subsection{Melonic Schwinger-Dyson equation and counter-terms}\label{App1}

\begin{figure}[h!]
\begin{center}
\includegraphics[scale=0.8]{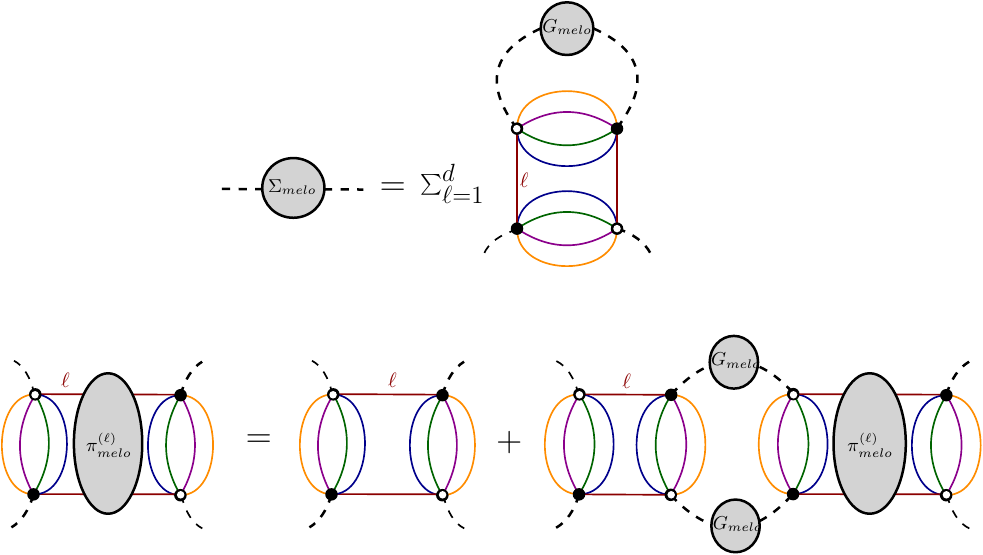}
\end{center}
\caption{Closed relations between the $2$ and $4$ point functions obtained from the Schwinger-Dyson equations in the melonic sector, as the "melo" index recalls.}\label{SDE}
\end{figure}

Schwinger-Dyson equations in quantum field theory are relations between observable, generally taking the form of self-consistency equations for 1PI functions. For the TFGT that we considered in this paper, two of them are especially relevant in the melonic sector, for $2$ and $4$-point functions, and have the structure pictured in Figure \ref{SDE} (see \cite{samary2014closed} for more details). Note that we focus on $2$ and $4$ point functions because the quartic model is just-renormalizable, and both these relations are sufficient to constrain the counter-terms.

Note that, in the second figure, we assumed that the $1$PI $4$-point function has the connected quartic melonic for boundary and $\pi_{\text{melo}}^{(\ell)}$ is defined by:
\begin{equation}
\Gamma_{\bar{M} M \bar{M} M}^{(4),(\ell)}= 2\pi^{(\ell)}_{\text{melo}}(p_{1\ell}^2,p_{3\ell}^2)\left(\mathcal{W}^{(\ell)}_{\bm p_1,\bm p_2,\bm p_3,\bm p_4} +\bm p_2 \leftrightarrow \bm p_4\right)\,.\label{rencondGamma4eq}
\end{equation}
Moreover $\Sigma_{\text{melo}}$ designates the self energy, related to the full $2$-point function $G_{\text{melo}}$ and the bare propagator $C$ by the Dyson equation:
\begin{equation}
G_{\text{melo}}=\frac{1}{C^{-1}-\Sigma_{\text{melo}}}\,,
\end{equation}
where following the definitions of section \ref{renth},
\begin{equation}
C^{-1}(\bm p)=Z_{-\infty} \bm p^2+m^2\,.
\end{equation}
Translated in equations, the first closed relations pictured in Figure \ref{SDE} reads:
\begin{equation}
\Sigma_{\text{melo}}(\bm p)= -2Z_\lambda \lambda_r \sum_{\ell=1}^d \sum_{\bm q\in \mathbb{Z}^d} \delta_{p_\ell q_\ell} \, \frac{1}{Z_{\infty}\bm q^2+m^2-\Sigma_{\text{melo}}(\bm q)}\,.
\end{equation}
This relation means that $\Sigma_{\text{melo}}(\bm p)=:\sum_\ell \sigma(p_\ell^2)$, with:
\begin{equation}
\sigma(p^2)=-2Z_\lambda \lambda_r\sum_{\bm q\in \mathbb{Z}^d} \delta_{p q_1} \, \frac{1}{Z_{\infty}\bm q^2+m^2-\Sigma_{\text{melo}}(\bm q)}\,.\label{closedsigma}
\end{equation}
Accordingly with the renormalization condition \eqref{2pointeqren}, $Z_{infty}$ and $Z_m$ are such that (see \cite{Lahoche:2018oeo} and references therein):
\begin{equation}
Z_{\infty}-\sigma^{\prime}(0)=1\,,\qquad Z_m m_r^2-d\times \sigma(0)=m_r^2\,.
\end{equation}
Hence:
\begin{equation}
\frac{1}{Z_{\infty}\bm q^2+m^2-\Sigma_{\text{melo}}(\bm q)} = \frac{1}{\bm q^2+m^2_r+\sum_{\ell=1}^d \sigma_r(q_\ell^2)}\,,
\end{equation}
where $\sigma_r(q_\ell^2)=\mathcal{O}(q_\ell^4)$, with zero and first derivative equal to zero. Differentiating relation \eqref{closedsigma} with respect to $p^2$, and setting $p=0$, we get:
\begin{equation}
-\sigma^\prime(0)\equiv 1-Z_\infty=-2Z_\lambda \lambda_r A_\infty \,,\label{relationone}
\end{equation}
where $A_\infty$ has been defined in \eqref{Ainfty}. Now, let us consider the second relation, pictured in Figure \ref{SDE}. In the equation this relation reads, setting all the external momenta to zero:
\begin{equation}
4\pi^{(\ell)}_{\text{melo}}(0,0)=4 Z_\lambda \lambda_r-8Z_\lambda \lambda_r \pi^{(\ell)}_{\text{melo}}(0,0) A_\infty\,.
\end{equation}
If we use the standard renormalization condition (see \cite{ZinnJustinBook2,peskin2018introduction}):
\begin{equation}
\pi^{(\ell)}_{\text{melo}}(0,0)=:\lambda_r\,,
\end{equation}
the previous relation simplifies as:
\begin{equation}
\boxed{
Z_\lambda^{-1}=1-2\lambda_r A_\infty\,.}\label{relationtwo}
\end{equation}
Moreover, from \eqref{relationone},
\begin{equation}
Z_\lambda^{-1}-\frac{Z_\infty}{Z_\lambda}=-2Z_\lambda \lambda_r A_\infty\,,
\end{equation}
and using \eqref{relationtwo}, we obtain finally:
\begin{equation}
\boxed{Z_\infty=Z_\lambda\,.}
\end{equation}

\subsection{Nonperturbative RG in the non-branching sector}\label{App2}
In this section, we summarize some aspects of the nonperturbative RG in the symmetric phase for the melonic non-branching sector using EVE. All details can be found in \cite{Lahoche_2020b}, as recalled at the beginning of this section. The derivation of the equation follows essentially the same strategy as explained in section \ref{floweqsection}. Hence, in the deep UV regime $1\ll k\ll \Lambda$, and using the Litim regulator:
\begin{equation}
r_k(\bm p)= Z(k) (k^2-\bm p^2)\theta(k^2-\bm p^2)\,,
\end{equation}
the resulting $\beta$-functions read:
\begin{align}
\left\{
\begin{array}{ll}
\beta_m&=-(2+\eta)\bar{m}^{2}-10 \bar{\lambda}\,\frac{\pi^2}{(1+\bar{m}^{2})^2}\,\left(1+\frac{\eta}{6}\right)\,, \\
\beta_{\lambda}&=-2\eta \bar{\lambda}+4\bar{\lambda}^2 \,\frac{\pi^2}{(1+\bar{m}^{2})^3}\,\left(1+\frac{\eta}{6}\right)\Big[1-6\pi^2\bar{\lambda}\left(\frac{1}{(1+\bar{m}^{2})^2}+\left(1+\frac{1}{1+\bar{m}^{2}}\right)\right)\Big]\,, \label{syst3}
\end{array}
\right.
\end{align}
where the anomalous dimension $\eta$ is given by:
\begin{equation}
\eta=4\bar{\lambda}\pi^2\frac{(1+\bar{m}^{2})^2-\bar{\lambda}\pi^2(2+\bar{m}^{2})}{(1+\bar{m}^{2})^2\Omega(\bar{\lambda},\bar{m}^{2})+2\frac{(2+\bar{m}^{2})}{3}\bar{\lambda}^2\pi^4}\,\label{eta1}
\end{equation}
and
\begin{equation}
\Omega(\bar m^2,\bar \lambda):=(\bar m^2+1)^2-\pi^2\bar \lambda\,.
\end{equation}
To obtain these equations, we closed the hierarchy using the same method as discussed in the section \ref{structureeq}, by expressing $\Gamma^{(6)}_k$ in the expression of $\beta_\lambda$ in terms of $\bar{\lambda}$ and $\bar{m}^2$. Moreover, we used the Ward identities to compute the derivative of the $4$-point vertex $\Gamma^{(4)}_k$ with respect to external momenta, which plays a role in the computation of the anomalous dimension $\eta$. This additional contribution, of order $\bar{\lambda}^2$ is not a small correction for a pure local potential approximation disregarding such a contribution. Indeed, taking into account this term push forward the singularity of the denominator of $\eta$ down the singularity $\bar{m}^2=-1$, coming from our restriction to the symmetric phase, and thus maximally extends the investigated portion of the full phase space. In the computation of loops involved both in the expression for $\Gamma^{(6)}_k$ and the derivative of $\Gamma^{(4)}_k$, we used the derivative expansion for $2$-point function, the same approximation used for the computation of flow equations. In \cite{Lahoche_2020b} and reference therein, it has been pointed out that such an approximation makes sense for the computation of superficially convergent integrals, and remains in agreement with Ward identities. Hence, the RG flow described by equations \eqref{syst3} satisfies the Ward identities. Moreover, the computer program Mathematica is not able to find any physically relevant fixed point for that system, and the Gaussian fixed point is the only UV-relevant fixed point, at least in this regime.
\medskip

The quartic model (considered as an initial condition for the RG flow) is endowed with an additional amazing specificity. Indeed, the Ward identities impose a constraint between $4$ and $2$-point functions that can be translated locally along the RG flow as a non-trivial relation between $\beta$-functions for relevant couplings:
\begin{equation}
\boxed{
\beta_\lambda+\eta\bar{\lambda}\, \frac{\Omega(\bar{m}^2,\bar{\lambda})}{(1+\bar{m}^2)^2}-\frac{2\pi ^2\bar{\lambda}^2}{(1+\bar{m}^2)^3}\beta_m=0\,.} \label{constraint}
\end{equation}
The flow equation for $\bar{m}^2$ given by \eqref{syst3} is exact in the melonic sector, as we restrict ourselves to the connected interactions. Hence, equation \eqref{constraint} defines the function $\beta_\lambda$. On the other hand, the flow equation for $\lambda$ involves $\Gamma^{(6)}_k$. Therefore, equalizing the two expressions for $\beta_\lambda$ provides a non-trivial expression for $\Gamma^{(6)}_k$ (with zero external momenta). Note that this contribution may involve in principle non-connected contributions, but we discard them from our analysis. Hence, we can use the resulting expression for $\Gamma^{(6)}_k$ in the Ward identity expressing the derivative of $4$-point functions for external momenta, as discussed in section \eqref{derivlambda}. Finally, this expression allows computing the anomalous dimension:
\begin{equation}
\eta=\frac{4 \pi ^2\bar{\lambda} \left(\frac{\pi ^2 \bar{\lambda} }{5 (1+\bar{m}^2)^3}+1\right)}{(1+\bar{m}^2)^2-\Omega_1(\bar{m}^2,\bar{\lambda})}\,, \label{stateeta}
\end{equation}
where:
\begin{equation}
\Omega_1(\bar{m}^2,\bar{\lambda}) :=\frac{6 \pi ^2\bar{\lambda} }{5}-\frac{4 \pi ^4 \bar{\lambda}^2}{(1+\bar{m}^2)^3}-\frac{12 \pi ^2 \bar{\lambda} \bar{m}^2}{5 (1+\bar{m}^2)}-\frac{4 \pi^2 \bar{\lambda} }{5 (1+\bar{m}^2)}\,.
\end{equation}
This strategy is expected to provide a non-trivial improvement for the previous system \eqref{syst3}, because no additional approximations are required to compute loops involved in the structure equations, as we had to do in the section \ref{structureeq}. But the conclusions are essentially the same: The theory is asymptotically free in the UV, and no physically relevant additional fixed point in found\footnote{Indeed, the constraint \eqref{constraint} imposes $\eta=0$ for any non-trivial fixed point. But the solution \eqref{stateeta} shows that $\bar{\lambda}=0$ is the only non-negative solution for $\bar{m}^2>-1$.}. Near the Gaussian fixed point,
\begin{equation}
\boxed{
\eta \approx 4\pi^2 \bar{\lambda}\,,\qquad \beta_\lambda \approx -\eta \bar{\lambda}\,.}
\end{equation}

\section{Proof of relation \eqref{relationregulator}}\label{AppC}

From time reversal symmetry, $\Delta S_k$ transforms as:
\begin{align}
\nonumber\Delta S_k[\bm q,\bm \chi]&\to \sum_{\bm p \in \mathbb{Z}^d}\int dt dt^\prime \,\bigg( \bar{\chi}_{\bm p}(t) R_{k}^{(2)}(\bm p,t^\prime-t){\chi}_{\bm p}(t^\prime)+i \bar{\chi}_{\bm p}(t) R_k^{(1)}(\bm p,t^\prime-t) {T}_{\bm p}(t^\prime)\\\nonumber
&+i \bar{T}_{\bm p}(t^\prime) R_k^{(1)}(\bm p,t^\prime-t) {\chi}_{\bm p}(t)-\frac{2}{\Omega}\dot{\bar{T}}_{\bm p}(t) R_k^{(1)}(\bm p,t^\prime-t) {T}_{\bm p}(t^\prime) \\\nonumber
&-\frac{2}{\Omega}{\bar{T}}_{\bm p}(t^\prime) R_k^{(1)}(\bm p,t^\prime-t) \dot{T}_{\bm p}(t) +\frac{2i}{\Omega}R_{k}^{(2)}(\bm p,t^\prime-t)\big(\dot{\bar{T}}_{\bm p}(t){\chi}_{\bm p}(t^\prime)+\bar{\chi}_{\bm p}(t)\dot{T}_{\bm p}(t^\prime)\big) \\
&-\frac{4}{\Omega^2}\dot{\bar{T}}_{\bm p}(t)R_{k}^{(2)}(\bm p,t^\prime-t)\dot{T}_{\bm p}(t^\prime) \bigg)\,.
\end{align}
Then, integrating by part, we find that the invariance condition of $\Delta S_k$ by time reversal is written, up to a total derivative:
\begin{align}
\nonumber0&\equiv \sum_{\bm p \in \mathbb{Z}^d}\int dt dt^\prime \,\bigg(i \bar{\chi}_{\bm p}(t) \bigg(R_k^{(1)}(\bm p,t^\prime-t)-R_k^{(1)}(\bm p,t-t^\prime)+\frac{2}{\Omega}\dot{R}_{k}^{(2)}(\bm p,t-t^\prime)\bigg) {T}_{\bm p}(t^\prime)\\\nonumber
&-i \bar{T}_{\bm p}(t^\prime) \bigg(R_k^{(1)}(\bm p,t^\prime-t)-R_k^{(1)}(\bm p,t-t^\prime)-\frac{2}{\Omega}\dot{R}_{k}^{(2)}(\bm p,t^\prime-t)\bigg) {\chi}_{\bm p}(t) \\\nonumber
&-\frac{2}{\Omega}\dot{\bar{T}}_{\bm p}(t) \left(R_k^{(1)}(\bm p,t^\prime-t) -\frac{1}{\Omega}\dot{R}_{k}^{(2)}(\bm p,t^\prime-t)\right) {T}_{\bm p}(t^\prime) \\
&-\frac{2}{\Omega}{\bar{T}}_{\bm p}(t^\prime) \left(R_k^{(1)}(\bm p,t^\prime-t)+\frac{1}{\Omega}\dot{R}_{k}^{(2)}(\bm p,t^\prime-t) \right) \dot{T}_{\bm p}(t)\bigg)\,.
\end{align}
Finally, exploiting the fact that $R^{(2)}_k$ is a symmetric function, the two last terms can be rewritten as follows:
\begin{align}
\nonumber0&\equiv \sum_{\bm p \in \mathbb{Z}^d}\int dt dt^\prime \,\bigg(i \bar{\chi}_{\bm p}(t) \bigg(R_k^{(1)}(\bm p,t^\prime-t)-R_k^{(1)}(\bm p,t-t^\prime)+\frac{2}{\Omega}\dot{R}_{k}^{(2)}(\bm p,t-t^\prime)\bigg) {T}_{\bm p}(t^\prime)\\\nonumber
&-i \bar{T}_{\bm p}(t^\prime) \bigg(R_k^{(1)}(\bm p,t^\prime-t)-R_k^{(1)}(\bm p,t-t^\prime)-\frac{2}{\Omega}\dot{R}_{k}^{(2)}(\bm p,t^\prime-t)\bigg) {\chi}_{\bm p}(t) \\
&-\frac{2}{\Omega}\dot{\bar{T}}_{\bm p}(t) \left(R_k^{(1)}(\bm p,t^\prime-t)-R_k^{(1)}(\bm p,t-t^\prime)-\frac{2}{\Omega}\dot{R}_{k}^{(2)}(\bm p,t^\prime-t)\right) {T}_{\bm p}(t^\prime)\,.
\end{align}
These relations show that a sufficient condition to avoid breaking the time reflection symmetry along the RG flow is to impose \eqref{relationregulator}.

\section{Melonics WT identities for $2$ and $4$ point vertices}\label{AppB}
We give in this appendix a technical complement on Ward identities, focusing on the relation between 4-point and 2-point functions. Relations between $1PI$ functions can be obtained by taking successive derivatives for the sources, but vanishing them at the end of the computation. Alternatively, one can derive with respect to the classical fields, and this is what we do in the following sections.

\paragraph{Another proof of heteroclicity.} Let us start by considering the second functional derivative on both sides of the equation \eqref{Ward1} with respect to the classical fields $M_{\bm p}({\omega})$ and $\bar{M}_{\bm p}({\omega})$. We apply the operator $\partial^2/\partial M_{\bm q}({\omega}_1)\partial \bar{M}_{\bm{\bar{q}}}({\bar{\omega}}_1)$ on equation \eqref{Ward1}, and because definition \eqref{defeffectiveaction}, we have:
\begin{equation}
\frac{\partial M_{\bm p}({\omega})}{\partial J_{\bm p^\prime}({\omega}^\prime) }= \frac{\partial^2 W_k}{\partial \bar{J}_{\bm p}({\omega}) \partial J_{\bm p^\prime}({\omega}^\prime)}\,,
\end{equation}
and:
\begin{equation}
\frac{\partial \Gamma_k}{\partial M_{\bm p}({\omega})}= \bar{J}_{\bm p}({\omega})-i R_k^{(1)}(\bm p,\omega) \bar{\sigma}_{\bm p}(\omega)\,,
\end{equation}
\begin{equation}
\frac{\partial \Gamma_k}{\partial \sigma_{\bm p}({\omega})}=\bar{\jmath}_{\bm p}({\omega})- \big[ \bar{\sigma}_{\bm p}(\omega) R_k^{(2)}(\bm p,\omega)+iR_k^{(1)}(\bm p,-\omega) \bar{M}_{\bm p}(\omega) \big]\,.
\end{equation}
We introduce the notations:
\begin{equation}
\delta p^2:=p^2-(p^\prime)^2\,,\quad R_k^{(I)}(\bm p,\omega)-R_k^{(I)}(\bm p^\prime,\omega):=\delta R_k^{(I)}(\bm p,\omega)\,,
\end{equation}
and we get:
\begin{align}
\nonumber0&=\int d{\omega}\sum_{\bm p, \bm p^\prime} \prod_{j\neq i} \delta_{p_jp_j^\prime} \bigg[\left(iZ_{\infty}\delta p^2+i\delta R_k^{(1)}(\bm p,\omega)\right)\frac{\partial^2G_{k,\bar{\sigma}M}^{(1;\bar{1})}(\bm p^\prime,{\omega};\bm p,{\omega})}{\partial M_{\bm q}({\omega}_1)\partial \bar{M}_{\bm{\bar{q}}}({\bar{\omega}}_1)} \\\nonumber
&+\left(iZ_{\infty}\delta p^2+i\delta R_k^{(1)}(\bm p,-\omega)\right)\frac{\partial^2G_{k,\bar{M}\sigma}^{(\bar{1};1)}( \bm p,{\omega};\bm p^\prime,{\omega})}{\partial M_{\bm q}({\omega}_1)\partial \bar{M}_{\bm{\bar{q}}}({\bar{\omega}}_1)}\\\nonumber
&+ \delta R_k^{(2)}(\bm p,\omega) \times \frac{\partial^2G_{k,\bar{\sigma}\sigma}^{(0;{1}+\bar{1})}(\bm p^\prime,{\omega};\bm p,{\omega})}{\partial M_{\bm q}({\omega}_1)\partial \bar{M}_{\bm{\bar{q}}}({\bar{\omega}}_1)} \\
&-\frac{\partial \bar{J}_{\bm p}({\omega})}{\partial \bar{M}_{\bm{\bar{q}}}(\bar{\omega}_1)} \delta_{\bm p^\prime \bm q} \delta(\omega-\omega_1)+ \frac{\partial J_{\bm p^\prime}({\omega})}{\partial M_{\bm q}(\omega_1)}\delta_{\bm p\bm{\bar{q}}}\delta(\omega-\bar{\omega}_1)
\bigg] \delta_{p_ip}\delta_{p_i^\prime p^\prime}\,.\label{Ward2}
\end{align}
Derivatives to sources can be easily computed, leading to:
\begin{equation}
\frac{\partial \bar{J}_{\bm p}({\omega})}{\partial \bar{M}_{\bm{\bar{q}}}(\bar{\omega}_1)}=\frac{\partial^2 \Gamma_k}{\partial \bar{M}_{\bm{\bar{q}}}(\bar{\omega}_1)\partial M_{\bm p}({\omega})}\,,\qquad \frac{\partial J_{\bm p^\prime}({\omega})}{\partial M_{\bm q}(\omega_1)}=\frac{\partial^2 \Gamma_k}{\partial M_{\bm q}(\omega_1)\partial \bar{M}_{\bm p^\prime}({\omega})}\,.
\end{equation}
In the same way,
\begin{equation}
\frac{\partial \bar{\jmath}_{\bm p}({\omega})}{\partial \bar{M}_{\bm{\bar{q}}}(\bar{\omega}_1)}=\frac{\partial^2 \Gamma_k}{\partial \bar{M}_{\bm{\bar{q}}}(\bar{\omega}_1)\partial \sigma_{\bm p}({\omega})}+iR_k^{(1)}(\bm{\bar{q}},-{\omega})\delta_{\bm p\bm{\bar{q}}} \delta(\omega-\bar{\omega}_1)\,,
\end{equation}
and:
\begin{equation}
\frac{\partial \jmath_{\bm p^\prime}({\omega})}{\partial M_{\bm q}(\omega_1)}=\frac{\partial^2 \Gamma_k}{\partial M_{\bm q}(\omega_1)\partial \bar{\sigma}_{\bm p^\prime}({\omega})}+iR_k^{(1)}(\bm{{q}},{\omega})\delta_{\bm p^\prime\bm{{q}}} \delta(\omega-{\omega}_1)\,.
\end{equation}
The derivatives of the $2$-point functions can be rewritten as follows. Note that because $\partial \bar{J}_{\bm p}({\omega})/\partial \bar{M}_{\bm{\bar{q}}}(\bar{\omega})=0$, we must have (using the short notation $p\equiv (\bm p,\omega)$),
\begin{equation}
\sum_{p_1} \frac{\partial \jmath_{p_1}}{\partial M_{q}} \frac{\partial \sigma_p}{\partial \jmath_{p_1}}=\sum_{p_1} \frac{\partial \jmath_{p_1}}{\partial M_{q}} \langle \bar{\chi}_{p_1} {\chi}_p \rangle = 0\,,\label{equation0}
\end{equation}
because $G_{k,\bar{\chi}\chi}=0$ (equation \eqref{conditionG}). We made use of \eqref{equation0} to obtain \eqref{Ward2}. The functional derivatives can be easily computed following the method described in the previous section to obtain the flow equations. Because we focus on the symmetric phase and assume that effective vertices must have only one component along the response field and that $G_{k,\bar{\sigma}\sigma}=0$, we have:
\begin{align}
\nonumber\frac{\partial^2G_{k,\bar{\sigma}M}^{(1;\bar{1})}(\bm p^\prime,{\omega};\bm p,{\omega})}{\partial M_{\bm q}({\omega}_1)\partial \bar{M}_{\bm{\bar{q}}}({\bar{\omega}}_1)}=&- \sum_{p_1,p_1^\prime} G_{k,\bar{\sigma}M}(p^\prime,p_1^\prime) \Gamma_{k,M\bar{M}M \bar{\sigma}}^{(4)}(q,\bar{q},p_1^\prime,p_1) G_{k,\bar{\sigma} M}(p_1,p)\\
&-\sum_{p_1,p_1^\prime} G_{k,\bar{\sigma}M}(p^\prime,p_1^\prime) \Gamma_{k,M\bar{M}M \bar{M}}^{(4)}(q,\bar{q},p_1^\prime,p_1) G_{k,\bar{M} M}(p_1,p)\,.
\end{align}
In the same way:
\begin{align}
\nonumber\frac{\partial^2G_{k,\bar{M}\sigma}^{(\bar{1};1)}( \bm p,{\omega};\bm p^\prime,{\omega})}{\partial M_{\bm q}({\omega}_1)\partial \bar{M}_{\bm{\bar{q}}}({\bar{\omega}}_1)}=&- \sum_{p_1,p_1^\prime} G_{k,\bar{M}\sigma}(p,p_1) \Gamma_{k,M\bar{M}\sigma \bar{M}}^{(4)}(q,\bar{q},p_1^\prime,p_1) G_{k,\bar{M} \sigma}(p_1^\prime,p^\prime)\\
&-\sum_{p_1,p_1^\prime} G_{k,\bar{M}M}(p,p_1) \Gamma_{k,M\bar{M}M \bar{M}}^{(4)}(q,\bar{q},p_1^\prime,p_1) G_{k,\bar{M} \sigma}(p_1^\prime,p^\prime)\,,
\end{align}
and:
\begin{equation}
\frac{\partial^2G_{k,\bar{\sigma}\sigma}^{(0;{1}+\bar{1})}(\bm p^\prime,{\omega};\bm p,{\omega})}{\partial M_{\bm q}({\omega}_1)\partial \bar{M}_{\bm{\bar{q}}}({\bar{\omega}}_1)}= - \sum_{p_1,p_1^\prime} G_{k,\bar{\sigma}{M}}(p,p_1) \Gamma_{k,M\bar{M} M \bar{M}}^{(4)}(q,\bar{q},p_1^\prime,p_1) G_{k,\bar{M} {\sigma}}(p_1^\prime,p^\prime)\,.
\end{equation}
Following the definition \eqref{rencondGamma4}, we introduce: $\Gamma_{k,M\bar{M} M \bar{M}}^{(4)}=:\sum_\ell \Gamma_{k,M\bar{M} M \bar{M}}^{(4),(\ell)}$, and the decomposition:
\begin{equation}
\Gamma_{k,M\bar{M} M \bar{M}}^{(4),(\ell)}(p_1,p_2,p_3,p_4)=:\frac{i}{\pi} \varpi^{(2)}_k(p_{1\ell}^2,p_{3\ell}^2)\left(\mathcal{W}^{(\ell)}_{\bm p_1,\bm p_2,\bm p_3,\bm p_4} +\bm p_2 \leftrightarrow \bm p_4\right)\delta(\hat{\omega}_1-\hat{\omega}_2+\hat{\omega}_3-\hat{\omega}_4)\,,\label{rencondGamma42}
\end{equation}
such that Ward identity \eqref{Ward2} reads as follows:
\begin{align}
\nonumber &\sum_{\ell=1}^d\left(\vcenter{\hbox{\includegraphics[scale=0.8]{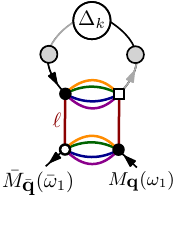}}}+\,\vcenter{\hbox{\includegraphics[scale=0.8]{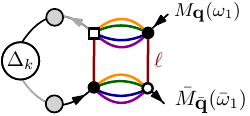}}}+\vcenter{\hbox{\includegraphics[scale=0.8]{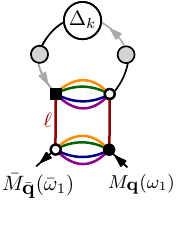}}}+\,\vcenter{\hbox{\includegraphics[scale=0.8]{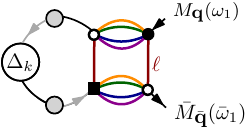}}}\right) \\\nonumber
&\sum_{\ell=1}^d\left(\vcenter{\hbox{\includegraphics[scale=0.8]{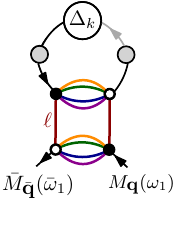}}}+\,\vcenter{\hbox{\includegraphics[scale=0.8]{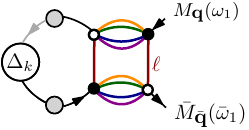}}}+\vcenter{\hbox{\includegraphics[scale=0.8]{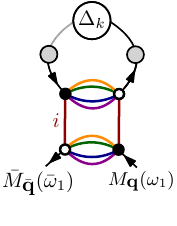}}}+\,\vcenter{\hbox{\includegraphics[scale=0.8]{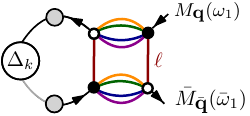}}}\right) \\\nonumber
&+ \sum_{\ell=1}^d\left(\vcenter{\hbox{\includegraphics[scale=0.8]{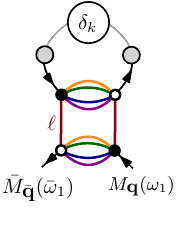}}}+\,\vcenter{\hbox{\includegraphics[scale=0.8]{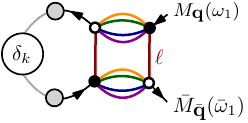}}} \right)-\sum_{\bm p, \bm p^\prime} \prod_{j\neq i} \delta_{p_jp_j^\prime}\bigg[\frac{\partial \bar{J}_{\bm p}({\omega_1})}{\partial \bar{M}_{\bm{\bar{q}}}(\bar{\omega}_1)} \delta_{\bm p^\prime \bm q}\\
&- \frac{\partial J_{\bm p^\prime}(\bar{\omega}_1)}{\partial M_{\bm q}(\omega_1)}\delta_{\bm p\bm{\bar{q}}}
\bigg] \delta_{p_ip}\delta_{p_i^\prime p^\prime}=0\label{Warddiag1}
\end{align}
Equation \eqref{Warddiag1} involves two kinds of diagrams. The first ones, corresponding to the first, third and fifth contributions to the left-hand side of \eqref{Warddiag1} create $(d-2)$ or $(d-1)$ faces, respectively for $\ell \neq i$ and $\ell=i$.
\begin{itemize}
\item For $\ell \neq i$ the contribution vanishes, because Kronecker deltas in $\mathcal{W}^{(\ell)}_{\bm p_1,\bar{\bm{p}}_2,\bm p_3,\bar{\bm{p}}_4}$ impose $p=p^\prime$.

\item For $\ell = i$, the contribution does not vanish, and is melonic following definition \ref{definitionMelon}: $F=d-1\,(=4)$, $L=V=1$ and $\rho=0$.
\end{itemize}
The second kind of diagram corresponds to the second, fourth, and sixth contributions to the right-hand side of \eqref{Warddiag1}. They create no more than $0$ or $1$ face, respectively, for $\ell \neq i$ and $\ell=i$.
\begin{itemize}
\item For $\ell \neq i$ the contribution vanishes because Kronecker deltas impose $p=p^\prime$.

\item For $\ell = i$, the contribution does not vanish, but it is not melonic ($\rho=3$).
\end{itemize}
We restrict ourselves to the melonic sector, which, as recalled in section \ref{themodel} is the most divergent one, and thus the most relevant for RG. From these observations, the leading order (melonics) contribution to identities \eqref{Warddiag1} reads as:
\begin{align}
\nonumber &\vcenter{\hbox{\includegraphics[scale=0.8]{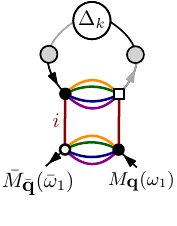}}}+\vcenter{\hbox{\includegraphics[scale=0.8]{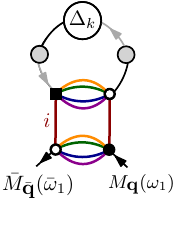}}}+\vcenter{\hbox{\includegraphics[scale=0.8]{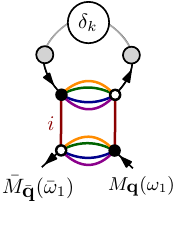}}}+\vcenter{\hbox{\includegraphics[scale=0.8]{OneLoopMassWI4PRIME.pdf}}}+\vcenter{\hbox{\includegraphics[scale=0.8]{OneLoopMassWI4SECOND.pdf}}} \\
&-\sum_{\bm p, \bm p^\prime} \prod_{j\neq i} \delta_{p_jp_j^\prime}\bigg[\frac{\partial \bar{J}_{\bm p}({\omega_1})}{\partial \bar{M}_{\bm{\bar{q}}}(\bar{\omega}_1)} \delta_{\bm p^\prime \bm q}- \frac{\partial J_{\bm p^\prime}(\bar{\omega}_1)}{\partial M_{\bm q}(\omega_1)}\delta_{\bm p\bm{\bar{q}}}
\bigg] \delta_{p_ip}\delta_{p_i^\prime p^\prime}= 0\quad\text{(Melonic order)}\,,\label{Warddiag2}
\end{align}
There are many options to interpret this equation. We know, from condition \eqref{conditionG} that the two last terms must vanish exactly. Hence, for the three first terms, we have essentially two kinds of integrals. The two first ones contributions of the involve loop integrals $\int d\omega\, G_{k,\bar{\sigma} M}^2(\omega)$ and $\int d\omega \,G_{k,\bar{M} \sigma}^2(\omega)$. If we assume causality, these integrals have to vanish for the same reason as we discussed in Remark \ref{remarkcausal} (equation \eqref{eqcausal}). Moreover, terms like $\int d\omega\, \delta R_k^{(1)}(\omega)G_{k,\bar{\sigma} M}^2(\omega)$ vanish for the same reason as the left-hand side of equation \eqref{conditiontrue} vanishes. With this argument, the last integral, which reads: $\int d\omega\, \delta R_k^{(2)}(\omega)G_{k,\bar{\sigma} M}(\omega)G_{k,\bar{M} \sigma}(\omega)$ does not vanish, and the Ward identity imposes 

\begin{equation}
\vcenter{\hbox{\includegraphics[scale=0.8]{OneLoopMassWI2BB.pdf}}}+\vcenter{\hbox{\includegraphics[scale=0.8]{OneLoopMassWI4PRIME.pdf}}}+\vcenter{\hbox{\includegraphics[scale=0.8]{OneLoopMassWI4SECOND.pdf}}}=0\,.
\end{equation}
In the deep IR, as $k\to 0$, regulators vanishes formally, and this condition then enforces $\varpi_0^{(2)}=0$, meaning that $\Gamma_{k=0}^{(2+\bar{2},0)}=0$. Hence, as the classical action (initial condition), the full effective action has no quartic interactions without response field and this can be show furthermore recursively for higher order interactions. Finally, for $k\neq 0$, using the truncation for $\Gamma_k^{(2)}$ for terms involving regulator, it is easy to check that the property also holds. This is expected because of the discussion of section \ref{sectioncausal}, where causality was assumed as well, but the fact that this condition comes from a constraint imposed by an internal symmetry is a non-trivial result. The origin of this phenomenon can be traced from the arguments discussed in \cite{Lahoche_2020b}, where authors pointed out a parallel between renormalization group equations and Ward identities. Indeed, if the flow equations dictate how the interactions change with the scale, the Ward identities dictate how the interactions deviate from ultralocality (i.e. from exact unit invariance). Thus, if in section \ref{sectioncausal} we were able to demonstrate the absence of response field independent interaction terms by an argument from the renormalization group, Ward's identities show that a local theory whose initial conditions correspond to the model \eqref{classicaction0} cannot deviate from locality by response field independent contributions.
\paragraph{Relation between $Z(k)$ and $\lambda(k)$.} In the same vein, but applying the operator $\partial^2/\partial M_{\bm q}({\omega}_1)\partial \bar{\sigma}_{\bm{\bar{q}}}({\bar{\omega}}_1)$ on the Ward identity \eqref{Ward1}, we obtain a relation between $\Gamma_{k,\sigma\bar{M} M \bar{M}}^{(4)}$ and $\Gamma_{k,\bar{\sigma} M}^{(2)}$. Using the same graphical representation as previously, we get (we introduce all the Kronecker and Dirac $\delta$ to be more clear):
\begin{align}
\nonumber &\Bigg(\vcenter{\hbox{\includegraphics[scale=0.8]{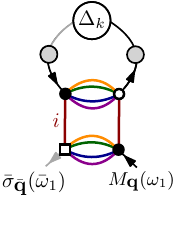}}}+\vcenter{\hbox{\includegraphics[scale=0.8]{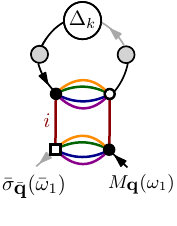}}}+\vcenter{\hbox{\includegraphics[scale=0.8]{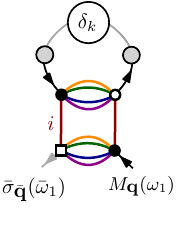}}}\Bigg)\delta_{p^\prime q_i} \delta_{p\bar{q}_i}\prod_{j\neq i}\delta_{q_j{\bar{q}}_j}\delta(\omega_1-\bar{\omega}_1)\\\nonumber
&-\sum_{\bm p, \bm p^\prime} \prod_{j\neq i} \delta_{p_jp_j^\prime} \Delta_k(\bm p,\omega_1) \delta_{\bm p\bm{\bar{q}}} \delta_{\bm q\bm p^\prime}\delta(\omega_1-\bar{\omega}_1) +\sum_{\bm p, \bm p^\prime} \prod_{j\neq i} \delta_{p_jp_j^\prime}\bigg[\gamma_{k,\bar{\sigma}M}^{(2)}(\bm p,\omega_1)- \gamma_{k,\bar{\sigma}M}^{(2)}(\bm p^\prime,\omega_1)
\bigg] \\
&\times \delta_{\bm p^\prime \bm q}\delta_{\bm p\bm{\bar{q}}} \delta_{p_ip}\delta_{p_i^\prime p^\prime}\delta(\omega_1-\bar{\omega}_1)+ \sum_{\bm p, \bm p^\prime} \prod_{j\neq i} \delta_{p_jp_j^\prime}\delta_{\bm p\bm{\bar{q}}}\delta_{\bm q\bm p^\prime} i\delta R_k^{(1)}(\bm p,\omega_1)\delta(\omega_1-\bar{\omega}_1)=0\,,
\end{align}
where we dropped out the non-melonic contributions and assumed them to be in the symmetric phase, using definition \eqref{decomp2points}. Because of the definition of $\Delta_k$, the second and fourth contributions simplify. We set $\bm q_\bot=\bm q_\bot^\prime$, $p^\prime=q_i$, $p=\bar{q}_i$, and $p=p^\prime+1$. In the deep UV regime, it is suitable to use a continuous approximation to compute finite differences. We introduce the nearly continuous variable $x:=p/\Lambda$, where $\Lambda$ denote some UV cut-off, such that, for any function $f(p^2)$ that can be expressed in terms of dimensionless quantities as $f(p^2)=\Lambda^r \tilde{f}(x^2)$, and ($x^\prime=x+1/\Lambda$):
\begin{equation}
f((p^\prime)^2)-f(p^2)= \Lambda^r(\tilde{f}((x^\prime)^2)-\tilde{f}(x^2))= \Lambda^{r-2}\left( \frac{d\tilde{f}}{dx^2}+\mathcal{O}(1/\Lambda^2) \right)\,.
\end{equation}
Hence, from the definition \eqref{defZY}, we have:
\begin{equation}
\big[\gamma_{k,\bar{\sigma}M}^{(2)}(\bm p,0)- \gamma_{k,\bar{\sigma}M}^{(2)}(\bm p^\prime,0)\big]\big\vert_{p_i=0} \approx iZ \delta p^2 + \mathcal{O}(\delta p^2)\,,\label{equationderiv}
\end{equation}
in agreement with equation \eqref{defZY}, accordingly with our choice $\Omega=1$. Finally, setting external momenta to zero, the Ward identity reads:
\begin{equation}
\vcenter{\hbox{\includegraphics[scale=0.8]{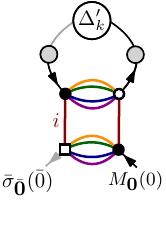}}}+\vcenter{\hbox{\includegraphics[scale=0.8]{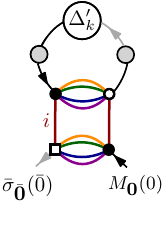}}}+\vcenter{\hbox{\includegraphics[scale=0.8]{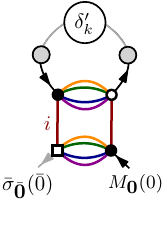}}}+i\left(Z-Z_{\infty}\right)=0\,,
\end{equation}
or explicitly:
\begin{align}
\nonumber\frac{2\lambda(k)}{\pi} \int d\omega & \sum_{\bm p \in \mathbb{Z}^{d-1}} \Bigg[2\left(iZ_{\infty}+i \frac{d}{dp_1^2}R_k^{(1)}(\bm p,\omega)\right)G_{k,\bar{M} M}(\bm p,\omega) G_{k,\bar{\sigma} M}(\bm p,\omega)\\
&+\frac{d}{dp_1^2}R_k^{(2)}(\bm p,\omega) G_{k,\bar{M} \sigma}(\bm p,\omega) G_{k,\bar{\sigma} M}(\bm p,\omega)\Bigg]\bigg\vert_{p_i=0}=-\left(Z-Z_{\infty}\right)\,.
\end{align}
This equation can be rewritten using dimensionless quantities as follows:
\begin{align}
\nonumber \frac{4i\bar{\lambda}(k)}{\pi} \int dy & \int_{\mathbb{R}^4} d\bm x \Bigg[\left( 1-\bar{Z}(k) \alpha \hat{\rho}(y) \theta(1-x)\right)\frac{1+\hat{\tau}(y)r(x)}{\hat{f}(x,y)\hat{f}^2(x,-y)}\\
&-\frac{1}{2}\alpha \bar{Z}(k) \hat{\tau}(y)\frac{ \theta(1-x) }{\hat{f}(x,y)\hat{f}(x,-y)}\Bigg]=1-\bar{Z}(k)\,,
\end{align}
where:
\begin{equation}
\bar{Z}(k):= \frac{Z(k)}{Z_{\infty}}\,.
\end{equation}
Note that, in these equations, $\hat{f}(x,y)$ is not expected to be of the form given by equation \eqref{truncationf}, except maybe for the terms involving the regulator. Indeed, for these terms, the selected windows of momenta are the same as for the flow equations. Hence, assuming the truncation \eqref{truncationf} for this contribution is not an additional assumption than assuming its validity for the computation of the flow equation themselves. Following the arguments given in \cite{Lahoche:2018oeo}, $Z^{-1}_\infty \sim \ln(\Lambda)$, and in the continuum limit, the previous Ward identity becomes:
\begin{align}
\nonumber \frac{4i\bar{\lambda}(k)}{\pi} \int dy & \int_{\mathbb{R}^4} d\bm x \Bigg[\left(Z_{\infty}{Z}^{-1}(k)-\alpha \hat{\rho}(y) \theta(1-x)\right)\frac{1+\hat{\tau}(y)r(x)}{\hat{f}(x,y)\hat{f}^2(x,-y)}\\
&-\frac{1}{2}\alpha \hat{\tau}(y)\frac{ \theta(1-x) }{\hat{f}(x,y)\hat{f}(x,-y)}\Bigg]\approx-1\,,\label{Wardcontinuum}
\end{align}

\printbibliography[heading=bibintoc]

@article{benedetti2015critical,
  title={Critical behavior in spherical and hyperbolic spaces},
  author={Benedetti, Dario},
  journal={Journal of Statistical Mechanics: Theory and Experiment},
  volume={2015},
  number={1},
  pages={P01002},
  year={2015},
  publisher={IOP Publishing}
}

@article{DePolsi:2022wyb,
    author = "De Polsi, Gonzalo and Wschebor, Nicol\'as",
    title = "{The regulator dependence in the functional renormalization group: a quantitative explanation}",
    eprint = "2204.09170",
    archivePrefix = "arXiv",
    primaryClass = "cond-mat.stat-mech",
    doi = "10.1103/PhysRevE.106.024111",
    month = "4",
    year = "2022"
}

@article{carrozza2016n,
    author = "Carrozza, Sylvain and Tanasa, Adrian",
    title = "{$O(N)$ Random Tensor Models}",
    eprint = "1512.06718",
    archivePrefix = "arXiv",
    primaryClass = "math-ph",
    doi = "10.1007/s11005-016-0879-x",
    journal = "Lett. Math. Phys.",
    volume = "106",
    number = "11",
    pages = "1531--1559",
    year = "2016"
}

@article{carrozza2022melonic,
    author = "Carrozza, Sylvain and Harribey, Sabine",
    title = "{Melonic Large $N$ Limit of $5$-Index Irreducible Random Tensors}",
    eprint = "2104.03665",
    archivePrefix = "arXiv",
    primaryClass = "math-ph",
    doi = "10.1007/s00220-021-04299-1",
    journal = "Commun. Math. Phys.",
    volume = "390",
    number = "3",
    pages = "1219--1270",
    year = "2022"
}

@article{carrozza2018large,
    author = "Carrozza, Sylvain",
    title = "{Large $N$ limit of irreducible tensor models: $O(N)$ rank-$3$ tensors with mixed permutation symmetry}",
    eprint = "1803.02496",
    archivePrefix = "arXiv",
    primaryClass = "hep-th",
    doi = "10.1007/JHEP06(2018)039",
    journal = "JHEP",
    volume = "06",
    pages = "039",
    year = "2018"
}

@article{benedetti20191,
    author = "Benedetti, Dario and Carrozza, Sylvain and Gurau, Razvan and Kolanowski, Maciej",
    title = "{The $1/N$ expansion of the symmetric traceless and the antisymmetric tensor models in rank three}",
    eprint = "1712.00249",
    archivePrefix = "arXiv",
    primaryClass = "hep-th",
    reportNumber = "LPT-Orsay-17-77, LPT-ORSAY-17-77",
    doi = "10.1007/s00220-019-03551-z",
    journal = "Commun. Math. Phys.",
    volume = "371",
    number = "1",
    pages = "55--97",
    year = "2019"
}

@article{lahoche2016renormalization,
    author = "Lahoche, Vincent and Oriti, Daniele",
    title = "{Renormalization of a tensorial field theory on the homogeneous space SU(2)/U(1)}",
    eprint = "1506.08393",
    archivePrefix = "arXiv",
    primaryClass = "hep-th",
    doi = "10.1088/1751-8113/50/2/025201",
    journal = "J. Phys. A",
    volume = "50",
    number = "2",
    pages = "025201",
    year = "2017"
}

@article{baratin2012group,
    author = "Baratin, Aristide and Oriti, Daniele",
    title = "{Group field theory and simplicial gravity path integrals: A model for Holst-Plebanski gravity}",
    eprint = "1111.5842",
    archivePrefix = "arXiv",
    primaryClass = "hep-th",
    reportNumber = "LPT-ORSAY-11-120",
    doi = "10.1103/PhysRevD.85.044003",
    journal = "Phys. Rev. D",
    volume = "85",
    pages = "044003",
    year = "2012"
}

@article{baratin2011quantum,
    author = "Baratin, Aristide and Oriti, Daniele",
    title = "{Quantum simplicial geometry in the group field theory formalism: reconsidering the Barrett-Crane model}",
    eprint = "1108.1178",
    archivePrefix = "arXiv",
    primaryClass = "gr-qc",
    reportNumber = "LPT-ORSAY-11-121",
    doi = "10.1088/1367-2630/13/12/125011",
    journal = "New J. Phys.",
    volume = "13",
    pages = "125011",
    year = "2011"
}

@article{jercher2022emergent,
    author = "Jercher, Alexander F. and Oriti, Daniele and Pithis, Andreas G. A.",
    title = "{Emergent cosmology from quantum gravity in the Lorentzian Barrett-Crane tensorial group field theory model}",
    eprint = "2112.00091",
    archivePrefix = "arXiv",
    primaryClass = "gr-qc",
    doi = "10.1088/1475-7516/2022/01/050",
    journal = "JCAP",
    volume = "01",
    number = "01",
    pages = "050",
    year = "2022"
}

@article{samary2014closed,
    author = "Ousmane Samary, Dine",
    title = "{Closed equations of the two-point functions for tensorial group field theory}",
    eprint = "1401.2096",
    archivePrefix = "arXiv",
    primaryClass = "hep-th",
    doi = "10.1088/0264-9381/31/18/185005",
    journal = "Class. Quant. Grav.",
    volume = "31",
    pages = "185005",
    year = "2014"
}

@article{samary2015correlation,
    author = "Ousmane Samary, Dine and P\'erez-S\'anchez, Carlos I. and Vignes-Tourneret, Fabien and Wulkenhaar, Raimar",
    title = "{Correlation functions of a just renormalizable tensorial group field theory: the melonic approximation}",
    eprint = "1411.7213",
    archivePrefix = "arXiv",
    primaryClass = "hep-th",
    doi = "10.1088/0264-9381/32/17/175012",
    journal = "Class. Quant. Grav.",
    volume = "32",
    number = "17",
    pages = "175012",
    year = "2015"
}

@article{pascalie2019large,
    author = "Pascalie, R. and P\'erez-S\'anchez, C. I. and Tanasa, A. and Wulkenhaar, R.",
    title = "{On the large $N$ limit of Schwinger-Dyson equations of a rank-3 tensor field theory}",
    eprint = "1810.09867",
    archivePrefix = "arXiv",
    primaryClass = "math-ph",
    doi = "10.1063/1.5080306",
    journal = "J. Math. Phys.",
    volume = "60",
    number = "7",
    pages = "7",
    year = "2019"
}

@article{lahoche2021no,
    author = "Lahoche, Vincent and Natta, B\^em-Bi\'eri Barth\'el\'emy and Ousmane Samary, Dine",
    title = "{No Ward-Takahashi identity violation for Abelian tensorial group field theories with a closure constraint}",
    eprint = "2108.10979",
    archivePrefix = "arXiv",
    primaryClass = "hep-th",
    doi = "10.1103/PhysRevD.104.106013",
    journal = "Phys. Rev. D",
    volume = "104",
    number = "10",
    pages = "106013",
    year = "2021"
}

@article{samary2014just,
    author = "Ousmane Samary, Dine and Vignes-Tourneret, Fabien",
    title = "{Just Renormalizable TGFT's on $U(1)^{d}$ with Gauge Invariance}",
    eprint = "1211.2618",
    archivePrefix = "arXiv",
    primaryClass = "hep-th",
    reportNumber = "1432-0916",
    doi = "10.1007/s00220-014-1930-3",
    journal = "Commun. Math. Phys.",
    volume = "329",
    pages = "545--578",
    year = "2014"
}

@article{lahoche2015renormalization,
    author = "Lahoche, Vincent and Oriti, Daniele and Rivasseau, Vincent",
    title = "{Renormalization of an Abelian Tensor Group Field Theory: Solution at Leading Order}",
    eprint = "1501.02086",
    archivePrefix = "arXiv",
    primaryClass = "hep-th",
    doi = "10.1007/JHEP04(2015)095",
    journal = "JHEP",
    volume = "04",
    pages = "095",
    year = "2015"
}

@article{Lahoche:2020pjo,
    author = "Lahoche, Vincent and Ousmane Samary, Dine",
    title = "{Reliability of the local truncations for the random tensor models renormalization group flow}",
    eprint = "2005.11846",
    archivePrefix = "arXiv",
    primaryClass = "hep-th",
    doi = "10.1103/PhysRevD.102.056002",
    journal = "Phys. Rev. D",
    volume = "102",
    number = "5",
    pages = "056002",
    year = "2020"
}

@article{Baloitcha:2020lha,
    author = "Baloitcha, Ezinvi and Lahoche, Vincent and Ousmane Samary, Dine",
    title = "{Flowing in discrete gravity models and Ward identities: a review}",
    eprint = "2001.02631",
    archivePrefix = "arXiv",
    primaryClass = "hep-th",
    doi = "10.1140/epjp/s13360-021-01823-z",
    journal = "Eur. Phys. J. Plus",
    volume = "136",
    number = "9",
    pages = "982",
    year = "2021"
}

@article{Lahoche:2019ocf,
    author = "Lahoche, Vincent and Ousmane Samary, Dine",
    title = "{Revisited functional renormalization group approach for random matrices in the large-$N$ limit}",
    eprint = "1909.03327",
    archivePrefix = "arXiv",
    primaryClass = "hep-th",
    doi = "10.1103/PhysRevD.101.106015",
    journal = "Phys. Rev. D",
    volume = "101",
    number = "10",
    pages = "106015",
    year = "2020"
}

@article{Lahoche:2018ggd,
    author = "Lahoche, Vincent and Ousmane Samary, Dine",
    title = "{Ward identity violation for melonic $T^4$-truncation}",
    eprint = "1809.06081",
    archivePrefix = "arXiv",
    primaryClass = "hep-th",
    doi = "10.1016/j.nuclphysb.2019.01.005",
    journal = "Nucl. Phys. B",
    volume = "940",
    pages = "190--213",
    year = "2019"
}

@article{Lahoche:2018oeo,
    author = "Lahoche, Vincent and Ousmane Samary, Dine",
    title = "{Nonperturbative renormalization group beyond melonic sector: The Effective Vertex Expansion method for group fields theories}",
    eprint = "1809.00247",
    archivePrefix = "arXiv",
    primaryClass = "hep-th",
    doi = "10.1103/PhysRevD.98.126010",
    journal = "Phys. Rev. D",
    volume = "98",
    number = "12",
    pages = "126010",
    year = "2018"
}

@book{Zinn-Justin:1989rgp,
    author = "Zinn-Justin, Jean",
    title = "{Quantum field theory and critical phenomena}",
    isbn = "978-0-19-850923-3, 978-0-19-883462-5",
    publisher = "Oxford University Press",
    series = "International Series of Monographs on Physics",
    volume = "77",
    month = "4",
    year = "2021"
}

@book{ZinnJustinBook2,
        author = {J. Zinn-Justin
},
        title = {From random walks to random matrices},
        publisher= {Oxford Graduate Texts},
        year = {2019}
}

@misc{https://doi.org/10.48550/arxiv.2112.02585,
  doi = {10.48550/ARXIV.2112.02585},
  
  url = {https://arxiv.org/abs/2112.02585},
  
  author = {Oriti, Daniele},
  
  title = {Tensorial Group Field Theory condensate cosmology as an example of spacetime emergence in quantum gravity},
  
  publisher = {arXiv},
  
  year = {2021},
  
  copyright = {arXiv.org perpetual, non-exclusive license}
}

@misc{https://doi.org/10.48550/arxiv.1807.04875,
  doi = {10.48550/ARXIV.1807.04875},
  
  url = {https://arxiv.org/abs/1807.04875},
  
  author = {Oriti, Daniele},
  
  title = {Levels of spacetime emergence in quantum gravity},
  
  publisher = {arXiv},
  
  year = {2018},
  
  copyright = {arXiv.org perpetual, non-exclusive license}
}

@misc{https://doi.org/10.48550/arxiv.2110.08641,
  doi = {10.48550/ARXIV.2110.08641},
  
  url = {https://arxiv.org/abs/2110.08641},
  
  author = {Oriti, Daniele},
  
  title = {The complex timeless emergence of time in quantum gravity},
  
  publisher = {arXiv},
  
  year = {2021},
  
  copyright = {arXiv.org perpetual, non-exclusive license}
}

@book{Seiberg_2007,
    author = "Nathan Seiberg",
    title = "{Emergent Spacetime}",
    doi="10.1142/9789812706768_0005",
    publisher = "World Scientific",
    series = "The Quantum Structure of Space and Time (proceeding)",
    year = "2007"
}

@misc{https://doi.org/10.48550/arxiv.1302.2849,
  doi = {10.48550/ARXIV.1302.2849},
  
  url = {https://arxiv.org/abs/1302.2849},
  
  author = {Oriti, Daniele},
  
  title = {Disappearance and emergence of space and time in quantum gravity},
  
  publisher = {arXiv},
  
  year = {2013},
  
  copyright = {arXiv.org perpetual, non-exclusive license}
}

@misc{https://doi.org/10.48550/arxiv.1001.3668,
  doi = {10.48550/ARXIV.1001.3668},
  
  url = {https://arxiv.org/abs/1001.3668},
  
  author = {Smolin, Lee},
  
  title = {Newtonian gravity in loop quantum gravity},
  
  publisher = {arXiv},
  
  year = {2010},
  
  copyright = {arXiv.org perpetual, non-exclusive license}
}

@book{Thiemann:2007pyv,
    author = "Thiemann, Thomas",
    title = "{Modern Canonical Quantum General Relativity}",
    doi = "10.1017/CBO9780511755682",
    isbn = "978-0-511-75568-2, 978-0-521-84263-1",
    publisher = "Cambridge University Press",
    series = "Cambridge Monographs on Mathematical Physics",
    year = "2007"
}

@book{rovelli_2004,
    author = "Rovelli, Carlo",
    editor = "Ashtekar, Abhay and Petkov, Vesselin",
    title = "{Quantum Spacetime}",
    booktitle = "{Springer Handbook of Spacetime}",
    doi = "10.1007/978-3-642-41992-8_36",
    pages = "751--757",
    year = "2014"
}

@article{Ashtekar_2021,
	doi = {10.1088/1361-6633/abed91},
  
	url = {https://doi.org/10.1088%2F1361-6633%2Fabed91},
  
	year = 2021,
	month = {mar},
  
	publisher = {{IOP} Publishing},
  
	volume = {84},
  
	number = {4},
  
	pages = {042001},
  
	author = {Abhay Ashtekar and Eugenio Bianchi},
  
	title = {A short review of loop quantum gravity},
  
	journal = {Reports on Progress in Physics}
}

@article{Perez_2013,
	doi = {10.12942/lrr-2013-3},
  
	url = {https://doi.org/10.12942%2Flrr-2013-3},
  
	year = 2013,
	month = {feb},
  
	publisher = {Springer Science and Business Media {LLC}},
  
	volume = {16},
  
	number = {1},
  
	author = {Alejandro Perez},
  
	title = {The Spin-Foam Approach to Quantum Gravity},
  
	journal = {Living Reviews in Relativity}
}

@article{Francesco_1995,
	doi = {10.1016/0370-1573(94)00084-g},
  
	url = {https://doi.org/10.1016%2F0370-1573%2894%2900084-g},
  
	year = 1995,
	month = {mar},
  
	publisher = {Elsevier {BV}},
  
	volume = {254},
  
	number = {1-2},
  
	pages = {1--133},
  
	author = {P.Di Francesco and P. Ginsparg and J. Zinn-Justin},
  
	title = {2D gravity and random matrices},
  
	journal = {Physics Reports}
}

@misc{https://doi.org/10.48550/arxiv.1510.04430,
  doi = {10.48550/ARXIV.1510.04430},
  
  url = {https://arxiv.org/abs/1510.04430},
  
  author = {Eynard, Bertrand and Kimura, Taro and Ribault, Sylvain},
  
  title = {Random matrices},
  
  publisher = {arXiv},
  
  year = {2015},
  
  copyright = {arXiv.org perpetual, non-exclusive license}
}

@article{dartois2013double,
    author = "Dartois, St\'ephane and Gurau, Razvan and Rivasseau, Vincent",
    title = "{Double Scaling in Tensor Models with a Quartic Interaction}",
    eprint = "1307.5281",
    archivePrefix = "arXiv",
    primaryClass = "hep-th",
    doi = "10.1007/JHEP09(2013)088",
    journal = "JHEP",
    volume = "09",
    pages = "088",
    year = "2013"
}

@article{bonzom2014double,
    author = "Bonzom, Valentin and Gurau, Razvan and Ryan, James P. and Tanasa, Adrian",
    title = "{The double scaling limit of random tensor models}",
    eprint = "1404.7517",
    archivePrefix = "arXiv",
    primaryClass = "hep-th",
    doi = "10.1007/JHEP09(2014)051",
    journal = "JHEP",
    volume = "09",
    pages = "051",
    year = "2014"
}

@article{Lahoche_2021,
   title={Large-d behavior of the Feynman amplitudes for a just-renormalizable tensorial group field theory},
   volume={103},
   ISSN={2470-0029},
   url={http://dx.doi.org/10.1103/PhysRevD.103.085006},
   DOI={10.1103/physrevd.103.085006},
   number={8},
   journal={Physical Review D},
   publisher={American Physical Society (APS)},
   author={Lahoche, Vincent and Ousmane Samary, Dine},
   year={2021},
   month=apr }

@article{bonzom2011critical,
    author = "Bonzom, Valentin and Gurau, Razvan and Riello, Aldo and Rivasseau, Vincent",
    title = "{Critical behavior of colored tensor models in the large N limit}",
    eprint = "1105.3122",
    archivePrefix = "arXiv",
    primaryClass = "hep-th",
    reportNumber = "PI-QG-224",
    doi = "10.1016/j.nuclphysb.2011.07.022",
    journal = "Nucl. Phys. B",
    volume = "853",
    pages = "174--195",
    year = "2011"
}

@article{bonzom2012random,
    author = "Bonzom, Valentin and Gurau, Razvan and Rivasseau, Vincent",
    title = "{Random tensor models in the large N limit: Uncoloring the colored tensor models}",
    eprint = "1202.3637",
    archivePrefix = "arXiv",
    primaryClass = "hep-th",
    reportNumber = "PI-QG-259",
    doi = "10.1103/PhysRevD.85.084037",
    journal = "Phys. Rev. D",
    volume = "85",
    pages = "084037",
    year = "2012"
}

@article{Gurau_2016,
	doi = {10.3842/sigma.2016.094},
  
	url = {https://doi.org/10.3842%2Fsigma.2016.094},
  
	year = 2016,
	month = {sep},
  
	publisher = {{SIGMA} (Symmetry, Integrability and Geometry: Methods and Application)},
  
	author = {Razvan Gurau and},
  
	title = {Invitation to Random Tensors},
  
	journal = {Symmetry, Integrability and Geometry: Methods and Applications}
}

@article{rivasseau2016random,
    author = "Rivasseau, Vincent",
    title = "{Random Tensors and Quantum Gravity}",
    eprint = "1603.07278",
    archivePrefix = "arXiv",
    primaryClass = "math-ph",
    doi = "10.3842/SIGMA.2016.069",
    journal = "SIGMA",
    volume = "12",
    pages = "069",
    year = "2016"
}

@book{guruau2017random,
  title={Random tensors},
  author={Gur{\u{a}}u, R{\u{a}}zvan Gheorghe},
  year={2017},
  publisher={Oxford University Press}
}

@article{Br_zin_1992,
	doi = {10.1016/0370-2693(92)91953-7},
  
	url = {https://doi.org/10.1016%2F0370-2693%2892%2991953-7},
  
	year = 1992,
	month = {aug},
  
	publisher = {Elsevier {BV}},
  
	volume = {288},
  
	number = {1-2},
  
	pages = {54--58},
  
	author = {Edouard Br{\'{e}}zin and Jean Zinn-Justin},
  
	title = {Renormalization group approach to matrix models},
  
	journal = {Physics Letters B}
}

@article{Eichhorn_2013,
	doi = {10.1103/physrevd.88.084016},
  
	url = {https://doi.org/10.1103%2Fphysrevd.88.084016},
  
	year = 2013,
	month = {oct},
  
	publisher = {American Physical Society ({APS})},
  
	volume = {88},
  
	number = {8},
  
	author = {Astrid Eichhorn and Tim Koslowski},
  
	title = {Continuum limit in matrix models for quantum gravity from the functional renormalization group},
  
	journal = {Physical Review D}
}

@article{Eichhorn_2014,
	doi = {10.1103/physrevd.90.104039},
  
	url = {https://doi.org/10.1103%2Fphysrevd.90.104039},
  
	year = 2014,
	month = {nov},
  
	publisher = {American Physical Society ({APS})},
  
	volume = {90},
  
	number = {10},
  
	author = {Astrid Eichhorn and Tim Koslowski},
  
	title = {Towards phase transitions between discrete and continuum quantum spacetime from the renormalization group},
  
	journal = {Physical Review D}
}

@article{Eichhorn_2019,
	doi = {10.1088/1361-6382/ab2545},
  
	url = {https://doi.org/10.1088%2F1361-6382%2Fab2545},
  
	year = 2019,
	month = {jul},
  
	publisher = {{IOP} Publishing},
  
	volume = {36},
  
	number = {15},
  
	pages = {155007},
  
	author = {Astrid Eichhorn and Johannes Lumma and Tim Koslowski and Antonio D Pereira},
  
	title = {Towards background independent quantum gravity with tensor models},
  
	journal = {Classical and Quantum Gravity}
}

@article{Eichhorn_2020,
	doi = {10.1007/jhep02(2020)110},
  
	url = {https://doi.org/10.1007%2Fjhep02%282020%29110},
  
	year = 2020,
	month = {feb},
  
	publisher = {Springer Science and Business Media {LLC}},
  
	volume = {2020},
  
	number = {2},
  
	author = {Astrid Eichhorn and Johannes Lumma and Antonio D. Pereira and Arslan Sikandar},
  
	title = {Universal critical behavior in tensor models for four-dimensional quantum gravity},
  
	journal = {Journal of High Energy Physics}
}

@article{samary2013beta,
    author = "Ousmane Samary, Dine",
    title = "{Beta functions of $U(1)^d$ gauge invariant just renormalizable tensor models}",
    eprint = "1303.7256",
    archivePrefix = "arXiv",
    primaryClass = "hep-th",
    doi = "10.1103/PhysRevD.88.105003",
    journal = "Phys. Rev. D",
    volume = "88",
    number = "10",
    pages = "105003",
    year = "2013"
}

@article{ben2011radiative,
    author = "Ben Geloun, Joseph and Bonzom, Valentin",
    title = "{Radiative corrections in the Boulatov-Ooguri tensor model: The 2-point function}",
    eprint = "1101.4294",
    archivePrefix = "arXiv",
    primaryClass = "hep-th",
    reportNumber = "PI-QG-208, ICMPA-MPA-002-2011",
    doi = "10.1007/s10773-011-0782-2",
    journal = "Int. J. Theor. Phys.",
    volume = "50",
    pages = "2819--2841",
    year = "2011"
}

@article{carrozza2015discrete,
    author = "Carrozza, Sylvain",
    title = "{Discrete renormalization group for SU(2) tensorial group field theory}",
    eprint = "1407.4615",
    archivePrefix = "arXiv",
    primaryClass = "hep-th",
    doi = "10.4171/aihpd/15",
    journal = "Ann. Inst. H. Poincare D Comb. Phys. Interact.",
    volume = "2",
    number = "1",
    pages = "49--112",
    year = "2015"
}

@article{carrozza2014renormalization2,
    author = "Carrozza, Sylvain and Oriti, Daniele and Rivasseau, Vincent",
    title = "{Renormalization of a SU(2) Tensorial Group Field Theory in Three Dimensions}",
    eprint = "1303.6772",
    archivePrefix = "arXiv",
    primaryClass = "hep-th",
    reportNumber = "LPT-ORSAY-13-25, AEI-2013-167",
    doi = "10.1007/s00220-014-1928-x",
    journal = "Commun. Math. Phys.",
    volume = "330",
    pages = "581--637",
    year = "2014"
}

@article{carrozza2014renormalization,
    author = "Carrozza, Sylvain and Oriti, Daniele and Rivasseau, Vincent",
    title = "{Renormalization of Tensorial Group Field Theories: Abelian U(1) Models in Four Dimensions}",
    eprint = "1207.6734",
    archivePrefix = "arXiv",
    primaryClass = "hep-th",
    reportNumber = "LPT-ORSAY-12-89, AEI-2012-079",
    doi = "10.1007/s00220-014-1954-8",
    journal = "Commun. Math. Phys.",
    volume = "327",
    pages = "603--641",
    year = "2014"
}

@article{https://doi.org/10.48550/arxiv.1701.03029,
    author = "Baloitcha, Ezinvi and Lahoche, Vincent and Ousmane Samary, Dine",
    title = "{Flowing in discrete gravity models and Ward identities: a review}",
    eprint = "2001.02631",
    archivePrefix = "arXiv",
    primaryClass = "hep-th",
    doi = "10.1140/epjp/s13360-021-01823-z",
    journal = "Eur. Phys. J. Plus",
    volume = "136",
    number = "9",
    pages = "982",
    year = "2021"
}

@article{Lahoche_2020b,
    author = "Lahoche, Vincent and Ousmane Samary, Dine",
    title = "{Pedagogical comments about nonperturbative Ward-constrained melonic renormalization group flow}",
    eprint = "2001.00934",
    archivePrefix = "arXiv",
    primaryClass = "hep-th",
    doi = "10.1103/PhysRevD.101.024001",
    journal = "Phys. Rev. D",
    volume = "101",
    number = "2",
    pages = "024001",
    year = "2020"
}

@article{Lahoche_2021c,
    author = "Lahoche, Vincent and Ousmane Samary, Dine",
    title = "{Large-$d$ behavior of the Feynman amplitudes for a just-renormalizable tensorial group field theory}",
    eprint = "1911.08601",
    archivePrefix = "arXiv",
    primaryClass = "hep-th",
    doi = "10.1103/PhysRevD.103.085006",
    journal = "Phys. Rev. D",
    volume = "103",
    number = "8",
    pages = "085006",
    year = "2021"
}

@article{Lahoche_2020d,
    author = "Lahoche, Vincent and Ousmane Samary, Dine and Pereira, Antonio D.",
    title = "{Renormalization group flow of coupled tensorial group field theories: Towards the Ising model on random lattices}",
    eprint = "1911.05173",
    archivePrefix = "arXiv",
    primaryClass = "hep-th",
    doi = "10.1103/PhysRevD.101.064014",
    journal = "Phys. Rev. D",
    volume = "101",
    number = "6",
    pages = "064014",
    year = "2020"
}

@article{Lahoche_2019a,
    author = "Lahoche, Vincent and Ousmane Samary, Dine",
    title = "{Ward-constrained melonic renormalization group flow for the rank-four $\phi^6$ tensorial group field theory}",
    eprint = "1908.03910",
    archivePrefix = "arXiv",
    primaryClass = "hep-th",
    doi = "10.1103/PhysRevD.100.086009",
    journal = "Phys. Rev. D",
    volume = "100",
    number = "8",
    pages = "086009",
    year = "2019"
}

@article{Lahoche_2019bb,
    author = "Lahoche, Vincent and Ousmane Samary, Dine",
    title = "{Ward identity violation for melonic $T^4$-truncation}",
    eprint = "1809.06081",
    archivePrefix = "arXiv",
    primaryClass = "hep-th",
    doi = "10.1016/j.nuclphysb.2019.01.005",
    journal = "Nucl. Phys. B",
    volume = "940",
    pages = "190--213",
    year = "2019"
}

@article{Carrozza_2017,
	doi = {10.1103/physrevd.96.066007},
  
	url = {https://doi.org/10.1103%2Fphysrevd.96.066007},
  
	year = 2017,
	month = {sep},
  
	publisher = {American Physical Society ({APS})},
  
	volume = {96},
  
	number = {6},
  
	author = {Sylvain Carrozza and Vincent Lahoche and Daniele Oriti},
  
	title = {Renormalizable group field theory beyond melonic diagrams: An example in rank four},
  
	journal = {Physical Review D}
}

@article{Carrozza_2017a,
	doi = {10.1088/1361-6382/aa6d90},
  
	url = {https://doi.org/10.1088%2F1361-6382%2Faa6d90},
  
	year = 2017,
	month = {may},
  
	publisher = {{IOP} Publishing},
  
	volume = {34},
  
	number = {11},
  
	pages = {115004},
  
	author = {Sylvain Carrozza and Vincent Lahoche},
  
	title = {Asymptotic safety in three-dimensional {SU}(2) group field theory: evidence in the local potential approximation},
  
	journal = {Classical and Quantum Gravity}
}

@article{Lahoche_2017bb,
    author = "Lahoche, Vincent and Ousmane Samary, Dine",
    title = "{Functional renormalization group for the U(1)-T$_5^6$ tensorial group field theory with closure constraint}",
    eprint = "1608.00379",
    archivePrefix = "arXiv",
    primaryClass = "hep-th",
    doi = "10.1103/PhysRevD.95.045013",
    journal = "Phys. Rev. D",
    volume = "95",
    number = "4",
    pages = "045013",
    year = "2017"
}

@article{Benedetti_2016,
	doi = {10.1088/0264-9381/33/9/095003},
  
	url = {https://doi.org/10.1088%2F0264-9381%2F33%2F9%2F095003},
  
	year = 2016,
	month = {apr},
  
	publisher = {{IOP} Publishing},
  
	volume = {33},
  
	number = {9},
  
	pages = {095003},
  
	author = {Dario Benedetti and Vincent Lahoche},
  
	title = {Functional renormalization group approach for tensorial group field theory: a rank-6 model with closure constraint},
  
	journal = {Classical and Quantum Gravity}
}

@article{Benedetti_2015,
	doi = {10.1007/jhep03(2015)084},
  
	url = {https://doi.org/10.1007%2Fjhep03%282015%29084},
  
	year = 2015,
	month = {mar},
  
	publisher = {Springer Science and Business Media {LLC}},
  
	volume = {2015},
  
	number = {3},
  
	author = {Dario Benedetti and Joseph Ben Geloun and Daniele Oriti},
  
	title = {Functional renormalisation group approach for tensorial group field theory: a rank-3 model},
  
	journal = {Journal of High Energy Physics}
}

@article{Geloun_2016,
	doi = {10.1103/physrevd.94.024017},
  
	url = {https://doi.org/10.1103%2Fphysrevd.94.024017},
  
	year = 2016,
	month = {jul},
  
	publisher = {American Physical Society ({APS})},
  
	volume = {94},
  
	number = {2},
  
	author = {Joseph Ben Geloun and Riccardo Martini and Daniele Oriti},
  
	title = {Functional renormalization group analysis of tensorial group field theories on $\mathbb{R}^d$.},
  
	journal = {Physical Review D}
}

@article{Ben_Geloun_2015,
	doi = {10.1209/0295-5075/112/31001},
  
	url = {https://doi.org/10.1209%2F0295-5075%2F112%2F31001},
  
	year = 2015,
	month = {nov},
  
	publisher = {{IOP} Publishing},
  
	volume = {112},
  
	number = {3},
  
	pages = {31001},
  
	author = {Joseph Ben Geloun and Riccardo Martini and Daniele Oriti},
  
	title = {Functional Renormalization Group analysis of a Tensorial Group Field Theory on $\mathbb{R}^3$.},
  
	journal = {{EPL} (Europhysics Letters)}
}

@misc{https://doi.org/10.48550/arxiv.1111.4997,
  doi = {10.48550/ARXIV.1111.4997},
  
  url = {https://arxiv.org/abs/1111.4997},
  
  author = {Geloun, Joseph Ben and Rivasseau, Vincent},
  
  title = {A Renormalizable 4-Dimensional Tensor Field Theory},
  
  publisher = {arXiv},
  
  year = {2011},
  
  copyright = {arXiv.org perpetual, non-exclusive license}
}

@article{Geloun_2018,
	doi = {10.1103/physrevd.97.126018},
  
	url = {https://doi.org/10.1103%2Fphysrevd.97.126018},
  
	year = 2018,
	month = {jun},
  
	publisher = {American Physical Society ({APS})},
  
	volume = {97},
  
	number = {12},
  
	author = {Joseph Ben Geloun and Tim A. Koslowski and Daniele Oriti and Antonio D. Pereira},
  
	title = {Functional renormalization group analysis of rank-3 tensorial group field theory: The full quartic invariant truncation},
  
	journal = {Physical Review D}
}

@article{Carrozza_2016ccc,
	doi = {10.3842/sigma.2016.070},
  
	url = {https://doi.org/10.3842%2Fsigma.2016.070},
  
	year = 2016,
	month = {jul},
  
	publisher = {{SIGMA} (Symmetry, Integrability and Geometry: Methods and Application)},
  
	author = {Sylvain Carrozza and   and   and},
  
	title = {Flowing in Group Field Theory Space: a Review},
  
	journal = {Symmetry, Integrability and Geometry: Methods and Applications}
}

@article{Carrozza_2015a,
	doi = {10.1103/physrevd.91.065023},
  
	url = {https://doi.org/10.1103%2Fphysrevd.91.065023},
  
	year = 2015,
	month = {mar},
  
	publisher = {American Physical Society ({APS})},
  
	volume = {91},
  
	number = {6},
  
	author = {Sylvain Carrozza},
  
	title = {Group field theory in dimension four minus epsilon},
  
	journal = {Physical Review D}
}

@article{Carrozza_2015aaa,
	doi = {10.4171/aihpd/15},
  
	url = {https://doi.org/10.4171%2Faihpd%2F15},
  
	year = 2015,
	publisher = {European Mathematical Society - {EMS} - Publishing House {GmbH}},
  
	volume = {2},
  
	number = {1},
  
	pages = {49--112},
  
	author = {Sylvain Carrozza},
  
	title = {Discrete renormalization group for {SU}(2) tensorial group field theory},
  
	journal = {Annales de l'Institut Henri Poincar{\'{e}} D}
}

@book{Carrozza_2014,
	doi = {10.1007/978-3-319-05867-2},
  
	url = {https://doi.org/10.1007%2F978-3-319-05867-2},
  
	year = 2014,
	publisher = {Springer International Publishing},
  
	author = {Sylvain Carrozza},
  
	title = {Tensorial Methods and Renormalization in Group Field Theories}
}

@misc{https://doi.org/10.48550/arxiv.1409.1450,
  doi = {10.48550/ARXIV.1409.1450},
  
  url = {https://arxiv.org/abs/1409.1450},
  
  author = {Dittrich, Bianca},
  
  title = {The continuum limit of loop quantum gravity - a framework for solving the theory},
  
  publisher = {arXiv},
  
  year = {2014},
  
  copyright = {arXiv.org perpetual, non-exclusive license}
}

@article{pawlowski2017physics,
    author = "Pawlowski, Jan M. and Scherer, Michael M. and Schmidt, Richard and Wetzel, Sebastian J.",
    title = "{Physics and the choice of regulators in functional renormalisation group flows}",
    eprint = "1512.03598",
    archivePrefix = "arXiv",
    primaryClass = "hep-th",
    doi = "10.1016/j.aop.2017.06.017",
    journal = "Annals Phys.",
    volume = "384",
    pages = "165--197",
    year = "2017"
}

@article{pawlowski2007aspects,
    author = "Pawlowski, Jan M.",
    title = "{Aspects of the functional renormalisation group}",
    eprint = "hep-th/0512261",
    archivePrefix = "arXiv",
    reportNumber = "HD-THEP-05-28",
    doi = "10.1016/j.aop.2007.01.007",
    journal = "Annals Phys.",
    volume = "322",
    pages = "2831--2915",
    year = "2007"
}

@article{Dupuis_2021,
	doi = {10.1016/j.physrep.2021.01.001},
  
	url = {https://doi.org/10.1016%2Fj.physrep.2021.01.001},
  
	year = 2021,
	month = {may},
  
	publisher = {Elsevier {BV}},
  
	volume = {910},
  
	pages = {1--114},
  
	author = {N. Dupuis and L. Canet and A. Eichhorn and W. Metzner and J.M. Pawlowski and M. Tissier and N. Wschebor},
  
	title = {The nonperturbative functional renormalization group and its applications},
  
	journal = {Physics Reports}
}

@incollection{Delamotte_2012,
	doi = {10.1007/978-3-642-27320-9_2},
  
	url = {https://doi.org/10.1007%2F978-3-642-27320-9_2},
  
	year = 2012,
	publisher = {Springer Berlin Heidelberg},
  
	pages = {49--132},
  
	author = {Bertrand Delamotte},
  
	title = {An Introduction to the Nonperturbative Renormalization Group},
  
	booktitle = {Renormalization Group and Effective Field Theory Approaches to Many-Body Systems}
}

@article{Berges_2002,
	doi = {10.1016/s0370-1573(01)00098-9},
  
	url = {https://doi.org/10.1016%2Fs0370-1573%2801%2900098-9},
  
	year = 2002,
	month = {jun},
  
	publisher = {Elsevier {BV}},
  
	volume = {363},
  
	number = {4-6},
  
	pages = {223--386},
  
	author = {Jürgen Berges and Nikolaos Tetradis and Christof Wetterich},
  
	title = {Non-perturbative renormalization flow in quantum field theory and statistical physics},
  
	journal = {Physics Reports}
}

@article{MORRIS_1994,
	doi = {10.1142/s0217751x94000972},
  
	url = {https://doi.org/10.1142%2Fs0217751x94000972},
  
	year = 1994,
	month = {jun},
  
	publisher = {World Scientific Pub Co Pte Lt},
  
	volume = {09},
  
	number = {14},
  
	pages = {2411--2449},
  
	author = {TIM R. MORRIS},
  
	title = {The exact renormalization group and approximate solutions
},
  
	journal = {International Journal of Modern Physics A}
}

@article{Morris_1994a,
	doi = {10.1016/0370-2693(94)90700-5},
  
	url = {https://doi.org/10.1016%2F0370-2693%2894%2990700-5},
  
	year = 1994,
	month = {aug},
  
	publisher = {Elsevier {BV}},
  
	volume = {334},
  
	number = {3-4},
  
	pages = {355--362},
  
	author = {Tim R. Morris},
  
	title = {On truncations of the exact renormalization group},
  
	journal = {Physics Letters B}
}

@article{Freidel_2005,
	doi = {10.1007/s10773-005-8894-1},
  
	url = {https://doi.org/10.1007%2Fs10773-005-8894-1},
  
	year = 2005,
	month = {oct},
  
	publisher = {Springer Science and Business Media {LLC}},
  
	volume = {44},
  
	number = {10},
  
	pages = {1769--1783},
  
	author = {L. Freidel},
  
	title = {Group Field Theory: An Overview},
  
	journal = {International Journal of Theoretical Physics}
}

@article{baratin2012ten,
    author = "Baratin, Aristide and Oriti, Daniele",
    editor = "Mena Marugan, Guillermo A. and Barbero, J. Fernando G. and Garay, Luis J. and Villasenor, Eduardo J. S. and Olmedo, Javier",
    title = "{Ten questions on Group Field Theory (and their tentative answers)}",
    eprint = "1112.3270",
    archivePrefix = "arXiv",
    primaryClass = "gr-qc",
    doi = "10.1088/1742-6596/360/1/012002",
    journal = "J. Phys. Conf. Ser.",
    volume = "360",
    pages = "012002",
    year = "2012"
}

@misc{https://doi.org/10.48550/arxiv.1110.5606,
  doi = {10.48550/ARXIV.1110.5606},
  
  url = {https://arxiv.org/abs/1110.5606},
  
  author = {Oriti, Daniele},
  
  title = {The microscopic dynamics of quantum space as a group field theory},
  
  publisher = {arXiv},
  
  year = {2011},
  
  copyright = {arXiv.org perpetual, non-exclusive license}
}

@misc{https://doi.org/10.48550/arxiv.gr-qc/0607032,
  doi = {10.48550/ARXIV.GR-QC/0607032},
  
  url = {https://arxiv.org/abs/gr-qc/0607032},
  
  author = {Oriti, Daniele},
  
  title = {The group field theory approach to quantum gravity},
  
  publisher = {arXiv},
  
  year = {2006},
  
  copyright = {Assumed arXiv.org perpetual, non-exclusive license to distribute this article for submissions made before January 2004}
}

@misc{https://doi.org/10.48550/arxiv.1210.6257,
  doi = {10.48550/ARXIV.1210.6257},
  
  url = {https://arxiv.org/abs/1210.6257},
  
  author = {Krajewski, Thomas},
  
  title = {Group field theories},
  
  publisher = {arXiv},
  
  year = {2012},
  
  copyright = {arXiv.org perpetual, non-exclusive license}
}

@misc{https://doi.org/10.48550/arxiv.1310.7786,
  doi = {10.48550/ARXIV.1310.7786},
  
  url = {https://arxiv.org/abs/1310.7786},
  
  author = {Oriti, Daniele},
  
  title = {Group field theory as the 2nd quantization of Loop Quantum Gravity},
  
  publisher = {arXiv},
  
  year = {2013},
  
  copyright = {arXiv.org perpetual, non-exclusive license}
}

@article{oriti2015group,
    author = {Oriti, Daniele and Ryan, James P. and Th\"urigen, Johannes},
    title = "{Group field theories for all loop quantum gravity}",
    eprint = "1409.3150",
    archivePrefix = "arXiv",
    primaryClass = "gr-qc",
    reportNumber = "AEI-2014-044",
    doi = "10.1088/1367-2630/17/2/023042",
    journal = "New J. Phys.",
    volume = "17",
    number = "2",
    pages = "023042",
    year = "2015"
}

@article{martinetti2003diamond,
    author = "Martinetti, Pierre and Rovelli, Carlo",
    title = "{Diamonds's temperature: Unruh effect for bounded trajectories and thermal time hypothesis}",
    eprint = "gr-qc/0212074",
    archivePrefix = "arXiv",
    doi = "10.1088/0264-9381/20/22/015",
    journal = "Class. Quant. Grav.",
    volume = "20",
    pages = "4919--4932",
    year = "2003"
}

@article{menicucci2011clocks,
    author = "Menicucci, Nicolas C. and Olson, S. Jay and Milburn, Gerard J.",
    title = "{Clocks and Relationalism in the Thermal Time Hypothesis}",
    eprint = "1108.0883",
    archivePrefix = "arXiv",
    primaryClass = "gr-qc",
    month = "8",
    year = "2011"
}

@article{rovelli2011thermal,
    author = "Rovelli, Carlo and Smerlak, Matteo",
    title = "{Thermal time and the Tolman-Ehrenfest effect: temperature as the 'speed of time'}",
    eprint = "1005.2985",
    archivePrefix = "arXiv",
    primaryClass = "gr-qc",
    doi = "10.1088/0264-9381/28/7/075007",
    journal = "Class. Quant. Grav.",
    volume = "28",
    pages = "075007",
    year = "2011"
}

@article{rovelli1993statistical2,
  title={The statistical state of the universe},
  author={Rovelli, Carlo},
  doi="10.1088/0264-9381/10/8/016",
  journal={Classical and Quantum Gravity},
  volume={10},
  number={8},
  pages={1567},
  year={1993},
  publisher={IOP Publishing}
}

@article{rovelli1993statistical,
  title={Statistical mechanics of gravity and the thermodynamical origin of time},
  author={Rovelli, Carlo},
  doi="10.1088/0264-9381/10/8/015",
  journal={Classical and Quantum Gravity},
  volume={10},
  number={8},
  pages={1549},
  year={1993},
  publisher={IOP Publishing}
}

@article{connes1994neumann,
    author = "Connes, A. and Rovelli, Carlo",
    title = "{Von Neumann algebra automorphisms and time thermodynamics relation in general covariant quantum theories}",
    eprint = "gr-qc/9406019",
    archivePrefix = "arXiv",
    doi = "10.1088/0264-9381/11/12/007",
    journal = "Class. Quant. Grav.",
    volume = "11",
    pages = "2899--2918",
    year = "1994"
}

@article{marchetti2021phase,
    author = {Marchetti, Luca and Oriti, Daniele and Pithis, Andreas G. A. and Th\"urigen, Johannes},
    title = "{Phase transitions in tensorial group field theories: Landau-Ginzburg analysis of models with both local and non-local degrees of freedom}",
    eprint = "2110.15336",
    archivePrefix = "arXiv",
    primaryClass = "gr-qc",
    doi = "10.1007/JHEP12(2021)201",
    journal = "JHEP",
    volume = "21",
    pages = "201",
    year = "2020"
}

@article{pithis2021no,
  title={(No) phase transition in tensorial group field theory},
  author={Pithis, Andreas GA and Th{\"u}rigen, Johannes},
  journal={Physics Letters B},
  volume={816},
  pages={136215},
  year={2021},
  publisher={Elsevier}
}

@article{pithis2020phase,
  title={Phase transitions in TGFT: functional renormalization group in the cyclic-melonic potential approximation and equivalence to O (N) models},
  author={Pithis, Andreas GA and Th{\"u}rigen, Johannes},
  journal={Journal of High Energy Physics},
  volume={2020},
  number={12},
  pages={1--54},
  year={2020},
  publisher={Springer}
}

@article{oriti2018black,
    author = "Oriti, Daniele and Pranzetti, Daniele and Sindoni, Lorenzo",
    title = "{Black Holes as Quantum Gravity Condensates}",
    eprint = "1801.01479",
    archivePrefix = "arXiv",
    primaryClass = "gr-qc",
    doi = "10.1103/PhysRevD.97.066017",
    journal = "Phys. Rev. D",
    volume = "97",
    number = "6",
    pages = "066017",
    year = "2018"
}

@article{gielen2018cosmological,
    author = "Gielen, Steffen and Oriti, Daniele",
    title = "{Cosmological perturbations from full quantum gravity}",
    eprint = "1709.01095",
    archivePrefix = "arXiv",
    primaryClass = "gr-qc",
    doi = "10.1103/PhysRevD.98.106019",
    journal = "Phys. Rev. D",
    volume = "98",
    number = "10",
    pages = "106019",
    year = "2018"
}

@article{de2017dynamics,
    author = "de Cesare, Marco and Oriti, Daniele and Pithis, Andreas G. A. and Sakellariadou, Mairi",
    title = "{Dynamics of anisotropies close to a cosmological bounce in quantum gravity}",
    eprint = "1709.00994",
    archivePrefix = "arXiv",
    primaryClass = "gr-qc",
    reportNumber = "KCL-PH-TH-2017-41",
    doi = "10.1088/1361-6382/aa986a",
    journal = "Class. Quant. Grav.",
    volume = "35",
    number = "1",
    pages = "015014",
    year = "2018"
}

@article{kegeles2018inequivalent,
    author = "Kegeles, Alexander and Oriti, Daniele and Tomlin, Casey",
    title = "{Inequivalent coherent state representations in group field theory}",
    eprint = "1709.00161",
    archivePrefix = "arXiv",
    primaryClass = "gr-qc",
    doi = "10.1088/1361-6382/aac39f",
    journal = "Class. Quant. Grav.",
    volume = "35",
    number = "12",
    pages = "125011",
    year = "2018"
}

@article{oriti2017universe,
    author = "Oriti, Daniele",
    title = "{The universe as a quantum gravity condensate}",
    eprint = "1612.09521",
    archivePrefix = "arXiv",
    primaryClass = "gr-qc",
    doi = "10.1016/j.crhy.2017.02.003",
    journal = "Comptes Rendus Physique",
    volume = "18",
    pages = "235--245",
    year = "2017"
}

@article{oriti2016horizon,
    author = "Oriti, Daniele and Pranzetti, Daniele and Sindoni, Lorenzo",
    title = "{Horizon entropy from quantum gravity condensates}",
    eprint = "1510.06991",
    archivePrefix = "arXiv",
    primaryClass = "gr-qc",
    doi = "10.1103/PhysRevLett.116.211301",
    journal = "Phys. Rev. Lett.",
    volume = "116",
    number = "21",
    pages = "211301",
    year = "2016"
}

@article{oriti2015generalized,
    author = "Oriti, Daniele and Pranzetti, Daniele and Ryan, James P. and Sindoni, Lorenzo",
    title = "{Generalized quantum gravity condensates for homogeneous geometries and cosmology}",
    eprint = "1501.00936",
    archivePrefix = "arXiv",
    primaryClass = "gr-qc",
    reportNumber = "AEI-2015-003",
    doi = "10.1088/0264-9381/32/23/235016",
    journal = "Class. Quant. Grav.",
    volume = "32",
    number = "23",
    pages = "235016",
    year = "2015"
}

@article{gielen2014quantum,
    author = "Gielen, Steffen and Oriti, Daniele",
    title = "{Quantum cosmology from quantum gravity condensates: cosmological variables and lattice-refined dynamics}",
    eprint = "1407.8167",
    archivePrefix = "arXiv",
    primaryClass = "gr-qc",
    reportNumber = "AEI-2014-032",
    doi = "10.1088/1367-2630/16/12/123004",
    journal = "New J. Phys.",
    volume = "16",
    number = "12",
    pages = "123004",
    year = "2014"
}

@article{gielen2014homogeneous,
    author = "Gielen, Steffen and Oriti, Daniele and Sindoni, Lorenzo",
    title = "{Homogeneous cosmologies as group field theory condensates}",
    eprint = "1311.1238",
    archivePrefix = "arXiv",
    primaryClass = "gr-qc",
    reportNumber = "AEI-2013-259",
    doi = "10.1007/JHEP06(2014)013",
    journal = "JHEP",
    volume = "06",
    pages = "013",
    year = "2014"
}

@article{marchetti2021effective,
    author = "Marchetti, Luca and Oriti, Daniele",
    title = "{Effective dynamics of scalar cosmological perturbations from quantum gravity}",
    eprint = "2112.12677",
    archivePrefix = "arXiv",
    primaryClass = "gr-qc",
    doi = "10.1088/1475-7516/2022/07/004",
    journal = "JCAP",
    volume = "07",
    number = "07",
    pages = "004",
    year = "2022"
}

@article{gielen2022effective,
  title={Effective cosmology from one-body operators in group field theory},
  author={Gielen, Steffen and Marchetti, Luca and Oriti, Daniele and Polaczek, Axel},
  doi="10.1088/1361-6382/ac5052",
  journal={Classical and Quantum Gravity},
  volume={39},
  number={7},
  pages={075002},
  year={2022},
  publisher={IOP Publishing}
}

@article{oriti2017bouncing,
  title={Bouncing cosmologies from quantum gravity condensates},
  author={Oriti, Daniele and Sindoni, Lorenzo and Wilson-Ewing, Edward},
  doi="10.1088/1361-6382/aa549a",
  journal={Classical and Quantum Gravity},
  volume={34},
  number={4},
  pages={04LT01},
  year={2017},
  publisher={IOP Publishing}
}

@article{oriti2016emergent,
  title={Emergent Friedmann dynamics with a quantum bounce from quantum gravity condensates},
  author={Oriti, Daniele and Sindoni, Lorenzo and Wilson-Ewing, Edward},
  doi="10.1088/0264-9381/33/22/224001",
  journal={Classical and Quantum Gravity},
  volume={33},
  number={22},
  pages={224001},
  year={2016},
  publisher={IOP Publishing}
}

@book{livi2017nonequilibrium,
  title={Nonequilibrium statistical physics: a modern perspective},
  author={Livi, Roberto and Politi, Paolo},
  year={2017},
  publisher={Cambridge University Press}
}

@article{wilson2019relational,
    author = "Wilson-Ewing, Edward",
    title = "{A relational Hamiltonian for group field theory}",
    eprint = "1810.01259",
    archivePrefix = "arXiv",
    primaryClass = "gr-qc",
    doi = "10.1103/PhysRevD.99.086017",
    journal = "Phys. Rev. D",
    volume = "99",
    number = "8",
    pages = "086017",
    year = "2019"
}

@article{li2017group,
    author = "Li, Yang and Oriti, Daniele and Zhang, Mingyi",
    title = "{Group field theory for quantum gravity minimally coupled to a scalar field}",
    eprint = "1701.08719",
    archivePrefix = "arXiv",
    primaryClass = "gr-qc",
    doi = "10.1088/1361-6382/aa85d2",
    journal = "Class. Quant. Grav.",
    volume = "34",
    number = "19",
    pages = "195001",
    year = "2017"
}

@misc{https://doi.org/10.48550/arxiv.2010.15445,
  doi = {10.48550/ARXIV.2010.15445},
  
  url = {https://arxiv.org/abs/2010.15445},
  
  author = {Kotecha, Isha},
  
  title = {On Generalised Statistical Equilibrium and Discrete Quantum Gravity},
  
  publisher = {arXiv},
  
  year = {2020},
  
  copyright = {arXiv.org perpetual, non-exclusive license}
}

@article{canet2011general,
    author = "Canet, Leonie and Chate, Hugues and Delamotte, Bertrand",
    title = "{General framework of the non-perturbative renormalization group for non-equilibrium steady states}",
    eprint = "1106.4129",
    archivePrefix = "arXiv",
    primaryClass = "cond-mat.stat-mech",
    doi = "10.1088/1751-8113/44/49/495001",
    journal = "J. Phys. A",
    volume = "44",
    pages = "495001",
    year = "2011"
}

@article{kubo1966fluctuation,
  title={The fluctuation-dissipation theorem},
  author={Kubo, Rep},
  journal={Reports on progress in physics},
  volume={29},
  number={1},
  pages={255},
  year={1966},
  publisher={IOP Publishing}
}

@article{marconi2008fluctuation,
  title={Fluctuation--dissipation: response theory in statistical physics},
  author={Marconi, Umberto Marini Bettolo and Puglisi, Andrea and Rondoni, Lamberto and Vulpiani, Angelo},
  journal={Physics reports},
  volume={461},
  number={4-6},
  pages={111--195},
  year={2008},
  publisher={Elsevier}
}

@article{aron2010symmetries2,
    author = "Aron, Camille and Biroli, Giulio and Cugliandolo, Leticia F.",
    title = "{Symmetries of generating functionals of Langevin processes with colored multiplicative noise}",
    eprint = "1007.5059",
    archivePrefix = "arXiv",
    primaryClass = "cond-mat.stat-mech",
    doi = "10.1088/1742-5468/2010/11/P11018",
    journal = "J. Stat. Mech.",
    volume = "1011",
    pages = "P11018",
    year = "2010"
}

@article{duclut2017frequency,
    author = "Duclut, Charlie and Delamotte, Bertrand",
    title = "{Frequency regulators for the nonperturbative renormalization group: A general study and the model A as a benchmark}",
    eprint = "1611.07301",
    archivePrefix = "arXiv",
    primaryClass = "cond-mat.stat-mech",
    doi = "10.1103/PhysRevE.95.012107",
    journal = "Phys. Rev. E",
    volume = "95",
    number = "1",
    pages = "012107",
    year = "2017"
}

@article{lahoche2021functional,
    author = "Lahoche, Vincent and Ousmane Samary, Dine and Ouerfelli, Mohamed",
    title = "{Functional renormalization group for multilinear disordered Langevin dynamics I: Formalism and first numerical investigations at equilibrium}",
    eprint = "2106.05690",
    archivePrefix = "arXiv",
    primaryClass = "hep-th",
    doi = "10.1088/2399-6528/ac61b3",
    journal = "J. Phys. Comm.",
    volume = "6",
    pages = "055002",
    year = "2022"
}

@article{synatschke2009flow,
  title={Flow equation for supersymmetric quantum mechanics},
  author={Synatschke, Franziska and Bergner, Georg and Gies, Holger and Wipf, Andreas},
  doi="10.1088/1126-6708/2009/03/028",
  journal={Journal of High Energy Physics},
  volume={2009},
  number={03},
  pages={028},
  year={2009},
  publisher={IOP Publishing}
}

@article{zappala2001improving,
    author = "Zappala, D.",
    title = "{Improving the renormalization group approach to the quantum mechanical double well potential}",
    eprint = "quant-ph/0108019",
    archivePrefix = "arXiv",
    doi = "10.1016/S0375-9601(01)00642-9",
    journal = "Phys. Lett. A",
    volume = "290",
    pages = "35--40",
    year = "2001"
}

@article{prokopec2018functional,
    author = "Prokopec, Tomislav and Rigopoulos, Gerasimos",
    title = "{Functional renormalization group for stochastic inflation}",
    eprint = "1710.07333",
    archivePrefix = "arXiv",
    primaryClass = "gr-qc",
    doi = "10.1088/1475-7516/2018/08/013",
    journal = "JCAP",
    volume = "08",
    pages = "013",
    year = "2018"
}

@article{wilkins2020functional,
  title={Functional Renormalisation Group for Brownian Motion I: The Effective Equations of Motion},
  author={Wilkins, Ashley and Rigopoulos, Gerasimos and Masoero, Enrico},
  eprint = "2008.00472",
    archivePrefix = "arXiv",
    primaryClass = "cond-mat.stat-mech",
  year={2020}
}

@article{wilkins2021functional2,
  title={Functional Renormalisation Group for Brownian Motion II: Accelerated Dynamics in and out of Equilibrium},
  author={Wilkins, Ashley and Rigopoulos, Gerasimos and Masoero, Enrico},
  eprint = "2102.04899",
    archivePrefix = "arXiv",
    primaryClass = "cond-mat.stat-mech",
  year={2021}
}

@article{aoki2002nonperturbative,
    author = "Aoki, Ken-Ichi and Horikoshi, Atsushi",
    title = "{Nonperturbative renormalization group approach for quantum dissipative systems}",
    eprint = "quant-ph/0205002",
    archivePrefix = "arXiv",
    reportNumber = "KANAZAWA-02-09",
    doi = "10.1103/PhysRevA.66.042105",
    journal = "Phys. Rev. A",
    volume = "66",
    pages = "042105",
    year = "2002"
}

@article{schoeller2009perturbative,
    author = "Schoeller, H.",
    editor = "Alkofer, Reinhard and Gies, Holger and Schaefer, Bernd-Jochen",
    title = "{A perturbative nonequilibrium renormalization group method for dissipative quantum mechanics}",
    eprint = "0902.1449",
    archivePrefix = "arXiv",
    primaryClass = "cond-mat.mes-hall",
    doi = "10.1140/epjst/e2009-00962-3",
    journal = "Eur. Phys. J. ST",
    volume = "168",
    pages = "179--266",
    year = "2009"
}

@article{jakobs2010nonequilibrium,
  title={Nonequilibrium functional renormalization group with frequency-dependent vertex function: A study of the single-impurity Anderson model},
  author={Jakobs, Severin G and Pletyukhov, Mikhail and Schoeller, Herbert},
  journal={Physical Review B},
 doi = "10.1103/PhysRevB.81.195109",
  volume={81},
  number={19},
  pages={195109},
  year={2010},
  publisher={APS}
}

@article{canet2003optimization,
    author = "Canet, Leonie and Delamotte, Bertrand and Mouhanna, Dominique and Vidal, Julien",
    title = "{Optimization of the derivative expansion in the nonperturbative renormalization group}",
    eprint = "hep-th/0211055",
    archivePrefix = "arXiv",
    doi = "10.1103/PhysRevD.67.065004",
    journal = "Phys. Rev. D",
    volume = "67",
    pages = "065004",
    year = "2003"
}

@article{canet2011nonperturbative,
    author = "Canet, Leonie and Chate, Hugues and Delamotte, Bertrand and Wschebor, Nicolas",
    title = "{Non-perturbative renormalisation group for the Kardar-Parisi-Zhang equation: general framework and first applications}",
    eprint = "1107.2289",
    archivePrefix = "arXiv",
    primaryClass = "cond-mat.stat-mech",
    doi = "10.1103/PhysRevE.84.061128",
    journal = "Phys. Rev. E",
    volume = "84",
    pages = "061128",
    year = "2011"
}

@article{gies2012introduction,
    author = "Gies, Holger",
    title = "{Introduction to the functional RG and applications to gauge theories}",
    eprint = "hep-ph/0611146",
    archivePrefix = "arXiv",
    doi = "10.1007/978-3-642-27320-9_6",
    journal = "Lect. Notes Phys.",
    volume = "852",
    pages = "287--348",
    year = "2012"
}

@article{zinn1975renormalization,
    author = "Zinn-Justin, Jean",
    editor = "Rollnik, H. and Dietz, K.",
    title = "{Renormalization of Gauge Theories}",
    reportNumber = "SACLAY-D.PH-T-74-88",
    doi = "10.1007/3-540-07160-1_1",
    journal = "Lect. Notes Phys.",
    volume = "37",
    pages = "1--39",
    year = "1975"
}

@article{mannella2022ito,
  title={It{\^o} versus Stratonovich: 30 years later},
  author={Mannella, Riccardo and McClintock, Peter VE},
  journal={The Random and Fluctuating World: Celebrating Two Decades of Fluctuation and Noise Letters},
  doi="10.1142/S021947751240010X",
  pages={9--18},
  year={2022},
  publisher={World Scientific}
}

@book{peskin2018introduction,
  title={An introduction to quantum field theory},
  author={Peskin, Michael E},
  year={2018},
  publisher={CRC press}
}
\end{document}